\documentclass[traditabstract,longauth]{aa}

\usepackage{graphicx}
\usepackage{txfonts}
\usepackage{natbib}
\usepackage[colorlinks=true,linkcolor=blue,citecolor=blue]{hyperref}

\usepackage{xspace}
\usepackage{longtable}
\usepackage{array}
\usepackage{lscape}
\usepackage{rotating}
\usepackage[labelformat=simple]{caption,subcaption}
\usepackage{gensymb}

\bibpunct{(}{)}{;}{a}{}{,}

\newcommand{\Teff}{\ensuremath{T_{e\!f\!f}}\xspace}

\newcommand{\degre}{\degree\xspace}
\newcommand{\mic}{$\muup$m\xspace}
\newcommand{\micsec}{$\muup$s\xspace}
\newcommand{\as}{\hbox{$^{\prime\prime}$}\xspace}
\newcommand{\lsd}{\hbox{$\lambda/D$}\xspace}

\begin{document}

\title{SPHERE: the exoplanet imager for the Very Large Telescope}

\author{
    J.-L. Beuzit\inst{1,2}  \and
    A. Vigan\inst{2} \and
    D. Mouillet\inst{1} \and
    K. Dohlen\inst{2} \and
    R. Gratton\inst{3} \and
    A. Boccaletti\inst{4} \and
    J.-F. Sauvage\inst{2,7} \and
    H. M. Schmid\inst{5} \and \\
    M. Langlois\inst{2,8} \and
    C. Petit\inst{7}
% ...
}

\institute{
    Univ. Grenoble Alpes, CNRS, IPAG, F-38000 Grenoble, France 
    \\ \email{\href{mailto:jean-luc.beuzit@lam.fr}{jean-luc.beuzit@lam.fr}} 
    \and
    Aix Marseille Univ, CNRS, CNES, LAM, Marseille, France 
    \and
    INAF - Osservatorio Astronomico di Padova, Vicolo della Osservatorio 5, 35122, Padova, Italy 
    \and
    LESIA, Observatoire de Paris, Université PSL, CNRS, Sorbonne Université, Univ. Paris Diderot, Sorbonne Paris Cité, 5 place Jules Janssen, 92195 Meudon, France 
    \and
    Institute for Particle Physics and Astrophysics, ETH Zurich, Wolfgang-Pauli-Strasse 27, 8093 Zurich, Switzerland 
    \and 
    Max Planck Institute for Astronomy, K\"onigstuhl 17, D-69117 Heidelberg, Germany 
    \and 
    ONERA (Office National d’Etudes et de Recherches Aérospatiales), B.P.72, F-92322 Chatillon, France 
    \and
    CRAL, UMR 5574, CNRS, Université Lyon 1, ENS, 9 avenue Charles André, 69561 Saint Genis Laval Cedex, France 
% ...
}

\date{Received 11 February 2019 / Accepted 20 September 2019}

\abstract{
    Observations of circumstellar environments that look for the direct signal of exoplanets and the scattered light from disks have significant instrumental implications. In the past 15 years, major developments in adaptive optics, coronagraphy, optical manufacturing, wavefront sensing, and data processing, together with a consistent global system analysis have brought about a new generation of high-contrast imagers and spectrographs on large ground-based telescopes with much better performance. One of the most productive imagers is the Spectro-Polarimetic High contrast imager for Exoplanets REsearch (SPHERE), which was designed and built for the ESO Very Large Telescope (VLT) in Chile. SPHERE includes an extreme adaptive optics system, a highly stable common path interface, several types of coronagraphs, and three science instruments. Two of them, the Integral Field Spectrograph (IFS) and the Infra-Red Dual-band Imager and Spectrograph (IRDIS), were designed to efficiently cover the near-infrared (NIR) range in a single observation for an efficient search of young planets. The third instrument, ZIMPOL, was designed for visible (VIS) polarimetric observation to look for the reflected light of exoplanets and the light scattered by debris disks. These three scientific instruments enable the study of circumstellar environments at unprecedented angular resolution, both in the visible and the near-infrared. In this work, we thoroughly present SPHERE and its on-sky performance after four years of operations at the VLT.
 }

\keywords{
    instrumentation: adaptive optics --
    instrumentation: high angular resolution --
    instrumentation: polarimeters --
    instrumentation: spectrographs --
    planets and satellites: detection
}

\maketitle

\section{Introduction}
\label{sec:introduction}

Even before the discovery of the first exoplanet, the study of circumstellar environments was already an important driver for designing instrumentation capable of detecting faint structures in the close vicinity of bright stars. It emphasized, from the beginning, the need for new observational capabilities. An emblematic example is that of the star $\beta$~Pictoris, where the detection of an infrared excess with IRAS \citep{Aumann1984} led to observations with early prototypes of stellar coronagraphs that enabled the discovery of a debris disk \citep{Smith1984}, extending up to several 100 au from the star. Another important step in the pioneering era was the detection of the cool brown dwarf Gl\,229~B thanks to high-angular resolution either from the ground with adaptive optics \citep{Nakajima1995} or from space \citep{Oppenheimer1995}. From these early observations, it was already clear that the path that would lead to significant progress in this field would be the combination of diffraction-limited large telescopes equipped with devices capable of suppressing or attenuating the starlight.

The performance of the first generation of adaptive optics-equipped, near-infrared (NIR) instruments on large ground-based telescopes like the VLT/NaCo \citep{Lenzen2003,Rousset2003} or Gemini/NIRI \citep{Hodapp2003,Herriot1998} was greatly improved as compared to previous systems on smaller telescopes \citep[e.g., ][]{Beuzit1997}. This prolific generation of instruments led to major discoveries from the exoplanet imaging point-of-view, such as the first direct image of a planetary-mass companion \citep{Chauvin2004}, to the first direct image of a multi-planet system \citep{Marois2008}, and the detection of a giant exoplanet in the disk surrounding $\beta$~Pictoris \citep{Lagrange2009}.

However, the high-contrast imaging limitations of these instruments were already foreseen during their design and early exploitation phases due to the limited number of degrees-of-freedom and cadence of their adaptive optics (AO) systems, and the limited contrast performance of their simple Lyot coronagraphs at very small angular separations. Major developments in coronagraphy and its interaction with adaptive optics in the early 2000s \citep{Rouan2000,Sivaramakrishnan2001,Sivaramakrishnan2002,Perrin2003,Soummer2003} as well as innovative observing strategies \citep{Racine1999,Marois2006} quickly paved the way toward a new generation of instruments entirely optimized for high-contrast observations. In particular, these studies highlighted the need for low wavefront errors (WFE) to minimize the residual speckles in the final focal plane.

With the objective of proposing a fully dedicated instrument for the ESO Very Large Telescope (VLT), two teams of European institutes led competitive phase A studies. This eventually resulted in a joint project, the Spectro-Polarimetic High contrast imager for Exoplanets REsearch (SPHERE). The instrument was developed by a consortium of eleven institutes\footnote{IPAG (Grenoble, France), MPIA (Heidelberg, Germany), LAM (Marseille, France), LESIA (Paris, France), Laboratoire Lagrange (Nice, France), INAF - Osservatorio di Padova (Italy), the department of astronomy of the University of Geneva (Switzerland), ETH Z\"urich (Switzerland), NOVA (Netherlands), ONERA (France), and ASTRON (Netherlands)} in collaboration with ESO. The development of SPHERE started in early 2006, with a preliminary design phase ending in September 2007 and a final design phase in December 2008. The assembly, integration, and testing of sub-systems at the various integration sites took almost three years until the fall of 2011, which wa followed by the final validation of the fully assembled instrument at IPAG until the end of 2013. SPHERE was finally shipped to the VLT in early 2014, saw its first light in May 2014, and was available to the ESO community in April 2015.

In parallel with the development of SPHERE for the VLT, two other important high-contrast imaging instruments were being developed for other eight~meter-class telescopes: the Gemini planet imager \citep[GPI;][]{Macintosh2006} and SCExAO for Subaru \citep{Guyon2010}. GPI is a facility instrument for Gemini South that was developed on a similar model as SPHERE, with more or less the same scientific goals and technical specifications for the AO and coronagraphy, but with different design choices partly due to observatory constraints (Cassegrain focus) and partly due to the available technologies upon its design and development phase. Both SPHERE and GPI followed a similar schedule, which ended in November 2013 with a first light of GPI \citep{Macintosh2014}, a few months in advance from SPHERE. GPI has since produced a wealth of discoveries, including a new directly imaged exoplanet around 51~Eri \citep[e.g., ][]{Macintosh2015,Hung2015,Konopacky2016}. SCExAO was conceived as a completely different facility; although, it also aimed to achieve better image quality and contrast, and a specific focus was placed on the innermost separations. Furthermore it was designed in a much more modular and incremental way, which enabled early on-sky validation of a variety of newly proposed techniques rather than a fixed design solution offered to a broad community \citep{Guyon2010,Guyon2011,Jovanovic2013,Jovanovic2016,Sahoo2018}. SCExAO has offered several instruments over the years. They were often offered on a shared risk basis and some of them have produced important scientific results like HiCIAO \citep{Hodapp2008,Tamura2009}. This instrument is currently available to the community with two focal plane science instruments \citep{Groff2017,Norris2015}, and it remains a flexible platform to quickly test new concepts.

It is also important to mention other instrumental developments on smaller telescopes or with slightly lower levels of specification that were done before or in parallel to SPHERE, GPI, and SCExAO. This includes the Lyot project \citep{Oppenheimer2004,Sivaramakrishnan2007}, the TRIDENT camera tested at the CFHT \citep{Marois2005}, Project P1640 for Palomar \citep{Oppenheimer2012,Hinkley2008,Hinkley2011}, the LBT first-light AO system \citep{Esposito2003}, the Keck AO system \citep{Wizinowich2000,vanDam2004} combined with the NIRC2 or OSIRIS instruments, the NICI instrument for Gemini \citep{Chun2008} or MagAO \citep{Close2013}. Each of these instruments or experiments can be considered as precursors in the maturation of technologies, concepts or processing algorithms that later demonstrated their full performance in SPHERE, GPI, and SCExAO.

In this paper we provide a detailed overview of the SPHERE instrument in its current state, following four years of operations at the ESO VLT. In Sect.~\ref{sec:tradeoffs} we first present the main trade-offs and design choices that drove the definition of the main functionalities of the instrument, in particular for what concerns the adaptive optics, the coronagraphs, and the science sub-systems. Then in Sect.~\ref{sec:global_system} we present the global system architecture and the performance of the common path interface (CPI), which is the heart of SPHERE and feeds all the science sub-systems. The two fundamental components required for high-angular resolution and high-contrast observations, namely the adaptive optics system and the coronagraphs, are described in Sect.~\ref{sec:saxo} and \ref{sec:coronagraphy} respectively. Then Sect.~\ref{sec:zimpol} is dedicated to ZIMPOL, the visible polarimetric imager of SPHERE, and Sect.~\ref{sec:ifs} and \ref{sec:irdis} are dedicated to IFS and IRDIS respectively, the two near-infrared instruments of SPHERE. Finally, the instrument control, the operations and the data reduction and handling are detailed in Sect.~\ref{sec:instrument_control_operations}. We conclude and propose some perspective for future upgrades in Sect.~\ref{sec:conclusions}.

\section{Main trade-offs and design choices}
\label{sec:tradeoffs}

The whole motivation and rationale behind the SPHERE project was to propose to a wide community on a large telescope, an instrument dedicated to high-image quality and high-contrast observation of bright targets. The primary scientific case is to study exoplanetary systems at large by offering imaging exploration capabilities of the outer giant planet population and circumstellar disks. This goal requires (i) a significant contrast performance improvement entering the detection capability in the planetary mass regime and (ii) the possibility to obtain such performance on a large target sample. Additionally, (iii) such an exquisite image quality should also be obtained over a field-of-view (FoV) sufficiently large to properly study circumstellar disks, and (iv) it should make possible to address many other secondary science goals from the same or derived observing modes. An underlying question immediately arises from this threefold main basis as to how ambitious the targeted performance should be. The answer is directly related to the project risk assessment in terms of technological readiness, complexity and system analysis of all the potential limitations in a new performance regime. It also requires a careful check of the compliance with items (ii), (iii) and (iv). 

In this section, we provide the major elements that have driven the SPHERE design in this context: once the primary goal is ensured, we present how the extension of observing modes in order to serve a wider case has been approached. We finally discuss a posteriori if these choices made at the time of the instrument design appeared to be validated after four years of operations on telescope.

\subsection{Key elements for high contrast capabilities}

The first major key for significant improvement with respect to previous generation imagers was to push the turbulence correction toward so called extreme AO. Lower residual wavefront variance can be obtained by increasing the number of actuators in the pupil and correcting at a faster rate. Technologies to progress on both fronts were not readily available at the time of the project design, but no showstoppers were identified. This motivation and state-of-the-art analysis motivated from the beginning a coordinated effort with industrial partners and ESO on the components identified as critical, namely the high-order deformable mirror (HODM), a new generation real-time computer (RTC), and fast and low-noise visible detector for wavefront sensing. The sensitivity of this sensor was critical since it directly impacted the AO sensitivity of the instrument.

We can see from this early basic dimensioning definition a necessary trade-off between the ultimate performance goal and sensitivity. The selected numbers of 41$\times$41 actuators at 1.5\,kHz servo-loop frequency provided an expected balanced WFE <60\,nm rms up to a target magnitude $R = 9$. This dimensioning was estimated to remain within a reasonable technological risk. Pushing the technological specifications further, in particular in terms of correction speed, would have entered a regime of higher development risk for an applicability restricted mostly to the brightest stars. Risk was not only technological: keeping the performance at this level of WFE imposes to align the specifications of all contributors of the WFE budget. Consistently, the design did include an AO calibration plan much stricter than for previous instruments. This plan included daily registration and updates in an automatic manner, and it also imposed severe accuracy for the conjugation between the AO-controlled corrective devices and sensors.

This led to the second important design driver for high-contrast: the instrument overall opto-mechanical stability. This stability is not only important for the calibration reliability, it is also directly required for efficient coronagraphy and differential imaging in order to distinguish any companion signature from the residual stellar halo. The specifications on these different aspects (AO correction, opto-mechanical stability, coronagraphy, differential imaging) are intimately coupled to define the resulting overall contrast performance. As an example, coronagraphy and spectral differential imaging in the integral field spectrograph (IFS) act together for high-contrast. However, while a wider spectral bandpass benefits the IFS differential imaging, it constrains and reduces the ultimate intrinsic performance of the coronagraph device, and finally both of them depend on the level of optical defects and their stability. 

Exploring the space of various specifications was approached by extensive numerical simulations around some first guesses of achievable optical quality and stability. These simulations did include a rough representation of data reduction to take into account at least the main dependencies of the differential imaging capabilities. This specification work resulted into a WFE budget distributed according to the location in the optical path (in particular with respect to the coronagraph), the aberration modes (from tilt and focus, to medium frequencies affecting all the stellar corrected halo, and up to high frequencies), stability over time, and chromaticity. A specific attention was devoted to the optical beam stability: it does not only guarantee on long timescales the alignment needed for coronagraphic extinction at any given time, it is also needed with an even finer accuracy within a typically 1-hour long observation sequence to guarantee the performance of speckle subtraction thanks to angular differential imaging \citep[ADI;][]{Marois2006}. Pupil stability during observation was ensured in rotation by a derotator early in the optical train, and in translation by a dedicated pupil tip-tilt mirror (TTM) actively controlled in closed-loop. Stability requirements also lead to add a dedicated sensor in a static setup very close to the coronagraphic focal plane operating at NIR wavelength. It monitors and corrects for any NIR image drift on the coronagraphic mask, either due to opto-mechanical variations or chromatic effects with respect to the visible WFS. No moving optical device is located before the coronagraph, except for the required derotator, the corrective mirrors (TTM and HODM), and the atmospheric dispersion correctors (ADCs).

Atmosphere refraction correction is first needed not to degrade the diffraction-limited resolution in broadband filters. More accurately, it is also mandatory to ensure a good on-coronagraph centering at any observing wavelength, and to guarantee that the beams hit the same optical footprints for every spectral channels involved in spectral differential imaging. This level of performance cannot be obtained with a single device from visible to NIR: two sets of ADCs are needed to cover the whole range. Whereas the goal of observing at high-image quality and stability both in Visible and NIR benefited from the same AO design choices, this requirement presents here its most important impact: the subsequent need for two ADCs induces some WFE which are unseen by the AO and variable in time. It additionally means that the ADCs cannot be located early in the optical train: the optical beam footprint on the surfaces before the ADC induces corresponding chromatic and variable WFE. For this reason, the requirement for observing capability in both visible and NIR induces a limit in ultimate performance. Its level was estimated to remain tolerable up to performance goals of typically 10$^{-6}$ but it would probably be a show-stopper for higher contrasts. 

Spectral differential imaging \citep[SDI;][]{Racine1999,Sparks2002,Thatte2007} is the additional processing step for high contrast in NIR. This critical step was included as a primary science requirement from the beginning, for speckle vs. companion discrimination in two main regimes: either in the comparison of nearby spectral channels selected close to molecular absorption features expected for the most interesting cool companions (T-type), or, probably less easily but for a wider range of companions, over a spectral range wide enough to identify the speckle separation shift with wavelength. This requirement drives the chromatic WFE budget. Within this budget, defects on optical surfaces far from the pupil plane (in comparison to the Talbot length) translate through Fresnel propagation into chromatic phase and amplitude defects. They eventually appear as wavelength-dependent artifacts on the final images. A dedicated analysis was first performed in the GPI team with conclusions on the derived optical surface specifications \citep{Marois2008b}. As for the point made on ADCs before, such an effect would certainly become dominant for contrast goals better than 10$^{-6}$ but the combination of the pupil size, location of optical surfaces (very few far from the pupil) and optical surface quality remained within the range of the performance goal.

In order to implement SDI, two main approaches were considered: dual-band imaging and a micro-lens based integral field spectrograph (IFS). We did not retain potential alternatives like a slicer-based IFS, a pupil-dispersed multiple band imager, or an integral field Fourier-transform spectrograph. The IFS presents the strong advantages over dual-band imaging of both a richer spectral information and an image sampling in a common focal plane before dispersion. The latter prevents differential aberrations between spectral channels for better channel-to-channel comparison, assuming a clean lenslet to detector flux propagation and signal extraction. This option was then considered with a specific attention to the spatial and spectral sampling, and a low level of cross-talk, which led to the so-called BIGRE design \citep{Antichi2009}. With this design we reached a good trade-off between field of view (FoV) and required spectral information (spectral range times spectral resolution). This design is significantly different from the one adopted by GPI and SCEXAO which consist of an integral field unit made of a single array of micro-lenses with the option of a fine-tuned pin-hole mask at their foci \citep{Peters2012,Peters-Limbach2013}.

The dual-band imaging approach is conceptually simpler, and was already implemented earlier in previous high contrast imagers \citep[e.g., ][]{Marois2005}. Unlike IFS, DBI necessarily requires some distinct optical surfaces that will induce differential aberrations. Even though such aberrations are expected to be quite static and therefore accessible to calibration, their combination to upstream (variable) common defects leads to effects on the image that are very difficult to calibrate. The baseline assumption is an intrinsic performance limitation below 10$^{-6}$ for differential WFE $< 10$\,nm rms \citep{Dohlen2008b}.

Considering that both SDI concepts are not excluding each other, it was decided to include both of them on the Nasmyth platform in a mutually consolidating manner. On top of risk mitigation, their complementarity allows to finally obtain a larger FoV with IRDIS, the IFS spectral information in the inner part, and the multiplex advantage of a larger instantaneous spectral range with the simultaneous use of these instruments in complementary bands. This full advantage appears achievable up to 10$^{-6}$ contrast over a spectral range of an octave (e.g., 0.95--1.7\,\mic). Aiming at higher contrast would probably require to restrict the spectral band, for instance to select a more performing but also more restrictive coronagraph. An additional and important advantage of adding a NIR imaging beam (IRDIS) to the IFS is that it opens the possibility of additional observing modes at low cost. We will see how beneficial such modes can be  to a wide range of other science cases.

A needed addition for the visible, without degrading the AO sensitivity when observing in NIR, is the beam splitting between the WFS and science camera ZIMPOL for optimal photon share depending on the observing case. A moving part in between the corrective mirror and the sensor of the AO is a possible source of degradation of the AO performance if the corresponding HODM-to-WFS registration and non-common path aberrations (NCPA) were poorly controlled. Once well identified, this risk could be mitigated through the calibration plan, and some possible restriction on the operations. On top of providing a very high-angular resolution close to the diffraction limit in the visible (<20 mas), the visible camera also provides high-contrast capability for reflected light in polarimetric differential imaging (PDI) with a very high differential accuracy thanks to the CCD-based ZIMPOL principle \citep{Povel90,Schmid2018}. The most polarimetric-unfriendly component is definitely the unavoidable K-mirror derotator in the common path. Even though the primary goal for ZIMPOL polarimetry is differential measurements, the polarimetric impact of the derotator has been handled at first order by a dedicated polarimetric calibration scheme and half-wave plate. This component is retractable and has thus no impact on NIR observations. Finally, the operational limit was reached in considering how to optimize simultaneous observations from visible to NIR. Some observations could actually be obtained simultaneously in visible and NIR on the same source but many difficulties arise in such conditions, starting from contradictory centering constraints for coronagraphic observations, different photon sharing trade-offs for the WFS, very different observing duration, or derotator control. Such a VIS+NIR observing mode appears seldom useful for high-performance observations in both channels to be worth the significant operational complexity.

\subsection{Wide exploitation of the SPHERE image quality and contrast}

\begin{table*}
    \centering
    \caption{Comparison of instrument specifications with performance routinely observed on sky.}
    \label{tab:perf_goals}
    \begin{tabular}{llll} 
    \hline
    Parameter & Specification & On-sky experience  & Reference or comment \\
    & [Goal] & & \\
	\hline \hline
    \multicolumn{4}{c}{Driving specifications and goals} \\
    \hline
    \multicolumn{4}{l}{\emph{\textbf{NIR contrast on bright targets}}} \\
    DBI contrast at 0.5\as for H<8       & 13.25 [15.75]                    & typical $\Delta_{\mathrm{mag}}$=13.5 & Fig.~\ref{fig:IRDIS_IFS_contrast} \\
                                        &                                  & best $\Delta_{\mathrm{mag}}$=15.0    & \citet{Vigan2015} \\
    IFS contrast at 0.5\as for J<8       & 15.00 [20.00]                    & typical $\Delta_{\mathrm{mag}}$15    & Fig.~\ref{fig:IRDIS_IFS_contrast}, Appendix~\ref{sec:apdx:noise_model_ifs} \\
                                        &                                  & best $\Delta_{\mathrm{mag}}$=16.5    & Sect.~\ref{sec:ifs:on-sky_results};        \citet{Vigan2015} \\
    IFS spectral information            & 0.95--1.35~\mic                  & ok	                         & e.g., \citet{Zurlo2016}, \\
                                        & or 0.95-1.65~\mic                &                             & \citet{Samland2017}, ... \\
    \hline
    \multicolumn{4}{l}{\emph{\textbf{VIS polarimetric contrast}}} \\
    Contrast at 1\as in 4 hrs on I=2.5   & $\Delta_{\mathrm{mag}}^{\mathrm{pol}}$= 21.2 & 17 at 0.2\as      & Sect.~\ref{sec:zimpol_high_contrast_pol}; Hunziker et al. (in prep.)	\\
                                        &                                  & >20 at 1\as                   & \\
    \hline
    \multicolumn{4}{l}{\emph{\textbf{NIR survey capability on large samples}}} \\
    IRDIS+IFS simultaneity	            & required                         & used routinely               & \\
    AO ultimate performance             & up to R=9	                       & ok                           & Fig.~\ref{fig:SAXO_PerfMagnitud}	\\
    AO robust operations	            & up to R=12                       & overdone: R=14	              & Fig.~\ref{fig:SAXO_PerfMagnitud}	\\
    Throughput                          & >15\%                            & >20\%                        & Fig.~\ref{fig:transmission_bands} \\
    Elevation                           & 5\degre--60\degre                & ok	                          & \\
    \hline
    \multicolumn{4}{c}{Additional observation capabilities without impact on driving specifications} \\
    \hline
    \multicolumn{4}{l}{\emph{\textbf{Near-infrared}}} \\
    Imaging in various filters          & $Y$ to $K$, BB, NB               & ok                           & \\
    Dual-polarimetry imaging contrast   & 9.25 [11.3] at 0.1\as             & 13.0 at 0.2\as                & \citet{vanHolstein2017},  high-efficiency \\
                                        & 11.75 [14.25] at 0.5\as           & 15.0 at 0.5\as                & in $J$-band, but strong dependency on         \\
                                        &                                  &                              & derotator orientation        \\
    Long-slit spectroscopy              & LRS (R$\sim$50 $Y$--$Ks$)        & ok                           & Fig.~\ref{fig:PZtel_LSS} \\
                                        & MRS (R$\sim$350 $Y$--$Ks$)       & ok                           & Fig.~\ref{fig:HR3549_LSS}; \citet{Hinkley2015} \\
    \hline
    \multicolumn{4}{l}{\emph{\textbf{Visible}}} \\
    High-angular resolution imaging     & NB to very BB  & ok              & Sect.~\ref{sec:zimpol_hr_imaging}; \citet{Schmid2018} \\
    High-contrast imaging               & best effort    & > 6.5 at 0.1\as  & Sect.~\ref{sec:zimpol_hc_imaging} \\
    (non-polarimetric) \\
    \hline
    \end{tabular}
    \tablefoot{All contrast values are specified in magnitudes and at 5$\sigma$.}
\end{table*}

The main features of the design have been driven by a high-contrast capability in NIR and diffraction-limited polarimetric imaging in VIS. We did mention some design choices but each of them was associated to the technological and system risk assessment, or to the potential impact on sensitivity. On the opposite, the targeted performance was not restricted or reduced in order to fulfill secondary drivers. This justifies the qualification of SPHERE as an instrument dedicated to high-contrast. From this basis, we further explored if and how this baseline could also benefit a wider astronomical science case, a wider user community, with which observing modes, keeping in mind that it should not degrade or restrict the primary goals. We see that indeed a number of additional observing modes were relevant, usually at moderate cost (but essentially operational and control complexity). Some of them were offered as a side-product from the primary baseline without guarantee or full system analysis. 

The first question was the maximum spectral range in the NIR where high-contrast can be obtained, for both IRDIS and IFS. The initial and minimum baseline was a spectral coverage from $Y$- to $H$-band, with a survey mode with IFS operating in $YJ$ and IRDIS in the $H$-band. Thermal background becomes an issue in $K$-band. Observations in $K$-band were included within IRDIS observations with the condition that, if trade-offs were needed, shorter wavelengths were to be kept optimal. Observations at longer wavelengths ($L$-band) are obviously very interesting but not included because this would clearly imply major modifications to the instrument design as a whole, starting with a complete cryogenic environment. If it is clear that $L$-band observations would gain a lot from better AO correction and would provide great results for exoplanet studies, the derived system analysis would certainly lead to completely different challenges, trade-offs and solutions. Also, at the time of the instrument design, the high-contrast performance analysis was not so clear to quantify the performance gain with respect to existing $L$-band instrument with already high-image quality (Sr > 70 \%) and facing background and sensitivity issues. On the IFS side, an additional mode extending the simultaneous spectral range to Y-H was studied. While it was not possible to guarantee the simultaneous observation with IRDIS in $K$-band, the interest of IFS-only product was considered high enough to justify this mode: a continuous spectral coverage from 0.95 to 1.65\,\mic is obtained, with the corresponding interest for discrimination of speckles vs. companions down to a shorter separation, while keeping an acceptable background level and spectral resolution. This mode was added, with parallel observation with IRDIS in $K$-band but with no guarantee on performance.

For IRDIS, the primary observing mode is dual-band imaging \citep[DBI;][]{Vigan2010} to look very deep for differential flux between the two simultaneous images, over a $\sim$4.5\as circular FoV, and through filter pairs probing molecular absorption features in NIR from $Y$- to $K$- band \citep[e.g., ][]{Vigan2016}. The same set-up but no dual-band filter provides classical imaging capability, for either broadband (BB) or narrow band (NB) filters \citep{Dohlen2008,Langlois2010a}. BB imaging can also be very useful for high-contrast thanks to ADI in particular for faint targets and/or L-type companions without strong absorption features. Lower contrast is expected from NB imaging, but this mode is obtained at low cost and complexity as soon as the usual optical ghosts associated to the complex coatings of such filters are acceptable, with known locations and intensity in the field\footnote{Fully documented here for IRDIS: \url{https://www.eso.org/sci/facilities/paranal/instruments/sphere/inst/sphere-irdis-narrow-band-psfs-and-ghosts.html}}.

Another mode derived from DBI is dual-polarimetry imaging \citep[DPI;][]{Langlois2010b}, with a very high interest to detect (and discriminate from the zero or uniform polarization of the speckle halo) reflected light on faint circumstellar disks. The relatively large IRDIS FoV is very interesting for a number of disks. The easy part of this mode implementation is how similar it is to DBI: coronagraphy, image quality and stability, very low differential WFE between beams are all the same very beneficial to DPI, just replacing the dual-band filter by two orthogonal linear polarizers. This optical pair is completed by a rotating half-wave plate (HWP) to allow polarization orientation selection and swapping for a complete Stokes Q and U polarimetric cycle measurement. These minor additions offer high-contrast capabilities to detect small variations of polarized flux in the FoV, very appropriate for detection and morphology of faint disk in reflected light. Above such a polarimetric imaging capability, a complete analysis of the instrument polarimetric properties would have required much more effort. It should have included, in particular, the level of instrumental polarization and efficiency as a function of mode and pointing direction, second-order effects of its variability within a polarimetric cycle, cross-talk due to misalignment of components, or even absolute polarization accuracy. This work was out of reach of the team resource at the time of the design, and since the primary features of the instrument were already driven by DBI, and the addition of polarizers was identified as beneficial for relative (morphology) measurements, the mode was included as such and did indeed allow spectacular early results on disk morphologies \citep{Benisty2015,Ginski2016,Stolker2016,vanBoekel2017,Garufi2017}. The complete instrumental polarimetric analysis was performed later based on both internal and on-sky data (\citealt{vanHolstein2017}; van Holstein et al. in prep.; de Boer et al in prep.).

NIR coronagraphic long-slit spectroscopy \citep[LSS;][]{Vigan2008} also appeared to be interesting and fully compatible with SPHERE CPI and IRDIS baseline design. In particular, they already included the access to a stable focal plane for the slit (coronagraphic focal plane), the capability to finely control the field position and orientation, and the access to a cold pupil plane to insert a dispersive element within IRDIS. This mode could then be added with its specific optical components, and corresponding calibration scheme. This mode actually provided additional exoplanet characterization capabilities also demonstrated with very early observations \citep{Hinkley2015}.

Finally, a sparse aperture masking (SAM) mode was also implemented as an addition of the appropriate pupil masks in the Lyot stop wheels located in the IFS and IRDIS arms \citep{Cheetham2016}. Similarly to the DPI mode, this observing mode benefits from all the high-level specifications deriving from the DBI mode with very little impact on the hardware. The mode was not initially supported in the instrument software at first light, but was implemented and offered later on.

For visible observations, the baseline design has been driven by high-resolution and high-contrast differential polarimetric imaging. Priority was set to the observation capability of very bright sources, looking for tiny polarimetric signal from reflecting exoplanets (see section \ref{sec:zimpol}). This includes the ability to handle the huge photon rate of the star in broadband without neutral density, and very fine polarimetric differential accuracy. Such requirements had a direct impact on the detector design and readout modes, with a very fine spatial sampling, a corresponding huge well capacity and the fast charge-shifting synchronized with the polarization modulation. To keep all the photons after polarization selection, the camera is duplicated to collect also the other polarization. The two beams, providing simultaneously but independently consistent polarimetric measurements can provide either the redundant information in the same filters, independent information in different bands, or complementary image in contiguous filters with differential spectral imaging capability around emission lines. Variations of the detector readout modes are selected in the case of fainter targets (with a lower detector capacity and noise) or when polarimetric information is not needed. ZIMPOL then also provides an imaging mode, when retracting from the beam the polarization control components, with similar or distinct filter on the two beams, and in pupil or field-stabilization mode.

\subsection{A posteriori look-back on the main choices and risk management}

If the logic of the design trade-offs and choices, at the time of the instrument development, has been presented above, four years of operations on the telescope for a wide user community offers now the opportunity to discuss this initial approach.  

Generally speaking, the system at large did reach completion; it could be integrated on the telescope and within operation scheme efficiently (within four two-week commissioning runs), reaching the performance specifications. The system could be operated and maintained by the observatory team and, from the beginning, a wide community, not restricted to experts or the instrument builders, did propose a variety of observations and produce new science results \citep{Hardy2015,Hinkley2015,Kervella2015,Csepany2015,Xu2015}. The community also actually used the whole range of the instrument modes benefiting from the high image quality and stability. Table~\ref{tab:perf_goals} provides an overview of the high-level specifications and the corresponding actual on-sky experience.

More specifically, regarding the primary goals, the high contrast performance level was reached close to 10$^{-6}$ in the survey-efficient mode combining IFS+IRDIS simultaneously and with an AO-sensitivity even better than the specified $R = 9$ limit. The complex set of specifications defined earlier appears thus sufficient to the performance goal. On the other hand, we should also note a posteriori that none of the driving specifications could have been significantly relaxed without directly degrading the performance (AO general dimensioning, pupil control, derotator, optical WFE budget, opto-mechanical stability, NIR auxiliary sensor, calibration scheme, ...). 

In terms of contrast, the minimum specification level could be reached, but not quite the optimistic goal level. This goal was set in order, whenever possible, to allow for further future improvement. Correspondingly, some explorations and provisions were made not necessary, such as trying to approach an even higher level of NIR detector flat-field accuracy (up to 10$^{-4}$), achieving better on-coronagraph centering (sub-mas absolute accuracy) or time-variable registration, and the correction of NCPA. Such levels of specifications never appeared to be required as they were masked by other limitations, whereas the general specifications on optical quality, opto-mechanical stability, and turbulence corrections have obviously benefited to all observing modes. Concerning sensitivity, the good performance of AO on faint stars appeared to encourage a number of science programs on faint stars. If we had put more scientific weight on fainter targets, rather than pushing the performance limit on the brightest targets, we could have modified some aspects, such as the visible detector sampling and read-out modes, or the specification on NIR IFS thermal background.

Going further, the following question is whether the project approach could have been more ambitious toward contrast in the $10^{-7}$ to $10^{-8}$ range. A number of indicators show that no, a more ambitious goal at that time without earlier demonstrator would have likely lead to some failures and/or more restricted science output. First of all, we should note that if all the new technological developments (WFS detector, HODM, RTC, visible ZIMPOL detector read-outs) succeeded, each of them remained on the project schedule critical path. This added not only delays but also complexity in the assembly and test phase. The HODM also showed some dead actuators and some features on the shape-at-rest dependency with time and temperature, that required late mitigating adjustments on the instrument design (adjusted Lyot stop, and remotely controlled active toric mirror). This shows the limits of including the necessary new technologies for breakthrough performance within a fixed-design instrument delivered from a building consortium to an operating observatory such as ESO, rather than the incremental development scheme adopted by SCExAO. Second, apart from technology maturity, on the system analysis level, we mentioned how many specifications of the system design are intimately coupled for a given contrast level. They need to be pushed in a balanced manner. Entering a new regime for contrast a factor of 10 or 100 higher requires to deal with many new effects that could be neglected up to now. It is clear that this requires the experience feedback obtained now and could not have been safely addressed ten years ago. Two examples can illustrate this: (i) the local thermal exchanges between the telescope mechanics and air within the dome (so-called low wind effect) show a significant effect on image quality that were not at all expected, and (ii) the approaches for finer sensing of low level pre-coronagraph WFE have significantly improved in the last decade. On top of these two effects, entering a better contrast regime will also require to revisit the balance between AO temporal error, chromatism, coronagraph device performance and spectral bandpass and finally signal processing techniques. The system analysis was not mature enough at the time of the initial SPHERE design but can be re-considered now, based on the current experience, as will be mentioned in the last section of this paper. 

\section{Global system \& CPI}
\label{sec:global_system}

\subsection{High-level technical requirements}

\begin{figure}
    \centering
    \includegraphics[width=0.5\textwidth]{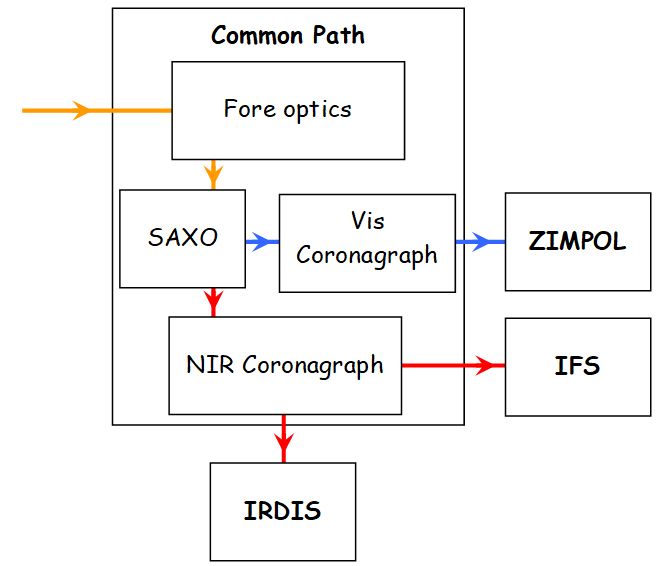}
    \caption{Global concept of SPHERE. The block diagram indicates the four sub-systems and main functionalities within the common path sub-system. Optical beams are indicated in orange for VIS+NIR, in blue for VIS and in red for NIR.}
    \label{fig:SPHEREconcept}
\end{figure}

The technical requirements for SPHERE are based on the top-level science requirements set to achieve the scientific goals related to exoplanet detection at the VLT: (i) detect faint objects such as planets and brown dwarfs in the close vicinity of bright stars, (ii) characterize their spectral features at low resolution through the $Y$-, $J$-, $H$- and $Ks$- atmospheric windows, covering wavelengths in the range 0.95 to 2.32\,\mic, (iii) detect and characterize light reflected off planets very close to some nearby bright stars, through accurate relative polarimetry in the visible, and (iv) study other circumstellar features such as disks around nearby bright stars.

In order to offer the required versatility and scientific complementarity, we have defined such an instrument to consist of four main parts: a common path system, including extreme AO and coronagraphs, a NIR differential imaging system, a NIR integral field spectrograph, and a visible polarimetric imager system. These functions define the four subsystems of SPHERE, the Common Path and Infrastructure (CPI), the Infrared Dual Imager and Spectrograph \citep[IRDIS;][]{Dohlen2008}, the Integral Field Spectrograph \citep[IFS;][]{Claudi2008}, and the Zurich Imaging Polarimeter \citep[ZIMPOL;][]{Schmid2018}, respectively, as illustrated in \ref{fig:SPHEREconcept} showing the top-level block diagram of SPHERE.

\subsection{Design}

\begin{figure*}
    \centering
    \includegraphics[width=0.8\textwidth]{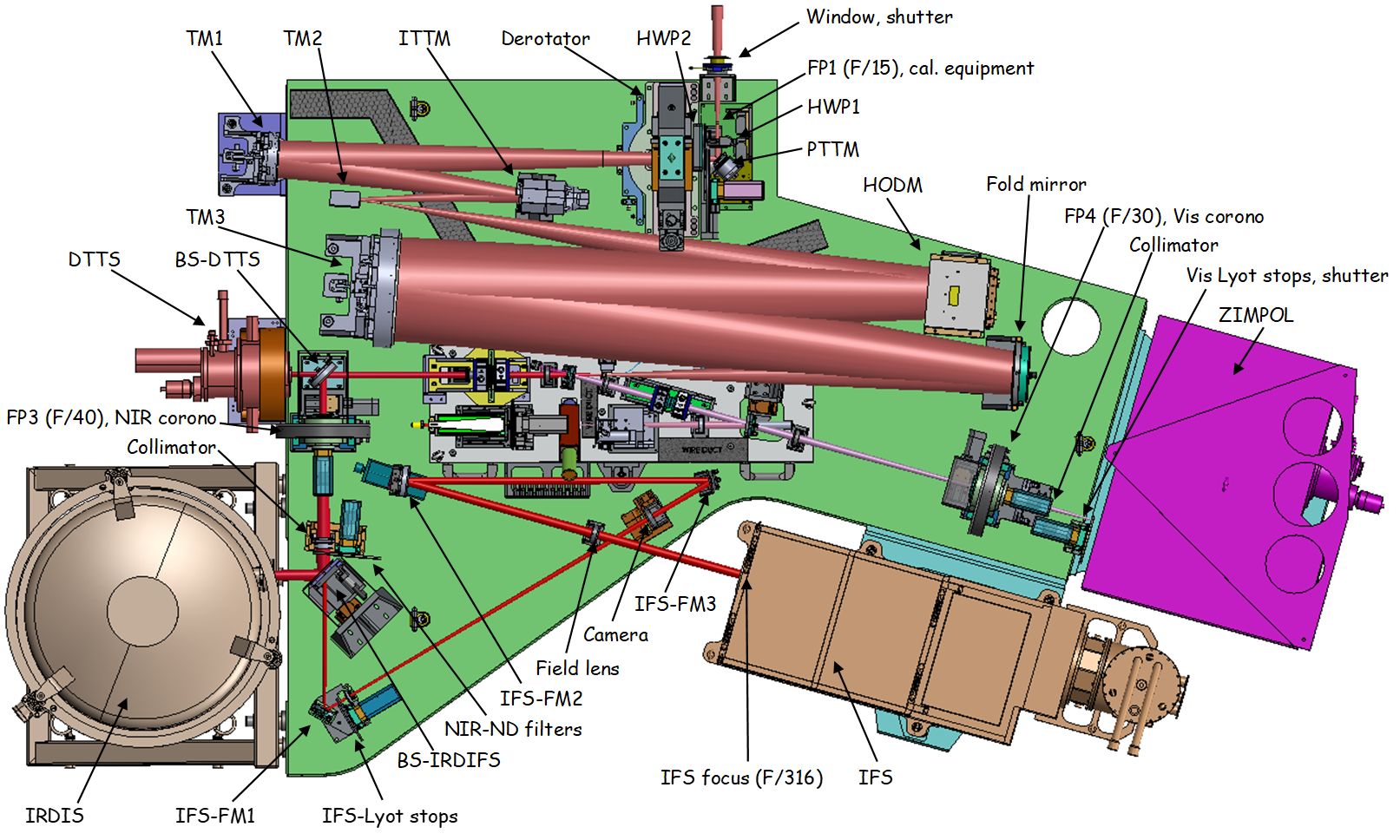}
    \caption{CAD top view of the SPHERE bench with most of the opto-mechanical components labeled. The CPI sub-bench that supports the VIS wavefront sensor and the ADCs is detailed in Fig.~\ref{fig:CPIsubbench}. The light from the telescope arrives from the top in this view.}
    \label{fig:SPHEREglobal}
\end{figure*}

\begin{figure}
    \centering
    \includegraphics[width=0.5\textwidth]{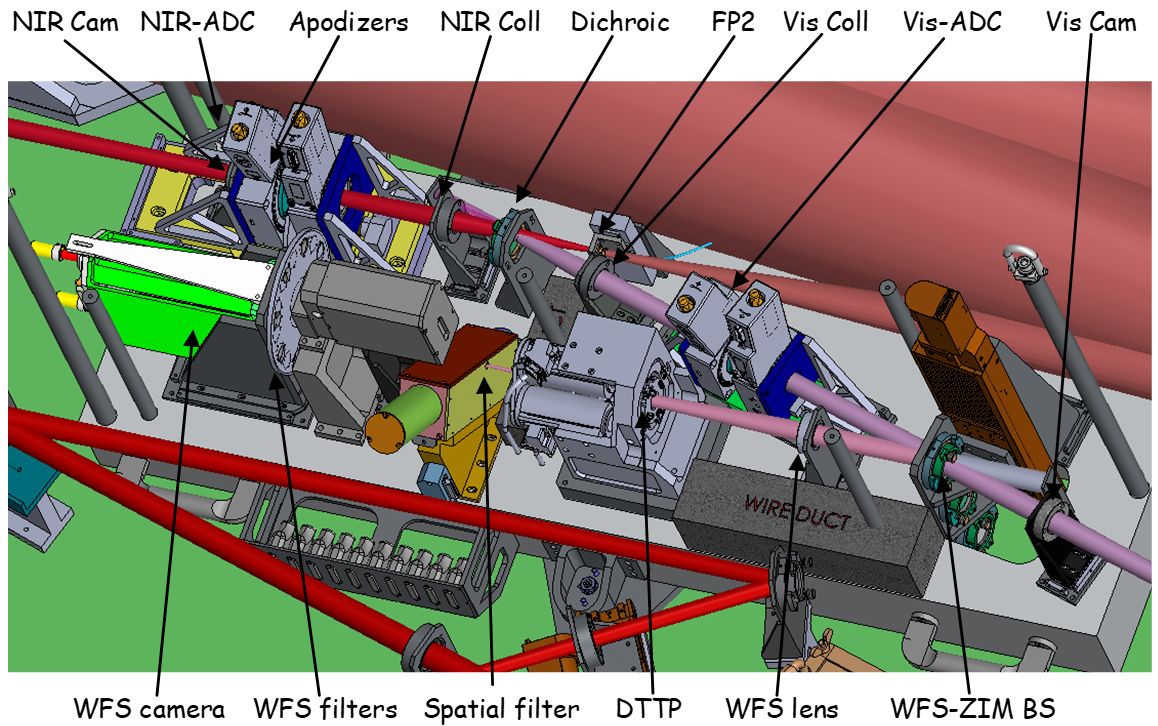}
    \caption{Detailed view of the CPI sub-bench that supports the spatially-filtered Shack-Hartman, the VIS-NIR dichroic beam splitter, the VIS and NIR ADCs, and the NIR apodizer wheel.}
    \label{fig:CPIsubbench}
\end{figure}

\begin{figure}
    \centering
    \includegraphics[width=0.5\textwidth]{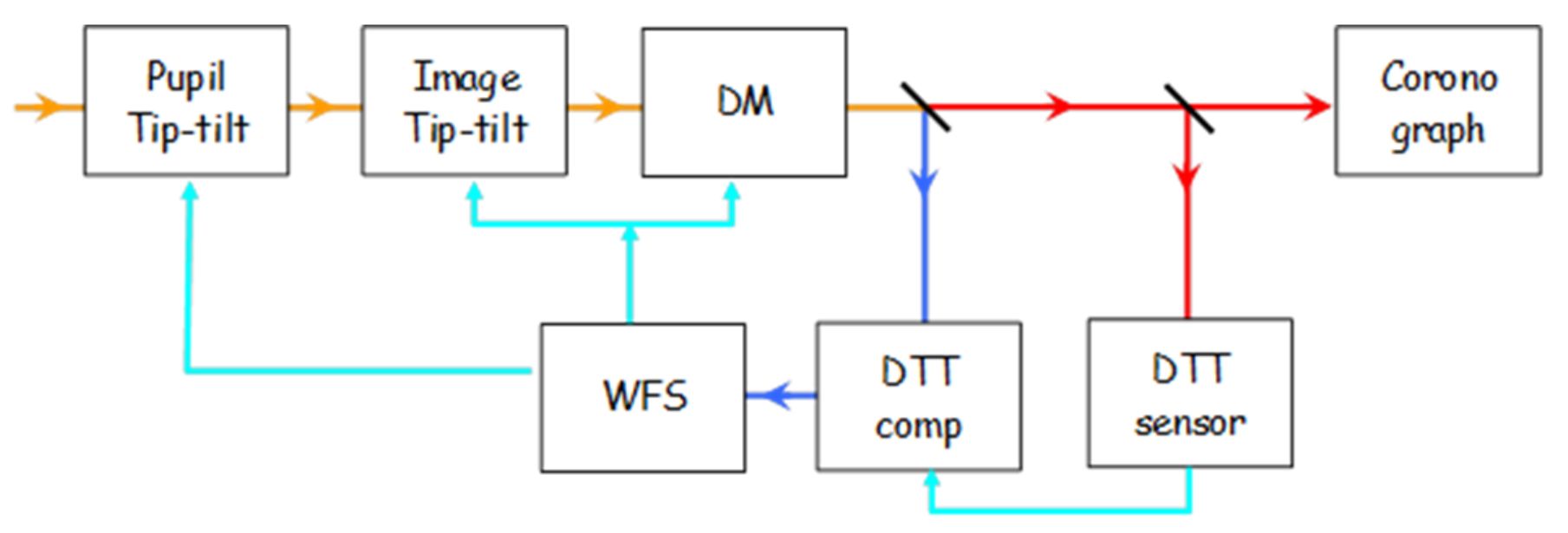}
    \caption{Block diagram of the AO components and loops in SPHERE. As in Fig.~\ref{fig:SPHEREconcept}, VIS+NIR light is in orange, VIS light is in blue and NIR light is in red. The control loops are in light-blue.}
    \label{fig:AOloops}
\end{figure}

A more detailed conceptual overview of the instrument is shown in Fig.~\ref{fig:SPHEREglobal}. After a busy section near the telescope focal plane, containing a pupil stabilization mirror (PTTM) and a derotator allowing either pupil or image stabilization, as well as calibration sources, the common path creates two successive pupil planes for the fast image tit-tilt mirror (ITTM) and the HODM by the aid of three toroidal mirrors, TM1, TM2, and TM3. These are produced using a stressed polishing technique \citep{Hugot2012} offering excellent surface polish and avoiding the mid-frequency surface errors associated with robotic polishing techniques. A second busy section is located around the focus at the exit of the relay (Fig.~\ref{fig:CPIsubbench}). A dichroic beam splitter separates visible light (reflected) from infrared light (transmitted). Visible light is transmitted through an atmospheric dispersion corrector (ADC) before hitting an exchangeable beam splitter separating light going to the wavefront sensor (WFS) from light going to the ZIMPOL polarimetric camera. Three positions are available: a mirror, sending all light toward the WFS for use during NIR observations, a gray beam splitter, sending 80\% of the light toward ZIMPOL, and an H$\alpha$ splitter sending only light around the H$\alpha$ line toward ZIMPOL. Reflected light enters the WFS, where it encounters a tiltable plane-parallel plate, the differential tip-tilt plate (DTTP), allowing fine control of wavefront tip-tilt, before arriving at a focus where an adjustable diaphragm allows for spatial filtering to minimize aliasing errors \citep{Poyneer2006}. Following this focus, the beam is collimated and a pupil projected onto the Shack-Hartmann microlens array.

For the visible beam, the light transmitted toward ZIMPOL passes through coronagraphic focal and pupil plane masks before entering the polarimetric camera (see Sect.~\ref{sec:zimpol}). The infrared beam, after passing through an infrared-optimized atmospheric dispersion corrector and coronagraphic masks, hits a second exchangeable beam splitter separating light between the IRDIS dual-band imaging camera and the IFS spectrograph. A mirror allows for an IRDIS only observations, and two dichroic separators allow observing simultaneously with IFS and IRDIS, the \texttt{DIC-H} sending the $YJ$-bands to the IFS while sending the $H$-band to IRDIS, and the \texttt{DIC-K}, sending the $YJH$-bands to the IFS and the $K$-band to IRDIS. It is important to note here that the \texttt{DIC-K} splitter was originally conceived for science observations with the IFS only, in coherence with the limited coronagraphic bandwidth, allowing IRDIS imaging only for navigation purposes. While this explains the relatively poor $K$-band efficiency of this mode, the mode was nevertheless commissioned and offered as an observing mode which has proven useful and successful.

Just before the infrared coronagraphic mask, a gray beam splitter separates a few percent of the light, which is sent to a technical camera, the differential tip-tilt sensor (DTTS), sensing precisely the position of the focal spot at a rate of 1\,Hz (see Sect.~\ref{sec:saxo}). This signal is fed back to the DTTP located in the wavefront sensor path, thus entering the AO loop (Fig.~\ref{fig:AOloops}) and allowing for the compensation of any slow movement between the coronagraph focus and the visible wavefront sensor due to thermal movements, residual differential dispersion, etc.

\subsection{Performance}

\begin{figure}
    \centering
    \includegraphics[width=0.5\textwidth]{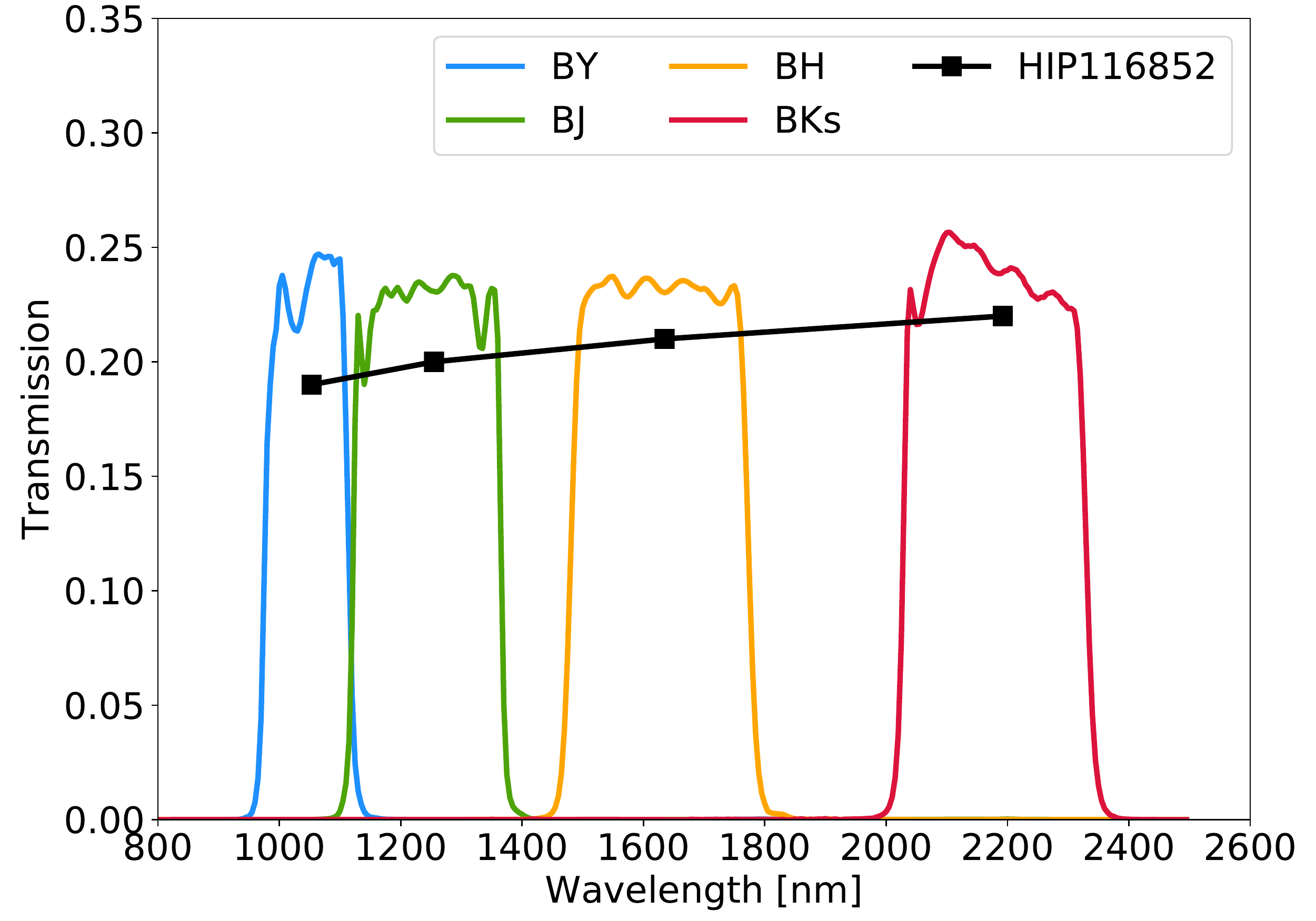}
    \caption{Comparison of predicted transmission curves with on-sky measurements. The predicted transmission curves are obtained from as-built measurements of all the SPHERE optics and filters. The on-sky measurements were obtained in open-loop on 22 August 2018 on HIP\,116852 without any coronagraph. The measurements are the average of the two IRDIS channels.}
    \label{fig:transmission_bands}
\end{figure}

\begin{figure}
    \centering
    \includegraphics[width=0.5\textwidth]{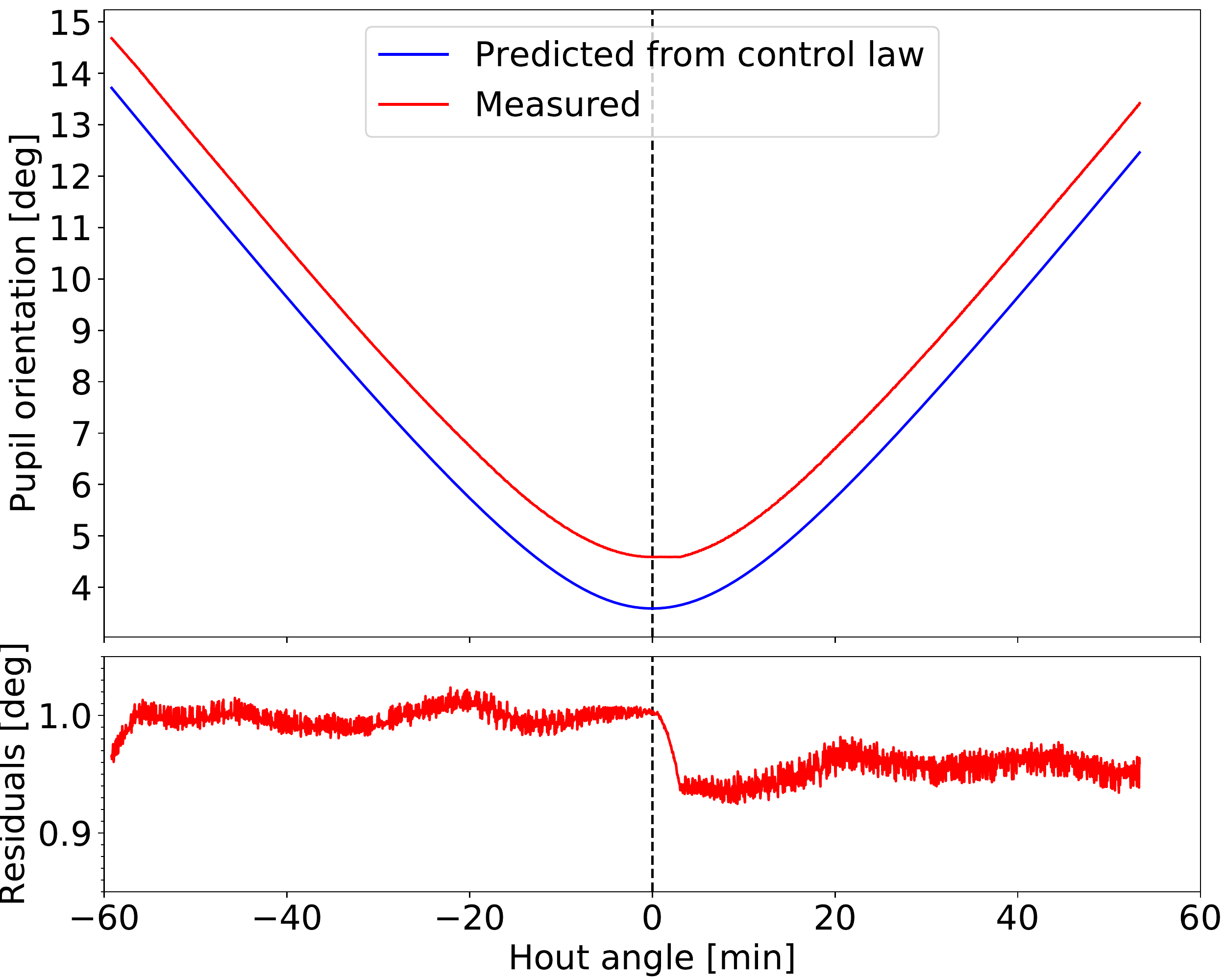}
    \caption{Comparison of pupil orientation in pupil-tracking mode predicted by the control law of the derotator with pupil orientation actually measured in a simulated observation using the distortion grid located in the SPHERE calibration unit. The top panel shows the absolute pupil orientation, while the bottom panel shows the residuals between the predicted and measured values. The average constant offset of $\sim$1\degre is not a problem and is simply caused by the arbitrary zero-point of the derotator. However, the drop $\sim$0.05\degre right after meridian passage is the result of a mechanical backlash of the derotator mechanism. The repeatability of this error has not been investigated in detail, but if it is stable it could be compensated for in the data processing of SPHERE pupil-tracking data.}
    \label{fig:drot_measured_info_ELEV}
\end{figure}

Comparison between budgeted, as-built performance predictions and on-sky data can be found in \citet{Dohlen2016}. This work shows good agreement in terms of attainable contrast as predicted by the power spectral density (PSD) based wavefront error budget, indicating margin for improvement through the implementation of more advanced non-common path aberrations (NCPA) than what is currently in operation \citep{N'Diaye2016}. Daily monitoring of the system with the ZELDA wavefront sensor \citep{N'Diaye2013} shows that the instrument presents an average of $\sim$50\,nm~rms of NCPA, with some daily fluctuations of 5 to 10\,nm. In addition to manufacturing errors of optical surfaces located in the non-common optical path, the NCPA budget also takes into account aberrations due to chromatic beam shift on optical surfaces upstream of the ADCs. These aberrations are particularly difficult to compensate because they are not seen during calibration on internal sources, and they vary with zenith angle. Further work on the issue of NCPA, their compensation and their temporal stability in SPHERE is reported in other publications (\citealt{Vigan2019}; Vigan et al. in prep).

Absolute transmission estimation of the instrument has been performed recently by observing standard stars in photometric atmospheric conditions. The measurements were performed on 22 August 2018 on HIP\,116852 in IRDIS classical imaging mode, without a coronagraph and with AO correction switched off (open loop). The four broad-band filters \texttt{B\_Y}, \texttt{B\_J}, \texttt{B\_H} and \texttt{B\_Ks} were used, and sky backgrounds were acquired immediately afterwards. The latter point is essential for the $H$- and $K$-band observations to avoid errors due to variations in temperature and therefore sky and thermal background. Comparing the observed flux to the one expected for the object as estimated at the entrance of the telescope allows estimating the instrumental throughput including the telescope. The resulting values, representing the average of the two IRDIS channels, are plotted as black squares in Figure \ref{fig:transmission_bands}. Overplotted on the same figure are modeled, as-built transmission curves for the broad-band filters representing the product of measurements of all the instrument's components, again representing the average throughput of the two IRDIS channels. In this estimation, the telescope is modeled by curve for bare aluminum mirrors with 18 month's dust coverage. While the measured transmission is within 20\% of the model for the $Y$-band, it is within 10\% for the $Ks$-band, indicating that the instrument's transmission model is well representative. It can be noted that the results shown here are more conclusive than the commissioning data reported in \citet{Dohlen2016}, for which observing conditions and instrumental setup were not optimal. 

Essential for the scientific exploitation of the instrument is its astrometric precision. While lateral astrometry is ensured only relative to the stellar object itself, the precision of the instrument must be trusted with the absolute rotational orientation and spatial scale of observations; the knowledge of true north, platescale, and distortion. A detailed report on the astrometric calibration of SPHERE is available in \citet{Maire2016}. During the first year of operations, a large and non-repeatable variation in true north was observed, leading to an investigation into the possible causes for such an error. The error was tracked down to a poor time synchronization of the instrument's computer, leading to up to 1\,min of error in derotator timing, which can have a severe impact on the control law of the derotator. Once this defect was corrected on 12 July 2016, true north was maintained to within its required value. Still, we observe a systematic error due to backlash in the derotator mechanism of $\sim$0.05\degre, as demonstrated in Fig.~\ref{fig:drot_measured_info_ELEV}. In pupil-stabilized mode, this leads to a $\sim$0.4\,pixel difference in the position of an object located at the edge of the IRDIS FoV on either side of the meridian. While this effect is currently deemed acceptable, it could easily be accounted for in data reduction.

\section{SAXO}
\label{sec:saxo}

\subsection{High-level technical requirements and design}

SAXO is the high-order extreme AO system of SPHERE. It is used to measure and correct any incoming wavefront perturbation (rapidly varying turbulence or quasi-static instrumental speckles) to ensure a high image quality. The system is fully specified in \citet{Fusco2006} and all the laboratory validations and performance are detailed in \citet{Sauvage2016} and \citet{Petit2012}. The results presented in this section correspond to the SAXO configuration validated during SPHERE commissioning and science verification periods. 

SAXO gathers advanced components and AO concepts. It incorporates a fast (800\,Hz bandwidth) tip-tilt mirror (ITTM), an active toric mirror, and a 41$\times$41 actuators (1377 active in the pupil) HODM. Wavefront sensing is based on a visible spatially-filtered Shack-Hartman concept \citep{Fusco2005}, using an amplified EMCCD detector running up to 1200\,Hz with less than 0.1\,$e^-$ of equivalent readout noise. Since the science verification period, the loop frequency has been increased to 1380\,Hz. The filtering pinhole, designed for removing aliasing effects, is automatically adjusted during the target acquisition as a function of the atmospheric conditions. It can also be changed manually by the instrument operator if conditions significantly change during the course of an observation.

The EMCCD characteristics, combined with dedicated centroiding techniques (e.g., weighted center of gravity), allow SAXO to have high-limiting magnitudes. Firstly the ultimate performance limit magnitude ($R$ = 9--10) for which the instrument meets its initial requirements in terms of wavefront correction. In this regime the Strehl ratio (SR) in $H$-band is higher or equal to 90\%. And secondly the classical limit magnitude ($R$ = 14--15, for which the AO system still provides a significant gain (typically a factor 5 to 10 with respect to the purely turbulent case).

In addition to the fast ITTM and the HODM, that represent SAXO main AO loop, two secondary loops are used in SPHERE to achieve the final performance. The first one is the differential tip-tilt loop, which ensures the final and accurate centering of the star on the coronagraph in the NIR. A small fraction of the stellar light is used to form a star image on the DTTS in $H$-band, and the DTTP located in the SAXO-WFS optical path is used to pre-compensate for this tip-tilt (see Sect.~\ref{sec:global_system}). The frame rate of this loop is 1\,Hz and its precision has been assessed in the integration phase and on-sky to be $<0.5$\,mas, which means 1/80th of the diffraction width in $H$-band. The second one is the pupil loop, which ensures the stabilization of the VLT-UT3 pupil in the system (at 0.1\,Hz). The sensor for this loop is based on intensity measurement in the Shack-Hartmann WFS edge sub-apertures \citep{Montagnier2007}, while the corrector is a pupil tip-tilt mirror (PTTM) situated in the entrance of the system out of a pupil plane. The precision of this loop is 1/10th of a sub-aperture and ensures that the pupil image is stabilized on the Lyot stop of the coronagraph.

The different controlled elements and loops of SAXO are presented in Fig.~\ref{fig:AOloops}. Finally, the real time computer, SPARTA, controls the ITTM and HODM with a final RTC latency of 80\,\micsec, corresponding to an overall 2.14 frame delay at 1200\,Hz. Since the science verification, the SAXO highest frame rate has been improved to 1380\,Hz, but the delay has not been remeasured.

\subsection{Control architecture and performance on sky}

The design of the SPHERE control architecture has been driven by the several major requirements, which are the accurate correction of tip-tilt, including filtering of potential vibrations, the accurate correction of higher order modes, the automatic adaptation to turbulence and vibration conditions, and the robustness to varying conditions during long integrations.

As a result an hybrid control architecture has been derived \citep{Petit2010} for SAXO main AO loop. It incorporates an optimized modal gain integrator for HODM modes. Control is performed in the HODM Karhunen Loève basis considering the high number of degrees of freedom and the good adequation of this basis with modal Signal to Noise Ratio \citep{Petit2008}. Anti-windup and garbage collection are implemented to handle effects of saturation. The ITTM is controlled thanks to a Linear-Quadratic-Gaussian (LQG) law specifically designed to automatically identify and filter out turbulence as well as up to ten vibration peaks per axis randomly spread between 10 and 300\,Hz. This last feature allows reaching a final residual jitter of less than 2\,mas rms (1/20th of the $H$-band diffraction), which is fundamental to ensure the optimal operation of the various SPHERE coronagraphs. Fine decoupling of the two control loops is ensured \citep{Petit2008}. Both control laws are regularly (typically every minute) and automatically updated to provide modal gain optimization for DM control on the one hand, and turbulence and vibration models for LQG on the other hand. Turbulence and vibrations are modeled through second order auto-regressive filters (Kalman filtering) based on the Shack-Hartmann measurements \citep{Petit2008b,Meimon2010}. 
\begin{figure}
    \centering
    \includegraphics[width=0.5\textwidth]{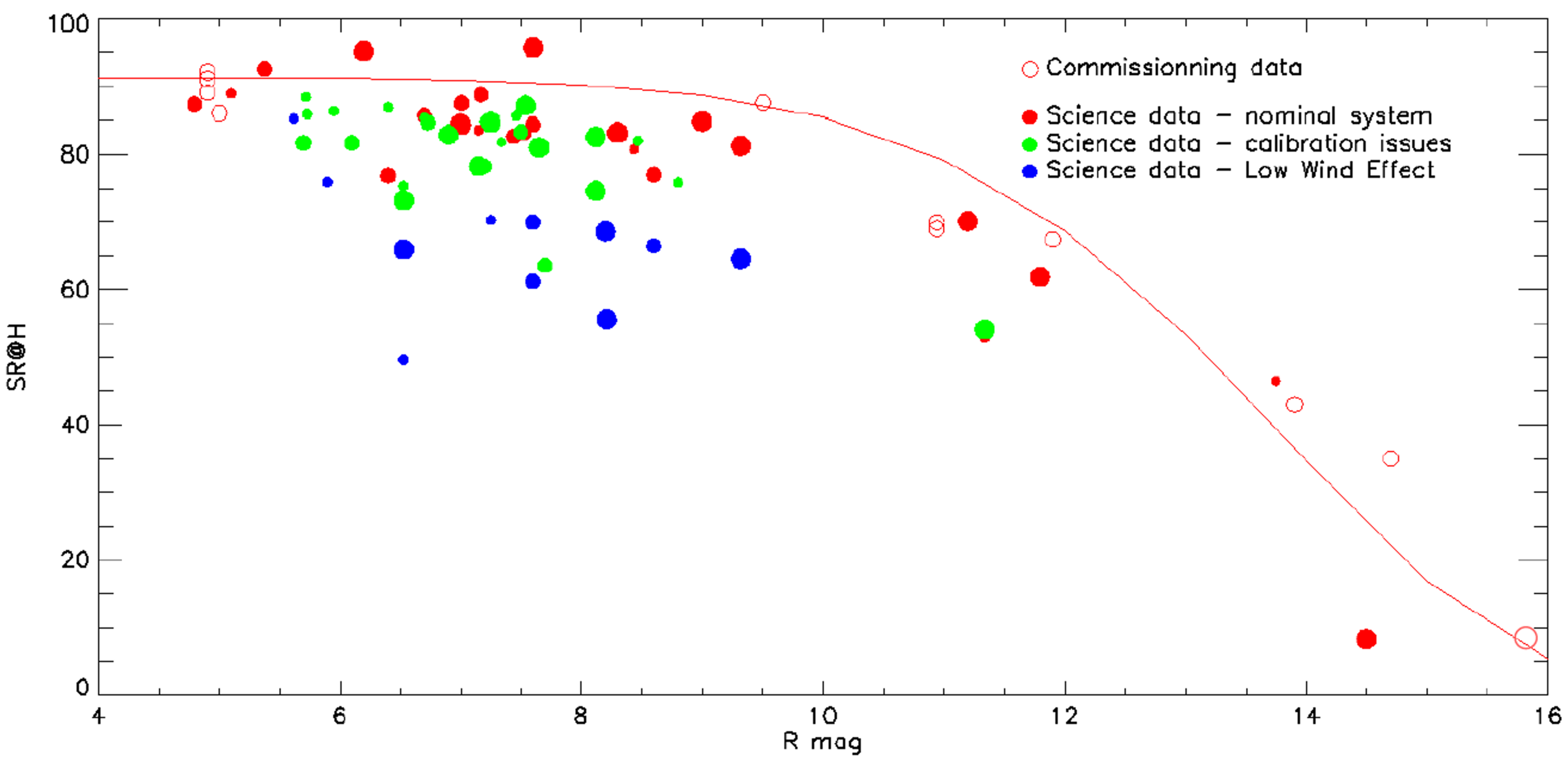}
    \caption{Strehl ratio in $H$-band, measured on-sky during commissioning and science verification. The nominal performance are plotted in red circles. Different problematic regimes are also plotted in green and blue circles. The nominal trend is over-plotted with a solid red line.}
    \label{fig:SAXO_PerfMagnitud}
\end{figure}

\begin{figure}
    \centering
    \includegraphics[width=0.5\textwidth]{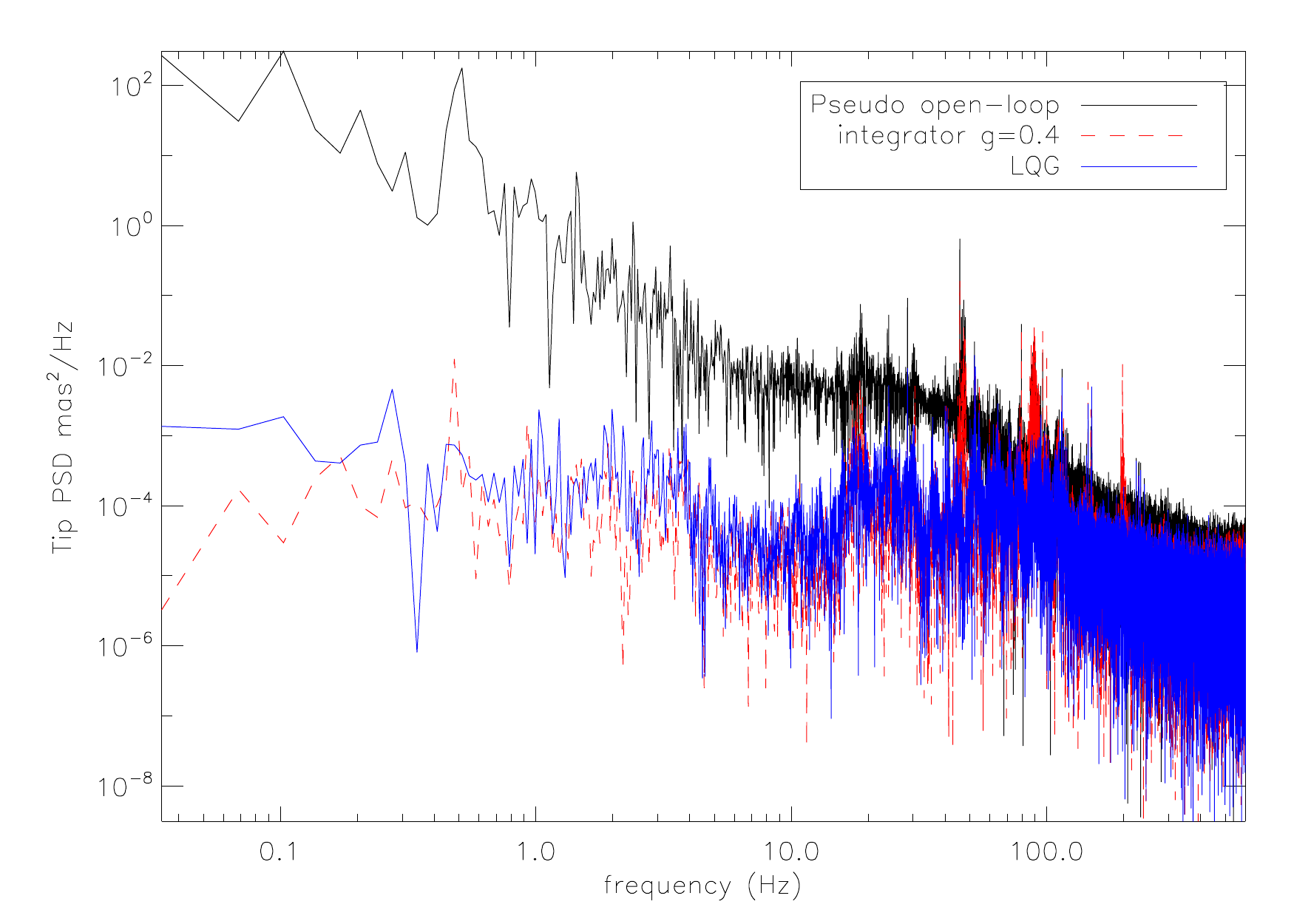}
    \caption{Example of tip mode power spectrum densities obtained on sky on HR\,7710, as derived in pseudo-open loop (black), integrator with 0.4 gain (red) and LQG (blue) with clear vibrations at 18, 46 and 90\,Hz, the origin of which is unknown. LQG dampens the vibration peaks, while the integrator amplifies the one at 90\,Hz.}
    \label{fig:SAXO_PSD}
\end{figure}

Figure~\ref{fig:SAXO_PerfMagnitud} demonstrates the high-level of performance of SAXO, with typical Strehl ratios of 90\% in $H$-band at high flux (see Sect.~\ref{sec:saxo_perf_vs_flux}), fully in agreement with in lab validations and specifications. Now considering more precisely LQG control of the ITTM, Fig.~\ref{fig:SAXO_PSD} shows an example of power spectrum density (PSD) of tip mode as obtained on sky on HR\,7710 during commissioning 4 (7 October 2014) in good conditions (seeing of 0.5\as on the star, integrated wind speed 2\,m/s), successively with LQG or integrator compared to open-loop. Open-loop data are built from closed-loop ones. As comparison cannot be performed synchronously, regular estimation of vibrations and performance with both control laws have been conducted over 30 minutes, demonstrating stable vibrations on both axis located at 18, 46 and 90Hz with weak and stable energy (2, 2.5, and 1.4 mas rms respectively). Performance with both control laws is stable and reproducible. Residual tip-tilt in average over the various acquisitions is 2\,mas rms with LQG compared to 2.9 mas rms with integrator (with a 0.1\,mas rms deviation for both), mostly due to vibration filtering (particularly the 90Hz vibration is dampened by LQG, amplified by integrator). This shows first that SPHERE is hardly affected by vibrations on tip-tilt, a conclusion regularly confirmed till now. Second, LQG indeed corrects for turbulence and filters vibrations leading to increased performance, even if in the SPHERE case the gain is reduced due to limited vibrations in the first place. This validates the use and gain of LQG  with regular updates of models on an operational system. Finally, the auxiliary loops (Differential Tip-Tilt and pupil stabilization) are controlled through simple integrators with standard saturation management (clipping). The pupil loop includes regular modifications of the control matrix to account for pupil rotation (PTTM is located before the derotator).

\subsection{Flux performance on sky}
\label{sec:saxo_perf_vs_flux}

The SAXO performance with respect to the guide star flux is an important driving parameter for science. Figure~\ref{fig:SAXO_PerfMagnitud} shows an estimation of this performance in various conditions obtained since the commissioning of SPHERE. The performance is estimated on long exposure stellar images acquired with IRDIS in $H$-band during different period of the instrument. The performance is estimated by a Strehl ratio computation, as detailed in \cite{Sauvage2007}. Three different regimes of the instrument are illustrated: (i) the nominal regime as during commissioning and nominal use of the system, (ii) a slightly reduced performance regime due to calibration issues during January 2015, and (iii) with a strongly reduced performance during episodes of low wind effect \citep{Milli2018}. The trend is added in solid line and shows the limit magnitude up to R=15. Of course, the performance also depends on the seeing and coherence time: for this figure, only seeing around 0.85\as (median value in Paranal) were selected to mostly remove the dependency \citep{Milli2017}. 

\subsection{Calibrations and operations}

An important aspect of the SPHERE design was to ensure optimal performance of the system, and in particular the AO, without significant adjustments by the nighttime astronomer in charge of operating the instrument. This relies on two separate aspects, which are a robust calibration scheme and a simplified set of adjustable parameters.

Obtaining the optimal performance and robustness of the SPHERE AO requires the following daily calibration sequence:

\begin{itemize}
    \item HODM health check: this procedure performs a sanity check on each of the actuators of the HODM. A voltage is sent on each of them, and the voltage applied is measured and compared to the command.
    \item Detector calibration: calibration of the wavefront sensor background, calibrated for each mode of the detector, that is for three different gains of 1, 30, 1000.
    \item Reference slopes calibration: this consists in calibrating the aberrations internal to the wavefront sensor path, by inserting a calibration source at the entrance of WFS. The impact of all optics in the WFS path are calibrated individually (spectral filter wheel for each position, beam splitter for each position, as well as spatial filter for each position). Depending on the AO mode selected for the night, a set of reference slopes is computed from adding the contribution of each element in the system. This calibration is performed only twice per month.
    \item Interaction matrix calibration: the calibration is made by Hadamard voltage oriented method \cite{Meimon2011}. The total calibration time is $2$ minutes;
    \item Offset voltage calibration: this procedure determines the voltages that flatten the HODM. They are computed by applying an internal closed loop operation, and averaging the voltages applied on the HODM.
    \item Internal quality check: a PSF is acquired on internal source, and its quality is assessed by a Strehl ratio measurement. 
\end{itemize}

Then the AO operations during the target acquisition sequence are described below. Unless specified otherwise, all these steps are performed automatically and without intervention from the astronomer:

\begin{itemize}
    \item Star pointing and acquisition: this step is mostly handled at the telescope level. However, after the telescope pointing and guiding have been performed the star can still be several arcseconds away from the center of the SPHERE FoV (depending on the accuracy of the VLT pointing model), which can sometimes be more than the 2\as FoV of the SPHERE wavefront sensor. The operating astronomer is therefore asked to check if spots are visible (in open loop) on the wavefront sensor detector. If they are, (s)he can validate and the acquisition proceeds automatically with the next steps. If not, the astronomer is asked to point the position of the star in an IRDIS ($\sim$11\as FoV) or ZIMPOL ($\sim$3.5\as FoV) field image and bring it close to the center of the FoV.
    \item Closed loop on tip-tilt mirror with low gain: check that the residual slopes are centered around zero and with a sufficiently small value. 
    \item Flux check for AO mode adjustment: a flux measurement is made on the WFS to check the received flux and eventually change the AO gain of the EMCCD if the flux is too low or too high. Possible gain values are 1, 30 and 1000.
    \item Closed loop on tip-tilt mirror and high order mirror: nominal 0.5 gain applied on all modes. 
    \item Closed loop on the differential tip-tilt loop, and check of closed loop quality. This loop is used for all stars up to H magnitude 10.5. For fainter stars, the loop is unable to properly stabilize the image. A possible and easy upgrade for SPHERE would be to define a more sensitive DTTS mode, with degraded frame rate.   
    \item Start of the atmosphere monitor: this functionality runs continuously during the operations and produces a regular (every 20\,sec) estimate of the wind equivalent velocity, $r_0$ and theoretical Strehl ratio based on a residuals slopes and applied voltages. 
    \item AO spatial filter adjustment: from the atmospheric monitor first estimate, the seeing is computed and the AO spatial filter size is chosen among three different choices: SMALL, MEDIUM, and LARGE. The SMALL size produces the best reduction of aliasing and hence the best dark hole, but is only robust with smallest seeings. Based on the automatic estimate, the astronomer has the option of validating the choice or opting for a different spatial filter size. This step can prove useful in very variable conditions where the automatic estimate can sometimes be inaccurate.
    \item Closed loop optimization: after a first estimation of optimal modal gain, the control matrix for high orders is updated and loaded into the system. After a first estimation of Kalman parameters, and a first estimation of vibration characteristics (frequency, amplitude, and phase), the tip-tilt control law is updated and optimized. This optimization process occurs every one minute during operations. 
\end{itemize}

During the observing sequence, the astronomer retains two possibilities to optimize SAXO operations. The first one is the possibility to adjust the size of the spatial filter based on the observing conditions. If conditions degrade, it can be useful to increase the size of the spatial filter to increase the stability of the turbulence correction. Consequently, if conditions become more stable the size of the spatial filter can be decreased. It is however important to keep in mind that the size of the spatial filter has a visible effect in the focal plane coronagraphic images \citep[e.g., ][]{Cantalloube2018}: from the point of view of data analysis techniques (ADI, SDI) it can be more important to keep the same setup and therefore have more stable images rather than optimizing the size of the spatial filter during the observing sequence. In any case this real-time adjustment is only possible for Visitor Mode observations.

The second one is the possibility to open the DTTS loop, which is the additional tip-tilt stabilization of the NIR PSF on the coronagraph. In the case of faint targets in the NIR (H $\gtrsim$ 10) or in the presence of clouds, the DTTS loop can sometimes become highly unstable because the PSF image of the star on the DTTS camera becomes invisible. In that case, the loop can diverge and drive the PSF image out of the coronagraphic mask or induce additional jitter that will decrease performance. The astronomer has therefore the possibility to open the DTTS loop to avoid any adverse impact on the performance. The status of any of the SAXO loops is in any case reported in the FITS headers of all the science files.

\subsection{Telescope limitations: low wind effect}

The main limitation identified in SPHERE, external to the instrument, is called the low wind effect (LWE). This effect, discovered during the commissioning has been understood and described in \cite{Sauvage2015LWEAO4ELT}, and a possible mitigating solution proposed in \cite{Sauvage2016LWESPIE}.

This effect is a step in the incoming wavefront created at the level of the spider by a thermal interaction between the local air and the cold spider. The metallic spider is cooled down by radiative loss and is therefore several degrees below the ambient air temperature. Due to thermal conductive transfer, the air is cooled by the spider when passing by, hence creating an optical path difference (OPD). This OPD is mainly seen as a phase step as sharp as the spider profile (5\,cm width). The order of magnitude of the OPD is approximately the wavelength: a 1\degree C temperature difference accumulated over a 1\,m-high spider creates 800\,nm of OPD. The lower the wind speed, the most efficient this conductive transfer, the higher the temperature difference, and finally the higher the OPD. The effect in the focal plane is catastrophic: each part of the pupil (four parts separated by the four spiders) generates its own PSF, evolving slowly with time. The Strehl Ratio drops down to less than 50\%, which makes the instrument unusable for high contrast imaging.

To mitigate the effect, ESO has applied between August and November 2017 a dedicated coating on the spiders to reduce the thermal transfer causing the LWE. This solution has showed a high gain in performance, decreasing the number of nights affected from 20\% down to 3\%. This gain in performance is consolidated by more than a year of use of the instrument with the solution implemented. The complete implementation, as well as the performance assessment have been detailed in \cite{Milli2018}. 

\section{Coronagraphy}
\label{sec:coronagraphy}

\subsection{High-level requirements}

Coronagraphy is intended to suppress or to attenuate the coherent part of a wavefront, that is the diffraction pattern of the star, to reveal the surrounding environment. Therefore, the net effect of a coronagraph is to reduce the impact of photon noise in the image which otherwise is the dominant source of noise at short angular separations. Since the original design of Bernard Lyot back in 1930's many ideas have been proposed and several were effectively implemented on the sky \citep{Mouillet1997,Boccaletti2004,Mawet2010}. The basic principle uses the combination of a mask installed at the focal plane, and a stop (the so-called Lyot stop) in the next pupil plane. The focal mask modifies the diffraction pattern of the on-axis point source (the star) which produces a specific diverging diffraction in the subsequent pupil plane. Instead of a uniform distribution of the intensity as in the entrance pupil, the beam, at first order, is diffracted outside the geometric pupil \citep{Malbet1996}. This can be regarded as a two-wave interference, which makes the pupil dark inside the geometric beam \citep{Aime2001}. The stellar light is then filtered out with the Lyot stop. An off-axis object, like a planet, will be much less affected by this mask plus stop combination. The optimization of a coronagraph relies on the trade-off between the rejection of the star light and the preservation of the planet light. Focal masks can be opaque, semi-transparent, or based on a phase pattern with various geometries, or a combination of amplitude and phase \citep[see][for a review]{Guyon2006}. The pupil stop geometry is often related to the telescope pupil shape. Coronagraphy can also take advantage of apodization in the input pupil plane to optimize the cancellation of the starlight in the coronagraphic pupil \citep{Aime2002}.

On a real telescope, the wavefront is affected by the atmospheric turbulence, which even if AO-corrected, is not a perfectly flat wavefront. The incoherent part of the wavefront will escape the effect of the coronagraph. As a rule of thumb, if the wavefront is corrected at a Strehl ratio of 90\%, then about 10\% will be left in the coronagraphic image, whatever the coronagraphic design. For SPHERE, the design of coronagraphs was driven by the main requirements, which are a large wavelength coverage and achievement of high-contrast at short angular separations \citep{Boccaletti2008a,Boccaletti2008b}. The former is quite constraining in SPHERE as the instrument is designed to operate from $Y$- to $K$-band in a single snapshot for what concern the NIR arm. The latter is also critical to reach the intensity level of young giant planets so we targeted an IWA\footnote{The inner working angle is the angular distance from the on-axis object (in general the star, which is positioned at the center of the mask) at which an off-axis object (the planet) will have a transmission of 50\%.} of $1-2\lambda/D$.  

\subsection{Design of the coronagraphs}

\begin{table}[t]
    \centering
    \caption{Maximum wavelength (in \mic) and corresponding filter for each APLC configurations.}
    \label{tab:aplcmask}
    \begin{tabular}{lcccc} \hline
    	 & \multicolumn{2}{c}{APO1} & \multicolumn{2}{c}{APO2} \\
    \hline \hline		
    ALC1 & 1.41 & $\sim$J  & 1.08 & $\sim$Y \\
    ALC2 & 1.79 & $\sim$H  & 1.38 & $\sim$J \\
    ALC3 & 2.33 & $\sim$Ks & 1.79 & $\sim$H \\
    \hline 
    \end{tabular}
\end{table}

\begin{figure}
    \centering
    \includegraphics[width=0.5\textwidth]{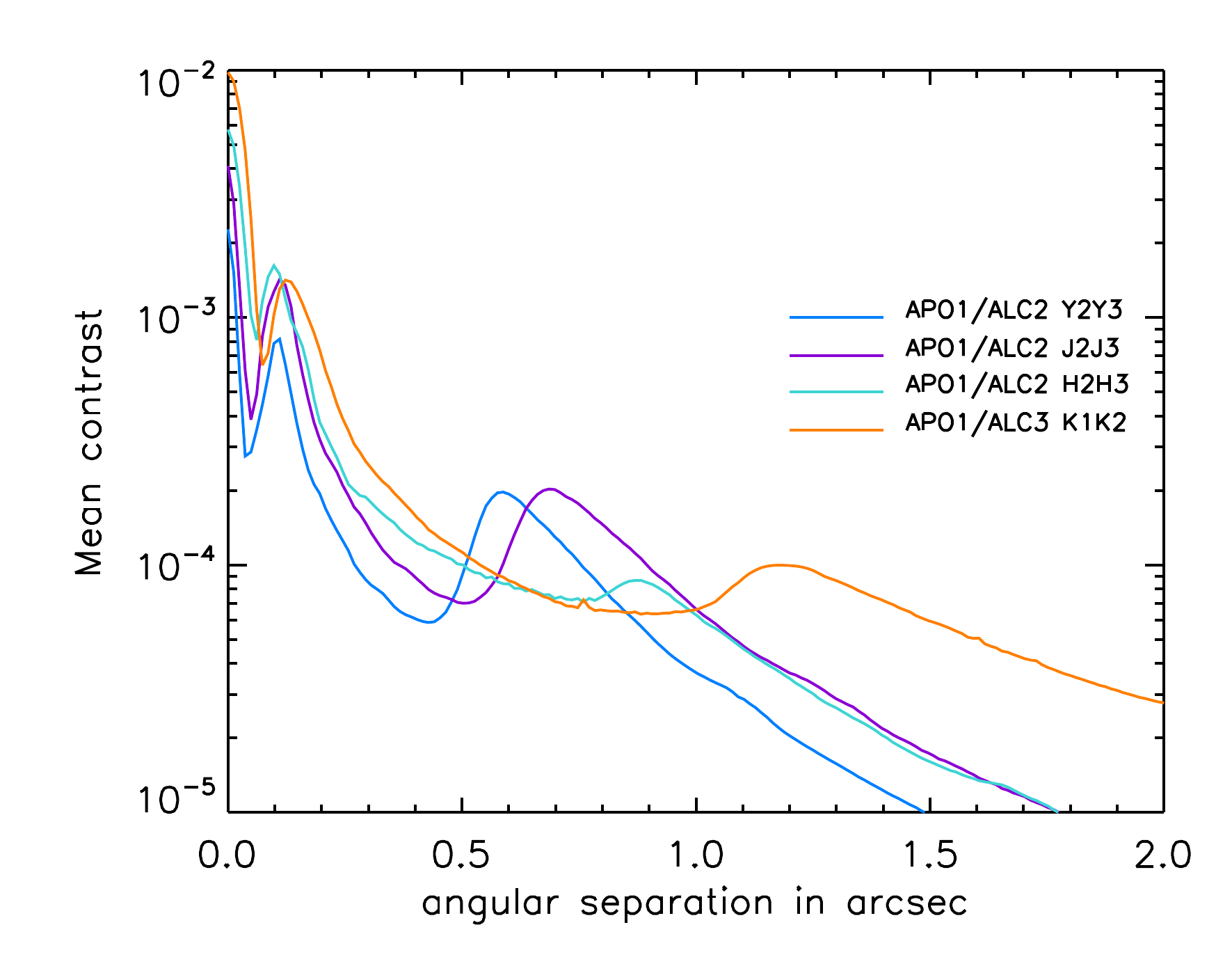}
    \caption{Near-infrared normalized contrast for two coronagraphic configurations (APO1/ALC2 and APO1/ALC3) in different filter pairs (Y23, J23, H23, K12) as measured on 15 May 2014 on HD\,140573.}
    \label{fig:contrast_aplc}
\end{figure}

\begin{figure*}
    \centering
    \includegraphics[width=0.49\textwidth]{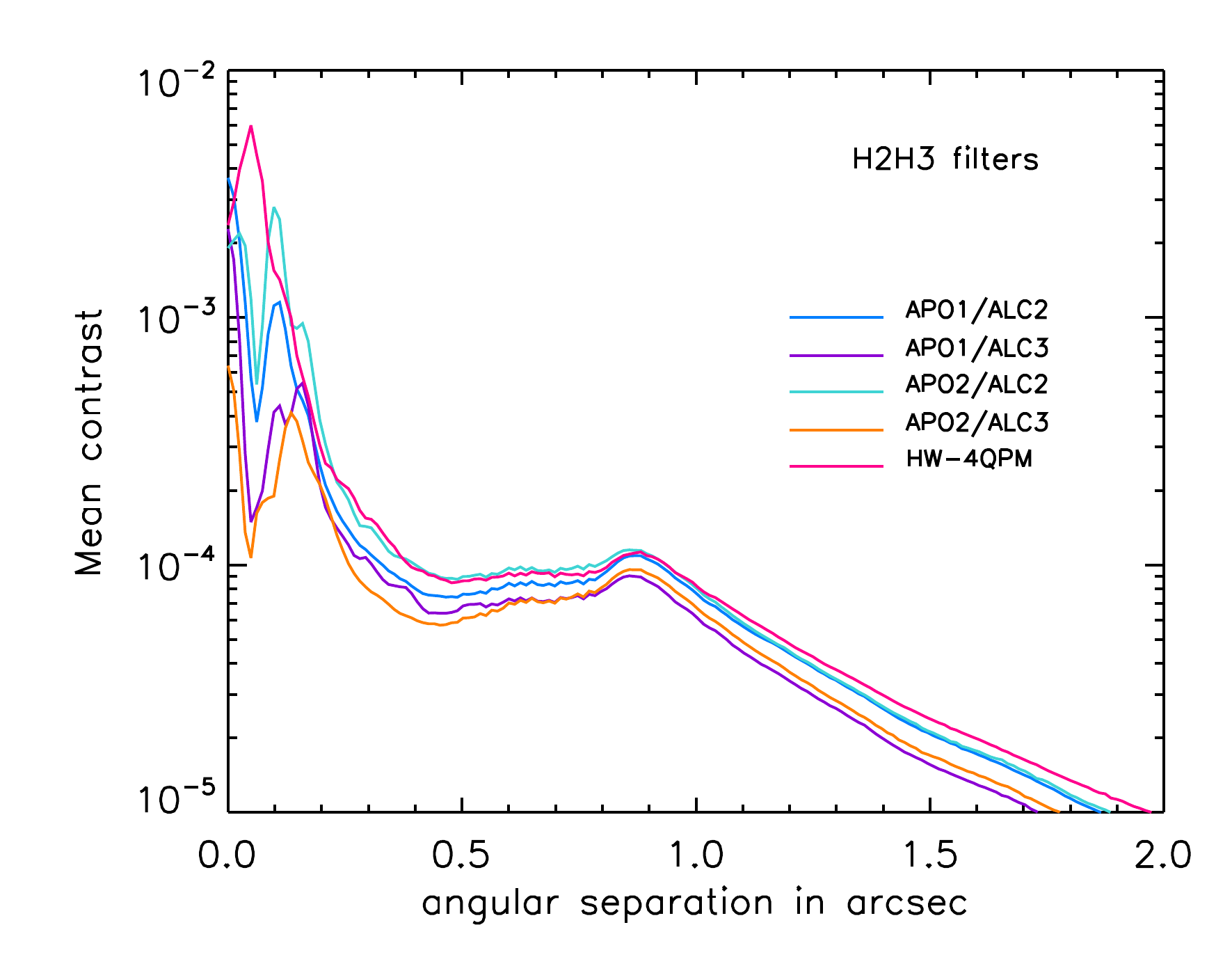}
    \includegraphics[width=0.49\textwidth]{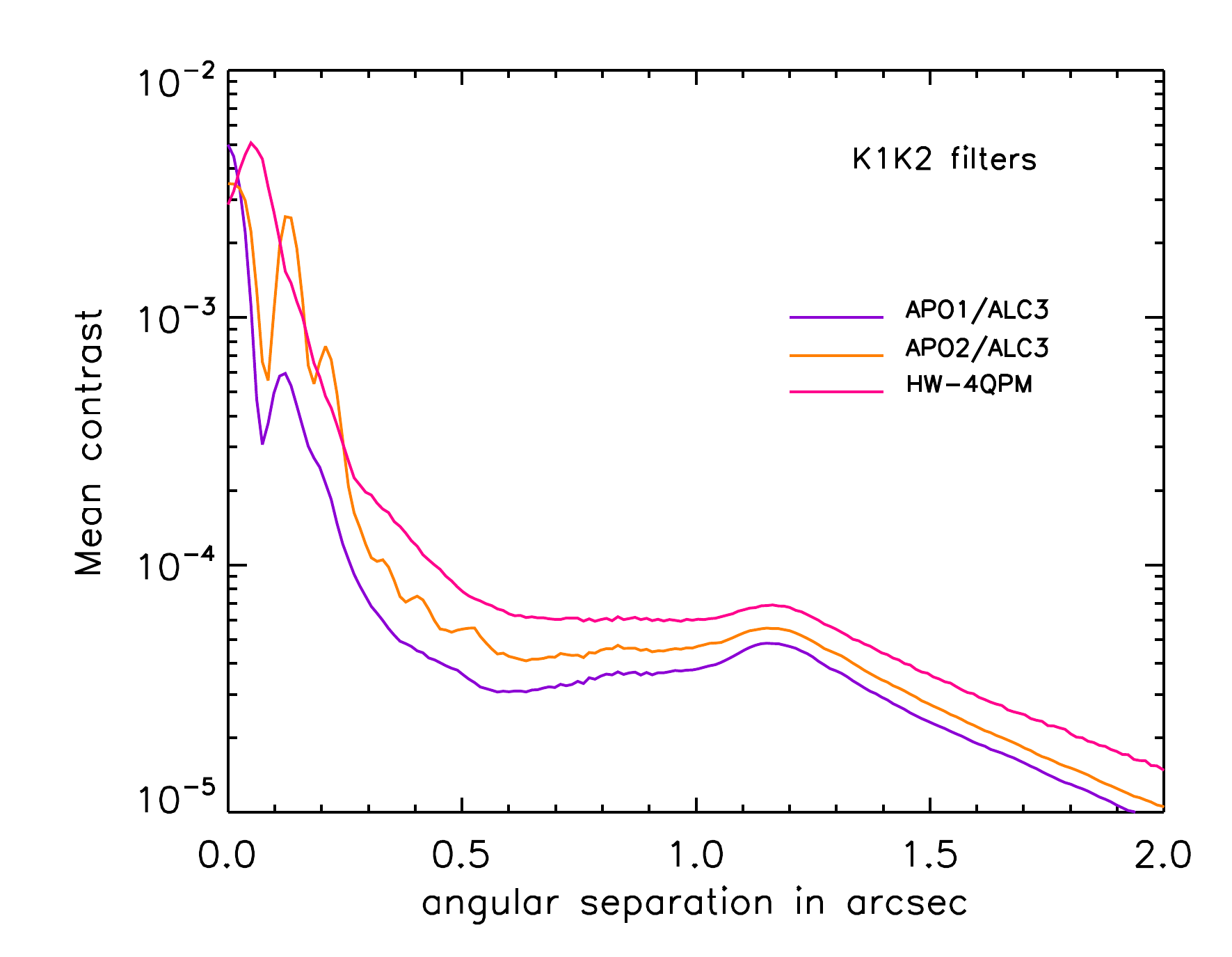}
    \caption{Near-infrared normalized contrast for several coronagraphic configurations in H23 (left) and K12 (right) as measured on 9 October 2014 on HR\,591.}
    \label{fig:contrast_all}
\end{figure*}

\begin{figure}
    \centering
    \includegraphics[width=0.5\textwidth]{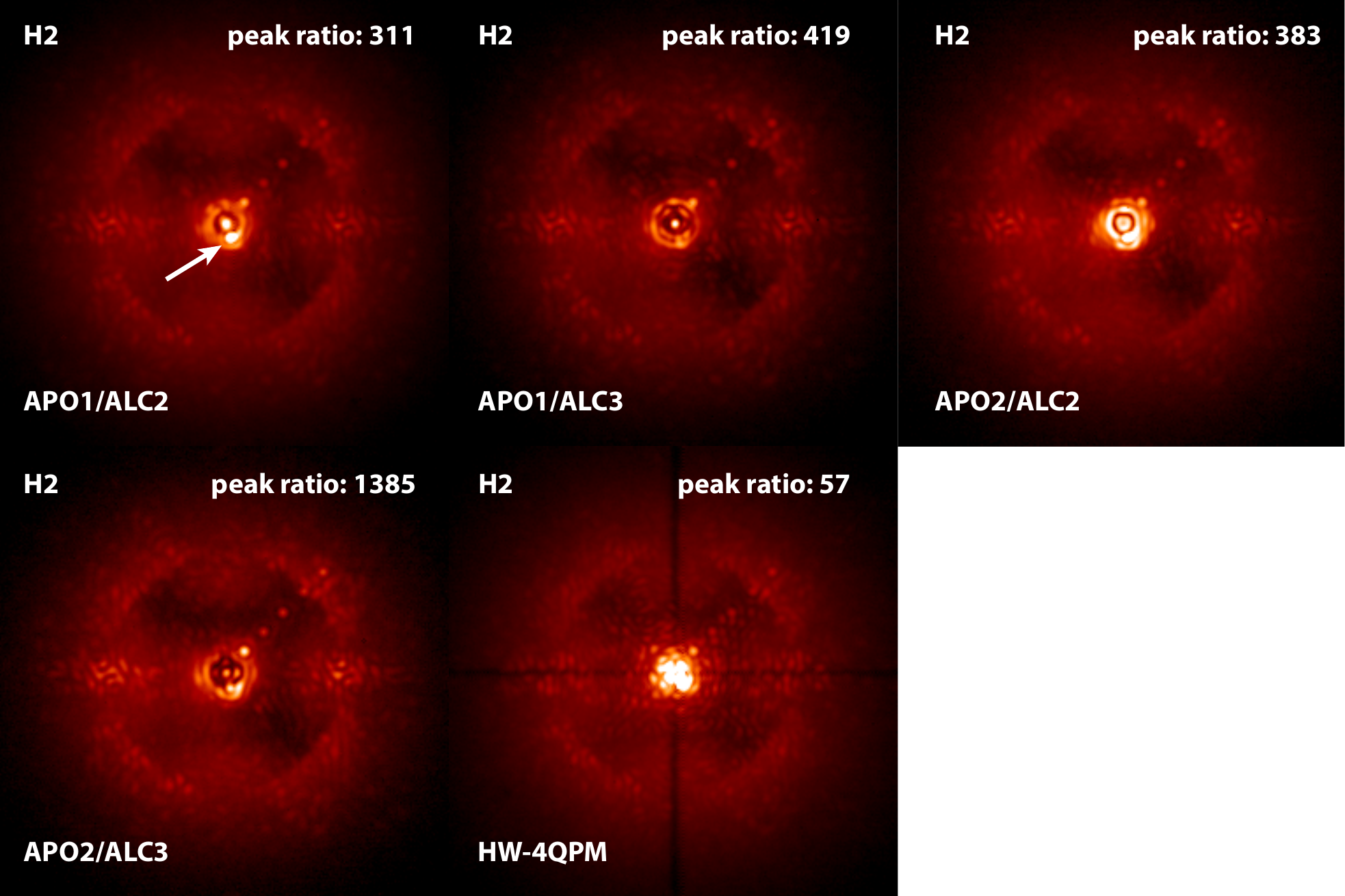}
    \includegraphics[width=0.5\textwidth]{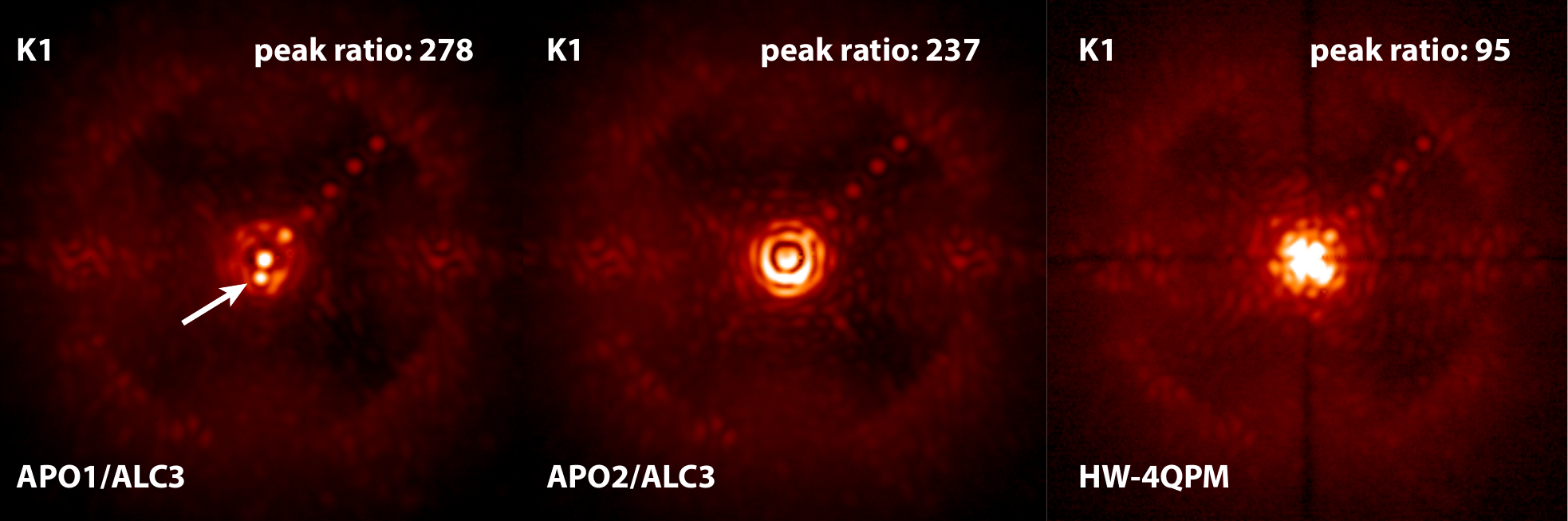}
    \caption{Near-infrared coronagraphic images (FoV = 3.2\as) of HR\,591 corresponding to the configurations of Fig. \ref{fig:contrast_all} for the H2 (top) and K1 (bottom) filters, together with fake planets at $10^{-3}$ and $2\times10^{-4}$ levels (see text). The stellar companion $\alpha$ Hyi B is indicated by an arrow.}
    \label{fig:corono_images}
\end{figure}

\begin{figure}
    \centering
    \includegraphics[width=0.5\textwidth]{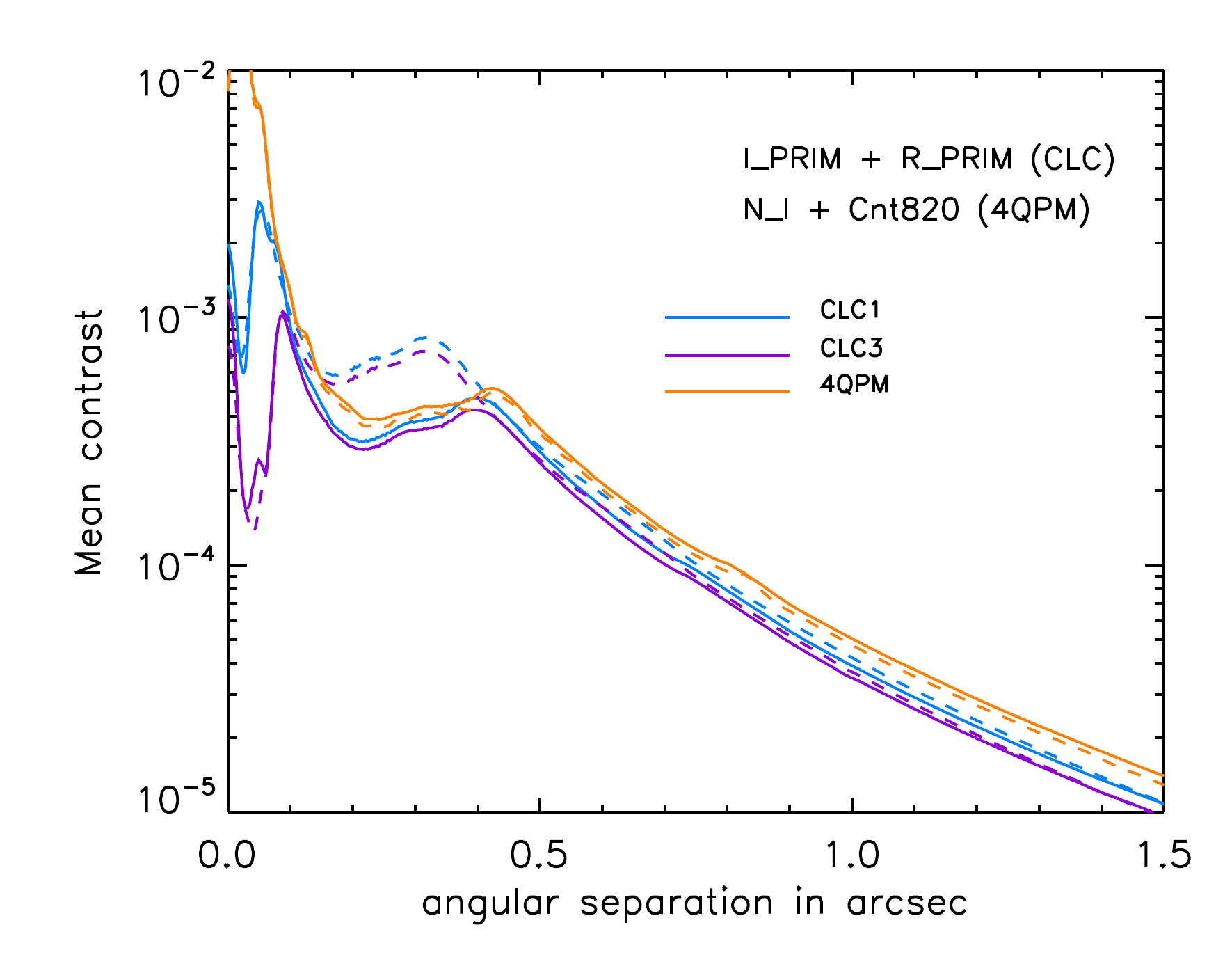}
    \caption{Visible normalized contrast for several coronagraphic configurations of ZIMPOL, as measured on 9 October 2014 on HR\,591.}
\label{fig:contrast_zim}
\end{figure}

\begin{figure}
    \centering
    \includegraphics[width=0.5\textwidth]{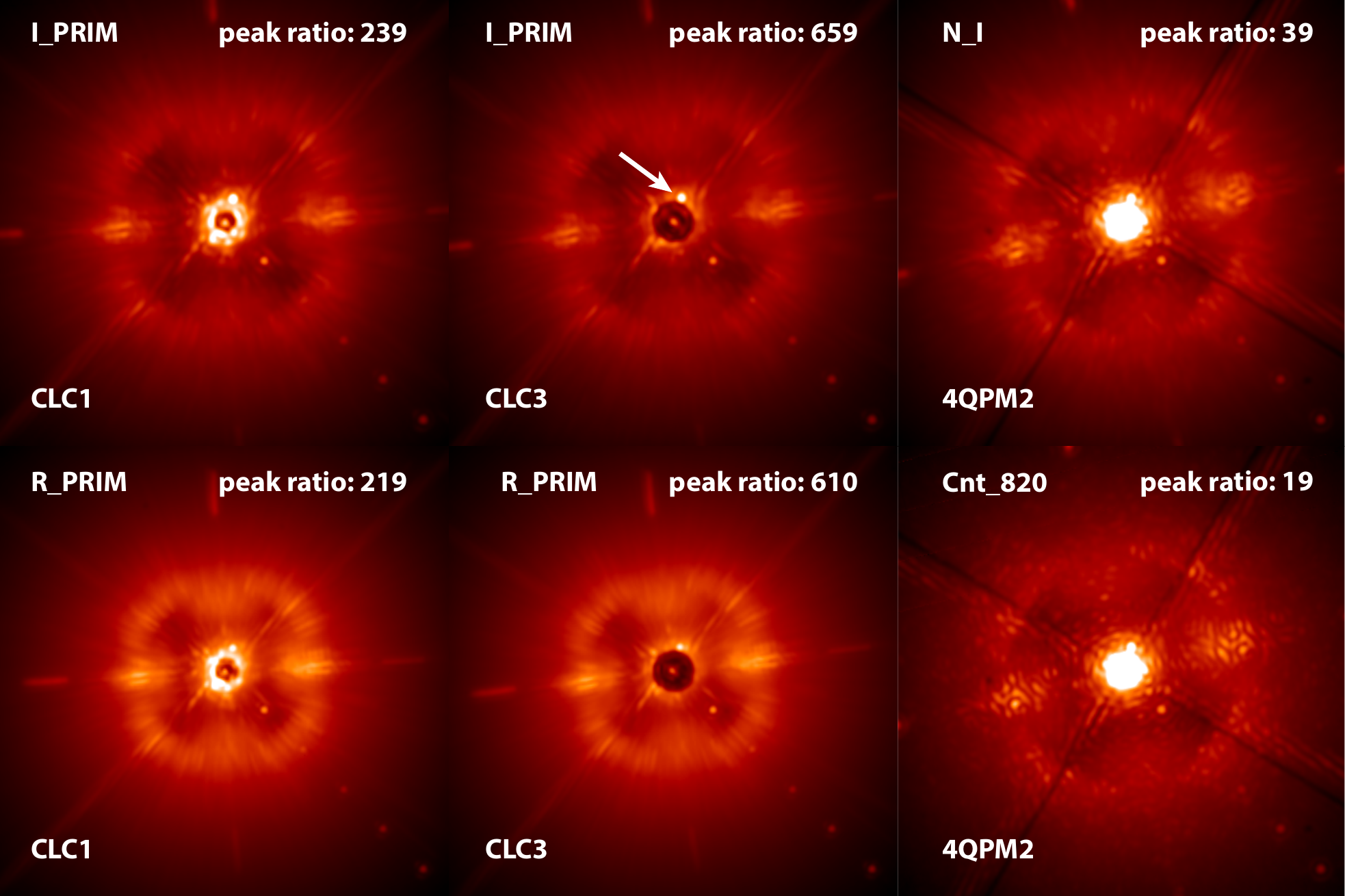}
    \caption{Visible coronagraphic images (FoV = 1.6\as) of HR\,591 corresponding to the configurations of Fig. \ref{fig:contrast_zim}, together with fake planets at $10^{-3}$ and $2\times10^{-4}$ levels. The stellar companion $\alpha$ Hyi B is indicated by an arrow.}
    \label{fig:corono_images_zim}
\end{figure}

During the study phase, several concepts of coronagraphs were considered and we chose to include a large fraction of these to allow flexibility with respect to different science cases and for the sake of redundancy. While the various designs are listed in the user manual, we are focusing here on the main configurations that are relevant to the highest performance in SPHERE, in particular for the NIR survey. The apodizers and focal plane masks are installed in the CPI, one series for the near IR channel and another one for the visible. The Lyot stops are implemented in each of the three instruments' pupil planes. The near IR channel offers two types of coronagraphs, the apodized-pupil Lyot coronagraph \citep[ALC, ][]{Soummer2005}, and the half-wave plate four-quadrant phase mask \citep[HW-4QPM][]{Rouan2000, Mawet2006}, both being designed to achieve very high-contrast at very short angular separations. The first prototypes are described in \citet{Boccaletti2008a}.

The APLC combines a pupil plane apodizer (APO\#), a focal plane mask (ALC\#) and a Lyot stop (STOP\#) fully described in \citet{Carbillet2011} and \citet{Guerri2011}. These three components were optimized to allow several configurations. The APLC is chromatic by design but chromaticity can be mitigated if the mask size is chosen for the largest operating wavelengths (shorter wavelength PSFs are simply blocked by the opaque mask more efficiently). SPHERE allows six configurations with two apodizers and three masks (Table~\ref{tab:aplcmask}). APO1 and APO2 are optimized respectively for a mask diameter of 4\,$\lambda/D$ (IWA = 2 $\lambda/D$) and 5.2\,$\lambda/D$ (IWA = 2.6 $\lambda/D$). Apodizers are manufactured with a microdot technology \citep{Martinez2009}. The corresponding transmissions are 63\% and 48\% with respect to the VLT pupil. The chromium-coated masks ALC1, ALC2 and ALC3 deposited onto a silica substrate have diameters of 145, 185 and 240\,mas. The maximum wavelength allowed with these configurations are provided in Table~\ref{tab:aplcmask}. One single stop is available slightly undersizing the telescope pupil (outer radius 96\%, inner radius 20\%, spider 2.5\%, relative to the geometric pupil size). A new stop including six patches of 5\% of the telescope pupil was manufactured before commissioning to block the diffraction of the HODM dead actuators. The size of these blockers results of an optimization between the throughput and the rejection of the light diffracted by dead actuators. The overall transmission (apodizer+stop) are about 58\% (APO1) and 45\% (APO2). For the SHINE survey, two configurations are used APO1/ALC2 (\texttt{N\_ALC\_YJH\_S}) and APO1/ALC3 (\texttt{N\_ALC\_Ks}) respectively optimized for the \texttt{IRDIFS} and \texttt{IRDIFS-EXT} modes. 

The original phase mask concept is also by nature chromatic but can be turned achromatic using a combination of two birefringent materials (quartz and MgF2) stacked on each side of a SiO2 substrate. Each side of this stack is obtained from the same piece of material, cut in four parts, two of which being rotated by 90\degre to flip the fast versus slow axis and mimic a HW-4QPM pattern. The alignment of the plates is made by hand with a lateral tolerance of 5\,\mic and 10\,arcmin in tip and tilt. The prototype was proven to achieve a very high degree of achromaticity at very high-contrast ($10^{-4}$ contrast at $2-3 \lambda/D$, $10^{-5}$ contrast at $6-8 \lambda/D$) on a test bench. These atmosphere-free performance were similar to those obtained with the APLC, while offering a twice smaller IWA \citep{Boccaletti2008a}. We initially manufactured two components: one optimized for the $YJH$-bands and the other for the $Ks$-band, only differing by the thickness of the birefringent plates used in the stack. Both were found to perform similarly across the whole spectral range of SPHERE. The stop of the HW-4QPM is more aggressive than the one used with the APLC (outer radius 90\%, inner radius 30\%, spider 2.5\%, relative to the geometric pupil size) and also requires larger blockers to mask the diffraction pattern of dead actuators (9\% of the pupil size). However, the total transmission (STOP+APO) is 68\%, which is slightly higher than for the APLC.

The suite of ZIMPOL coronagraphs implements simple Lyot masks (CLC) and two 4QPMs for achieving very small IWA at 656\,nm and 820\,nm. Because the Strehl ratios delivered by ZIMPOL are significantly lower than in the IR, the chromaticity of phase mask is not an issue, providing that the bandwidth is lower than 10-20\% \citep{Boccaletti2004}. Several diameters of Lyot masks are available, the smallest ones being deposited on substrate (93\,mas, and 155\,mas with an astrometric grid) and the largest being suspended with wires (155\,mas and 310\,mas). Associated Lyot stops achieve transmission of 56\% for the 93\,mas mask,  78\% for the 155\,mas mask and 73\% for the 4QPMs. Similarly to the IR channel, the dead actuators were masked with blockers 3\% of the pupil size.  
\subsection{Operations}

The most important setup relative to the coronagraph is the centering of the star onto the focal plane mask. In SPHERE this step is achieved in the acquisition template in a Lyot stop-less mode. In such a configuration, the image at the detector is linear with respect to tip and tilt, and more flux is leaking through the coronagraph. These two effects allow a higher sensitivity to determine the coronagraphic mask position. The masks' wheel is positioned to select the appropriate configuration, and the flat lamp is used to record a background image in which the mask appear as a dark disk (for APLC) or a dark cross (for the HW-4QPM). The pixel coordinates of the mask center is obtained from a correlation with a mask template allowing sub-pixel precision. Then, taking advantage of the internal source, the optimal alignment is searched with an iterative process, while minimizing the difference of flux in the horizontal and vertical axes. The iteration stops (maximum of 5) when the distance between the source and the mask reaches 0.5\,mas or better.  The optimal position defines the reference slopes for the tip-tilt mirror to form the image of the star at the center of the coronagraph. During the science exposure the DTTS takes care of maintaining this alignment within 0.5\,mas. The procedure must be repeated every time the masks' wheel is moved, that is when a new coronagraphic setup is used.

A similar procedure is also used to optimize the focus of the PSF on the coronagraphic mask. In this procedure, the HODM is used to introduce a ramp of focus that encompasses the true best focus. At each position of the focus ramp, the PSF is centered on the mask with the above procedure, then total flux within a pre-determined value in the coronagraphic image is computed and finally saved. After all the ramp positions have been covered, a parabolic fit is performed on the integrated flux vs. the introduced defocus, and the best focus is determined from the minimum of the parabola. The best focus is then applied on the HODM before performing the final fine centering described previously. Contrary to the centering that must be executed every time that the coronagraphic wheel is moved, the focus optimization is generally performed once at the beginning of the night and then saved, as it was shown not to vary over the course of the night.

\subsection{Performance \& limitations}

The coronagraphs were intensively tested during commissioning in various conditions. We mostly used IRDIS for these tests as it allows to measure the contrast at  large separations at all relevant wavelengths. Obviously the coronagraphic contrast is highly dependent on the residual phase behind the AO, so the tests obtained in different nights are not strictly comparable. An example of raw contrast dispersion, obtained with AU~Microscopii between 2014 and 2017, is shown in \citet{Boccaletti2018} for which we witnessed variations of a factor of $\sim$2 in the $H$-band and of $\sim$4 in the $J$-band at a separation of 0.4\as (mid-separation inside the control radius). 

On 16 May 2014 during the 1st commissioning, we observed HD\,140573 (R=1.82, H=0.238) with the main coronagraphic modes APO1/ALC2 (N\_ALC\_YJH\_S) in several dual-band filters Y23, J23, H23, and with APO1/ALC3 (N\_ALC\_Ks) in the K12 dual-band filter. The seeing varied from 0.86\as to 1.18\as along the sequence between 05:08\,UT and 06:36\,UT. As a consequence of a fixed mask size, the contrast increases progressively with shorter wavelengths in the range 0.1--0.5\as, since more flux from the star is blocked with smaller PSFs (Fig. \ref{fig:contrast_aplc}). At larger separations (0.5--1.2\as), the scaling of the control radius drives the achievable contrast. While the contrast curves beyond the control radius are nearly parallel in Y, J, and H, it is flatter for the $K$-band, mainly due to the increased impact of the thermal background. So the contrast in K12 is not directly comparable with those at shorter wavelengths. In fact we can expect a deeper contrast in the $K$-band (see next paragraph). 

The various coronagraph configurations were compared on 10 October 2014 during the fourth commissioning in the H23 and K12 filters. The star HR\, 591 ($\alpha$ Hyi, R=2.55, H=2.17), a very tight binary (0.091$\pm$0.003\as in separation) was observed from 05:11\,UT to 07:46\,UT. In the H23 filter pair, the largest the Lyot mask the highest the contrast for the same reason as above (Fig. \ref{fig:contrast_all}, left). The best contrast is achieved with the APO2/ALC3 configuration, which combines a large mask (240 mas) and sharper apodization. The worst contrasts are delivered by the HW-4QPM and the APO2/ALC2, the former because of a twice smaller IWA than APLC, which makes it more sensitive to low order aberrations, and the latter because this apodizer is not optimized for this mask size so the stellar diffraction is not sufficiently attenuated. Here, the default coronagraph (APO1/ALC2) for H23 is achieving contrast of $8\times10^{-5}$ at $\sim$0.5\as. Along this test the conditions were rather stable, if we rely on the contrast slopes measured beyond the control radius, but the seeing was in the 1.1--1.3\as range. In K12, APO1/ALC3 (default configuration) achieves a contrast as large as $3\times10^{-5}$ at $\sim$0.6\as (Fig. \ref{fig:contrast_all}, right). The HW-4QPM is clearly worse by a factor of about two inside the control radius. The apodizer APO2 combined to ALC3, although producing intermediate performance, creates diffraction rings around the mask edges, which degrades the contrast in the 0.10--0.25\as range to the same level as the HW-4QPM.  The presence of the stellar companion provides a visual assessment of the performances at very short separations, as displayed in Fig. \ref{fig:corono_images}. Fake companions were also inserted along the diagonal at separation of 0.2\as (ratio 1:1000), and 0.4\as, 0.6\as, 0.8\as, 1.0\as (ratio 1:5000). The stellar companion appears at different position angles, while the field of view rotates along this test. It lies very close to one of the HW-4QPM transition and is therefore significantly attenuated by the phase mask. An important number to consider when preparing observation to determine the DIT of coronagraphic sequence is the peak attenuation characterized by the ratio of the maximum intensity on the PSF to that of the coronagraphic image. These numbers are indicative and subject to variations (seeing, centering...) but some values are provided in Fig. \ref{fig:corono_images}. 

The very same target was observed also with ZIMPOL on 10 October 2014 (07:21\,UT to 08:44\,UT). While the various results are described at length in \citet{Schmid2018}, we focus here on the coronagraphs delivering the smallest IWA: the classical Lyot coronagraphs CLC1 (CLC\_S\_WF) and CLC3 (CLC\_MT\_WF), and the 4QPM2 (optimized for 820nm). Inner Working Angles are respectively 47\,mas, 78\,mas, and 1\,$\lambda/D$ (21\,mas at 820\,nm). The observations were obtained in the broad band filters I\_PRIM (790nm)+R\_PRIM (626nm) for CLCs, simultaneously in each arms of ZIMPOL, and with N\_I (817nm)+Cnt820 (817\,nm) for the 4QPM2. Contrast curves are displayed in Fig. \ref{fig:contrast_zim}. The contrasts are significantly better in the $I$-band than in the $R$-band by more than a factor of two inside the control radius (0.41\as and 0.32\as in I\_PRIM resp. R\_PRIM). The contrast profile inside the control radius is minimal at $\sim$0.2\as in I\_PRIM  or $\sim$0.15\as in R\_PRIM, which corresponds to an angular separation of 10 $\lambda/D$. For instance, a maximum contrast level of $3\times10^{-4}$ is measured with CLC3 at 0.2\as in I\_PRIM, which is not so different than the contrast achieved in the near IR. While in the near IR the contrast decreases (Fig.~\ref{fig:contrast_aplc}) or gets flat (Fig. \ref{fig:contrast_all}) all the way out to the control radius, the behavior is opposite in the visible with the contrast raising from 10 to 20 $\lambda/D$. Visible observations are necessarily more affected by temporal errors from the AO than in the near IR, hence brighter speckles are leaking inward of the control radius. Indicative values for the peak attenuation are provided in Fig. \ref{fig:corono_images_zim}. 

\section{ZIMPOL}
\label{sec:zimpol}

The Zurich IMaging POLarimeter (ZIMPOL) is the visible focal plane instrument of SPHERE covering the wavelength range from 510-900 nm and providing observational modes for polarimetric differential imaging, imaging, spectral differential imaging, and angular differential imaging. Previous publications give a general description of the ZIMPOL system and the on-sky performance \citep{Schmid2018}, detailed descriptions of the the opto-mechanical design \citep{Roelfsema10}, the optical alignment procedures \citep{Pragt12}, hardware test results at various phases of the project \citep{Roelfsema11,Roelfsema14,Roelfsema16}, the detectors \citep{Schmid12}, and the polarimetric calibrations \citep{Bazzon12}. 

\subsection{High-level scientific and technical requirements}

Polarimetric differential imaging (DPI) is very powerful for the speckle suppression in high-contrast imaging because two opposite polarization modes, for example $I_\perp$ and $I_\parallel$, can be measured simultaneously without differential chromatic aberrations. Light scattering by planets or by dust in circumstellar disks produces a strong, broad-band linear polarization of the light, while the light of most stars is unpolarized.

DPI is a very attractive mode for the polarimetric search of reflected light from extra-solar planets around the nearest, bright stars \citep{Schmid06}. There are about a dozen of good stellar systems around which giant planets could be detectable, if they are present \citep{Thalmann2008}. These systems include $\alpha$ Cen A + B, Sirius, $\epsilon$ Eri, $\tau$ Cet and Altair. The polarization signal of a reflecting planet is proportional to $\propto  R_p^2/d_p^2$, where $R_p$ is the radius of the planet and $d_p$ the separation. Therefore, only large planets $R_p\approx R_J$ at small separation $d_p< 2$\,au produce a detectable polarization contrast of $C_{\rm pol} = (p_p\times F_p) /F_S\approx 10^{-8}$ \citep[e.g., ][]{Buenzli09,Milli13}. The visual wavelength range is also favorable for such a polarimetric search because Rayleigh scattering and Rayleigh-like scattering by aerosols are in atmospheres of solar system objects the dominant processes for producing strong scattering polarization at short wavelengths $<1$\,\mic \citep{Tomasko82,Smith84,Stam04,Stam08,Schmid06,Schmid11,Bazzon13,Bazzon14}. 

The contrast performance of the AO system is less optimal in VIS because wavefront aberrations translate into larger phase aberrations at shorter wavelengths. This penalty for the visible range can be compensated with the CCD-based ZIMPOL technology \citep{Povel90}, which performs very high precision imaging polarimetry based on fast ($\sim 1$\,kHz) polarization modulation-demodulation. Sensitivities of $10^{-5}$ and below can be achieved if enough photons are collected, because (i) the fast modulations are faster than the atmospheric changes and therefore freeze the speckle variations, and (ii) the opposite polarization modes are measured simultaneously and pass in this single channel differential measurement strictly through the same optics \citep{Kemp81,Stenflo96}. Thus, the raw contrast of 10$^{-3}$ to 10$^{-4}$ from the AO system can be pushed with fast modulation DPI toward 10$^{-8}$ for the investigation of polarized sources around bright stars. 

For this task the key technical requirements for the ZIMPOL system design are AO-corrected high-resolution ($\approx 20$\,mas) observations in the wavelength range 550--900\,nm, in coronagraphic mode with the control of all important instrumental polarization effects from the telescope to the detector. The system must achieve high-photon count rates of the order $10^{6}$\,s$^{-1}$ per resolution element with a diameter of 20\,mas in the halo of a coronagraphic image at separations of 0.1\as-1.5\as. This allows for the brightest stars and broad-band observations ($\Delta\lambda\approx 100-300$\,nm) to reach the photon noise limit of about $\approx (N_\gamma=10^{10})^{-1/2}=10^{-5}$ with ZIMPOL polarimetry within half a night of observing time.

ZIMPOL polarimetry of circumstellar disks or dust shells around fainter stars requires a less demanding contrast performance because the required photon noise limits are less extreme, but one needs to adapt the system capabilities for lower photon flux conditions. This requirement is achieved with a slow modulation mode adapted for longer integrations and lower level of detector noise.

ZIMPOL includes also an imaging mode, which is using the system without polarimetric modulation and without polarimetric optical components. ZIMPOL is therefore just a dual-beam imager which benefits from the high contrast and high resolution capabilities provided by the wavefront correction of SAXO, and the image derotation, the atmospheric dispersion correction, and the visual coronagraph of the CPI. System modes for ADI and SDI are available to fulfill requirements for high-contrast capabilities. An important feature for differential imaging is the H$\alpha$ line, which is in many stellar objects the strongest, or one of the strongest, circumstellar emission line. Therefore, an key requirement for ZIMPOL was the implementation of appropriate H$\alpha$ line filters within its filter set. 

\begin{figure}
    \includegraphics[width=0.5\textwidth]{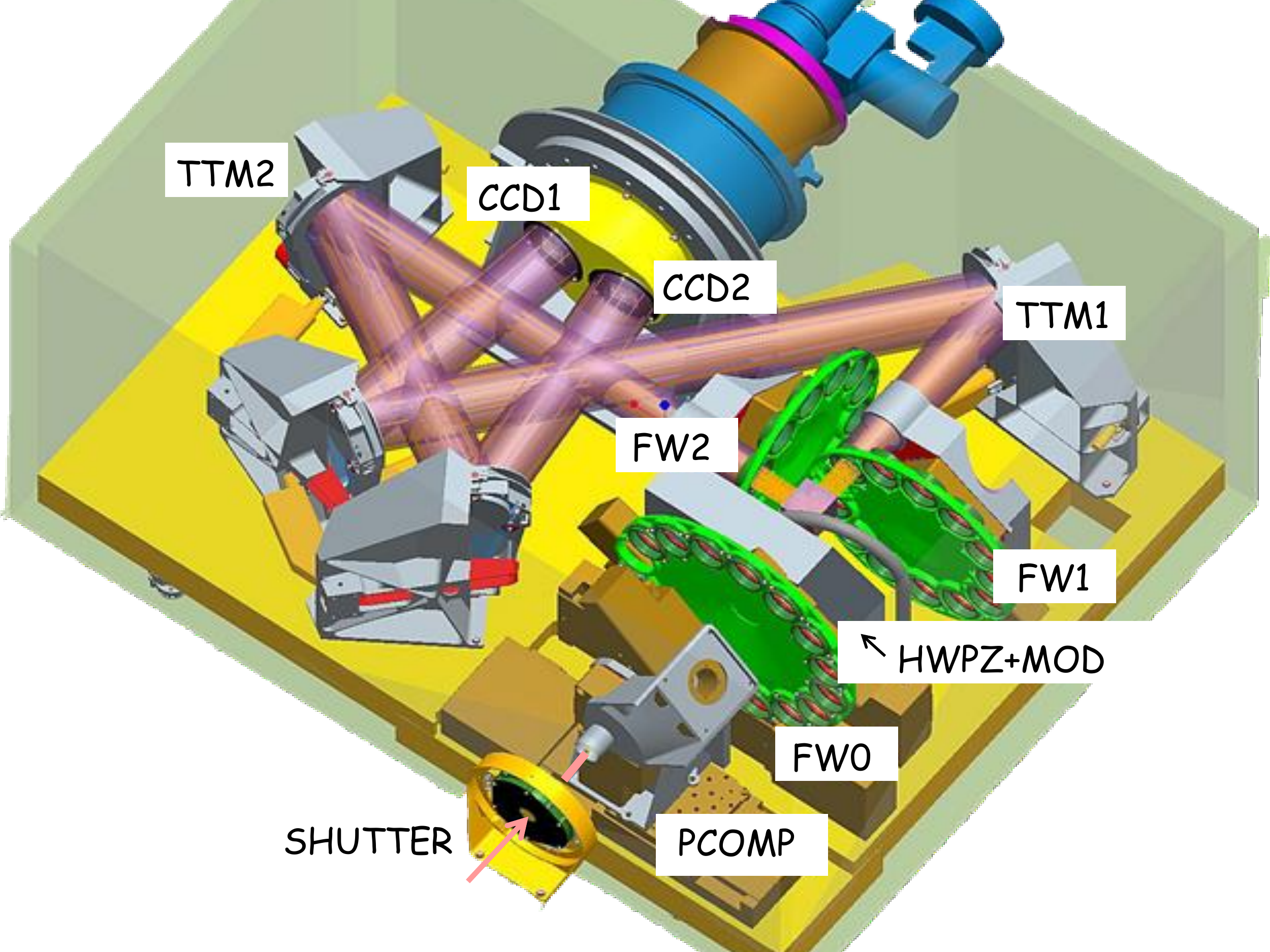}
    \caption{Opto-mechanical design of ZIMPOL with marked key components. PCOMP: polarization compensator, FW0: filter wheel in common path, HWPZ: ZIMPOL half wave plate, MOD: FLC modulator, FW1 and FW2 filter wheels in arm1 and arm2, TTM1 and TTM2: tip tilt mirrors, CCD1 and CCD2: detectors.}
    \label{FigZIMPOLa}
\end{figure}

\subsection{Sub-system description}

The ZIMPOL opto-mechanical model is shown Fig.~\ref{FigZIMPOLa} with key components identified. ZIMPOL consists of a collimated beam section with a filter wheel for bandpass selection and polarimetric calibration components, and polarization components, which can be inserted for polarimetric observations. Then follows a polarization beam splitter, which creates two arms, each with an own detector and own filter wheel. Using different filters in FW1 and FW2 allows for spectral differential imaging. The tip-tilt mirrors TTM1 and TTM2 enable  image dithering on the detectors for the removal of bad pixel effects and also for off-axis field observations. Some important characteristics of ZIMPOL are summarized in Table~\ref{TabZIMPOL}.

\begin{table*}
    \centering
    \caption{Key parameters for the ZIMPOL subsystem.}
    \label{TabZIMPOL}
    \begin{tabular}{lll}
    \hline \hline
    Observing modes       & imaging                      & Field stabilized or pupil stabilized \\
                          & polarimetry P1               & Instrument fixed (field and pupil rotating) \\
                          & polarimetry P2               & Field stabilized \\
    \hline
    Wavelength range      & 510--900 nm                  &  \\
    Filters               & 3 wheels                     & 6 broad band, 9 narrow band, 6 line, 3 neutral density filters \\
    \hline
    Detector modes        & imaging                      & High gain 10.5\,e$^-$/ADU \\
                          & fast polarimetry             & Modulation 968\,Hz, high gain 10.5\,e$^-$/ADU \\
                          & slow polarimetry             & Modulation 27\,Hz, low gain 1.5\,e$^-$/ADU \\
    Detectors             & e2v CCD 44-82                & Frame transfer CCD  \\
    Image area            & $1024\times1024$ pix         & Binned pixels with stripe mask \\
    Detector field        & $\approx 3.6 \times  3.6''$  & Same field for both detectors \\
    System field          & $8"$ diameter                & Accessible with field offsets \\        
    Pixel size            & $3.6 \times 3.6$ mas         & Illumination of every second row only \\
    Full well             & 670 ke$^-$/pix               & For high gain (fast polarimetry and imaging) \\
    Shortest DIT          & 1.1 s                        & For imaging and fast polarization \\
    \hline
    \end{tabular}
\end{table*}

The basic principle of high-precision polarimetry includes a fast polarization modulator with a modulation frequency in the kHz range, combined with a detector which demodulates the intensity signal in synchronisation with the polarization modulation \citep[e.g., ][]{Kemp81}. The polarization modulator and associated polarizer convert the degree-of-polarization signal into a fractional modulation of the intensity signal, which is then measured in a demodulating detector system by a differential intensity measurement between the two modulator states. Each detector element measures both the high and the low states of the intensity modulation and dividing by the sum of the two signals eliminates essentially all temporal changes during the measurement, notably the speckle variations introduced by the atmosphere and instrument drifts.

In ZIMPOL the modulator is a ferroelectric liquid crystal working at a frequency of about 1\,kHz, which is adapted for broad band measurements \citep{Gisler03}. The demodulator is a special ZIMPOL CCD camera that measures for each active pixel the intensity difference between the two modulation states. For this every second row of the CCD is masked so that charge packages created in the unmasked row during one half of the modulation cycle are shifted for the second half of the cycle to the next masked row, which is used as temporary buffer storage \citep{Povel90}. After many thousands of modulation periods the CCD is read out in about one second. The sum of the two images is proportional to the intensity while the normalized difference is the polarization degree of one Stokes component.

Key advantages of this technique are that images for the two opposite polarization modes are created practically simultaneously (the modulation is faster than the seeing variations), both images are recorded with the same pixels, while the storage in different buffer pixels can be calibrated with a phase-switch between subsequent images, there are only small (but still critical) differential aberrations between the two images, and the differential polarization signal is not affected by chromatic effects due to telescope diffraction or speckle chromatism.

ZIMPOL has proved to be an extremely precise technique for polarimetric imaging. It is probably the most precise differential imaging technique with array detectors available today. In solar applications ZIMPOL has routinely achieved a precision of better than 10$^{-5}$ in long-slit spectro-polarimetric mode \citep{Stenflo96,Gandorfer97}. 

A detailed description of the ZIMPOL subsystem is given in \citet{Schmid2018} and references therein. Here we highlight important system engineering aspects for the implementation of the ZIMPOL visible subsystem within SPHERE. ZIMPOL measures the polarization at the position of the fast modulator MOD. The measured polarization is composed by the polarization from the target (sky) and all the polarization effects introduced by the optical components in front of the modulator. Therefore, the modulator should be placed from a polarimetric point of view as early in the beam as possible. However, in SPHERE the concept for polarimetry had to be adapted in order to avoid any disturbing impact on the performance of the AO system, of the coronagraphic system and of the IR instruments. For this reason ZIMPOL includes an innovative polarimetric design to preserve the polarization signal from the sky throughout the system up to the ZIMPOL modulator \citep{Schmid2018,Bazzon12,Joos07}.  

The first problem is the strong instrument polarization and polarization cross-talks of the M3 mirror of the Nasmyth telescope which is actively compensated with the rotating HWP1 after the telescope mirror M3 and the following PTTM mirror (see Fig.~\ref{fig:SPHEREglobal}), which acts then as crossed mirror compensating the M3 polarization effects. For this reason the PTTM mirror had to be inclined by 45\degre and coated with aluminum, like the M3 mirror. This combination reduces the telescope polarization from about 4\% to 0.4\% \citep{Schmid2018}.

After PTTM follows HWP2 (Fig.~\ref{fig:SPHEREglobal}) which performs the standard polarization switching $Q^+$, $Q^-$, or $U^+$, $U^-$ by applying HWP2 position angle offsets of 0\degre, 45\degre, or 22.5\degre, 67.5\degre, respectively for the cancellation of the instrument polarization of all following components with a double difference (or double ratio) measuring procedure. In addition, HWP2 also rotates the polarization direction to be measured into the $Q_{\rm DROT}$-direction of the following derotator to minimize polarization cross-talk effects by the strongly inclined derotator mirrors. This $Q_{\rm DROT}$ polarization orientation is then rotated back with the half wave plate HWPZ within ZIMPOL into the $Q_Z$-orientation of the polarization modulator. The derotator also introduces instrument polarization $p_{\rm DROT}\approx 3$\,\% which is corrected with a co-rotating polarization compensation plate PCOMP within ZIMPOL. This $p_{\rm DROT}$ compensation only works well because the incidence angles for all the reflecting components in CPI for the visual beam are smaller than 15\degre and introduce essentially no additional instrument polarization (see beam geometry in Fig.~\ref{fig:SPHEREglobal}).

The visual HWPs in CPI are not compatible with IR science observations and all polarimetric components in ZIMPOL are adding additional aberrations and transmission losses for the ZIMPOL imaging mode. Therefore all the polarimetric components can be removed from the beam. Because HWP1 and HWP2 are in the diverging beam, introducing them changes the entrance focus of SPHERE which should coincide with the telescope focus. This is compensated for ZIMPOL by changing the telescope focus using M2 during on-sky observations of a target. For internal alignment and PSF calibrations the or the position of the point source in the SPHERE calibration unit can be moved for calibrations of the ZIMPOL polarization mode \citep{Wildi10}. In addition CPI and ZIMPOL have both exchange units with polarimetric calibration components for the initial adjustments, checks and calibration of the CPI and ZIMPOL instrument polarization \citep{Bazzon12}. On top of this, the telescope polarization needs to be determined with on-sky standard star measurements \citep{Schmid2018}. 

Thus, ZIMPOL polarimetry, with all its insertable and rotating components, and the requirements on the overall design and operation, adds very significantly to the SPHERE system complexity. Having a visual imaging system without high precision polarimetry would be a much simpler alternative, but one would loose one very important functionality. For this reason the scientific performance of ZIMPOL polarimetry is also of much interest for the planning of future high contrast systems with polarimetric modes.

\subsection{Performance and associated results}

\subsubsection{PSF characteristics}

The SAXO adaptive optics system delivers under good atmospheric conditions in the visual a PSF with a Strehl ratio of about 50\% (Sect.~\ref{sec:saxo}). This sounds like mediocre, because the Strehl is around 90\% in the near-IR, but one needs to consider the strong wavelength dependence of the diffraction limited PSF parameters. A PSF characterization parameter, which is much less wavelength dependent is the normalized peak surface brightness. This parameter yields for ZIMPOL about ${\rm SB}_0-m_{\rm star}\approx -6.5^m$\,arcsec$^{-2}$ in the $V$-, $R$-, and $I$-band \citep{Schmid2018}. ZIMPOL reaches routinely an angular resolution (FWHM) of about 22--27\,mas close to the diffraction limit \lsd of 17\,mas in the $R$-band and 21\,mas in the $I$-band. The PSF width degradation is mainly caused by residual atmospheric aberrations, telescope jitter, and quasi-static instrument aberrations. 

The PSF quality for visual ZIMPOL observations varies strongly with atmospheric conditions. The PSF peak surface brightness is often seen to vary by up to a factor of two (best to worst PSF), if atmospheric conditions are less than medium (seeing $\sim$1\as, $\tau_0$ $\sim$2\,ms). The strong PSF variations require for quantitative photometric and polarimetric measurements a careful observing and calibration strategy.

For faint sources, the AO performance is significantly worse for ZIMPOL when compared to the near-IR, because the wave front sensor (WFS) has to share the light with the ZIMPOL science channel. With the gray beam splitter between WFS and ZIMPOL only 21\% of the light is deflected to the WFS. Thus, the limiting magnitude for the WFS star is about $m_R\approx 8^m$ or about 1.75 magnitudes lower, when compared to the infrared channels, which use a mirror instead of a beam-splitter to feed the visual WFS. For fainter stars, the WFS receives not enough light, and the PSF peak brightness can be reduced by more than a factor of ten and the FWHM can be degraded to $>50$\,mas \citep{Schmid2018}. Alternatively, the available dichroic beam splitter can be used which transmits only the wavelength range 610--690\,nm to ZIMPOL and all other wavelengths, or 80\% of the light, to the WFS. This allows only for science observations in a narrow $R$-band (N\_R), and H$\alpha$ and O\_I line filter, but the AO system performs well for fainter stars up to about $R \approx 9.5^m$ because the WFS receives more light. With this approach, even good results were obtained for example for polarimeric observation of circumstellar disks of the rather faint star RX~J1604.3-2130 with $R = 11.8$ \citep{Pinilla15}.

The polarized Stokes intensities $Q$ and $U$ of a stellar PSF or coronagraphic image from an unpolarized source should in principle show no significant signal. However, the telescope introduces a polarization signal $p_{\rm tel}\cdot I$ in the form of a weak copy of the intensity signal. This was expected and a normalization between opposite polarization modes, for instance $I_0$ and $I_{90}$, or the multiplication of $Q$ or $U$ with a correction factor solves this problem 

Unfortunately we detected during the testing of ZIMPOL an unexpected beam-shift of about 0.5\,mas between opposite polarization modes because of the inclined mirrors in the VLT and SPHERE \citep{Schmid2018}. A shift of 0.5\,mas is very small, but it becomes the dominant problem for high contrast polarimetric imaging with ZIMPOL. The beam shift can be corrected with special procedures in the observing strategy and the data reduction so that the initially planned polarimetric contrast limits can still be reached (Hunziker et al. in prep.). 

\subsubsection{High-resolution imaging} 
\label{sec:zimpol_hr_imaging}

The high resolution of ZIMPOL was successfully used to separate close binaries, with separations less than 0.1\as, for example 45\,mas for R Aqr \citep{Schmid17}, or 38\,mas for AB Dor B \citep{Janson18}, or for resolving structures on objects with diameters smaller than 0.1\as, for example on nearby red giant stars \citep{Kervella16,Khouri16,Ohnaka16} or asteroids \citep{Vernazza18}. 
 
\subsubsection{Astrometry}

The high-resolution of ZIMPOL is of course useful for accurate ($\sim$1\,mas) relative astrometry between the bright central star and faint companions. For the AB~Dor~B binary it was possible to measure separations with a precision of $\pm$1\,mas \citep{Janson18}, and with the new, accurate astrometric calibration of ZIMPOL (Ginski et al. in prep.) even more accurate measurements should be achievable for easy targets. 

\subsubsection{Photometry}

Relative photometry between companion and central star is relatively easy and accurate, if the contrast is modest $f_{\rm comp}/f_{\rm star}> 0.001$ and the separation not too demanding >0.1\as, because then both objects can be measured simultaneously such as for PZ~Tel \citep{Maire16} or $\alpha$~Eri \citep{Schmid2018}. A more demanding contrast requires an observing strategy which overcomes the PSF variation problems. 

Selecting a narrow band filter in one channel and broad band filter in the other channel provides simultaneous flux measurements with ratios $f_{\rm comp}/f_{\rm star}< 0.001$ . However, for accurate relative measurements (better than $\pm 5^\circ$) one needs to take the polarization dependent throughput of the VLT, CPI, and ZIMPOL into account. 

Absolute photometric measurements for the central star or a target near a variable wave front probe is also possible with zero-point calibrations \citep[e.g., for R Aqr][]{Schmid17}. Certainly, SPHERE is not built for accurate photometry of bright stars, but of course a photometric calibration is useful for enhancing the value of high contrast data. Unfortunately, a detailed photometric characterization of ZIMPOL is still pending.  

\subsubsection{High-contrast imaging} 
\label{sec:zimpol_hc_imaging}

The high spatial resolution of SPHERE provides especially in the visual a small inner working angle for high contrast imaging. Best results published up to now are a contrast of $\Delta m=6.5^m$ at a separation of 91 mas for $\alpha$ Hyi B in the $R$-band \citep{Schmid2018}, or $\Delta m=7.3$ at 63\,mas for HD\,142527 B in the narrow CntHa-filter \citep{Cugno19}. 

There exist not many faint companion detections with separations $>100$\,mas with ZIMPOL. Usually faint companions are low mass objects which are much brighter and observationally less demanding objects in the near-IR. One detection based on a commissioning test measurements taken under mediocre condition is reported by \citet{Maire16}, who measured for PZ~Tel B a contrast of $\Delta m=9.8^m$ at 480\,mas in the $R$-band. 

High contrast limits for the intensity signal of faint companions were also determined for the very deep search of polarized light around the nearest stars (Hunziker et al. in prep.). For $\alpha$ Cen A in the $R$-band, $5\sigma$-contrast limits of about $\Delta m=12^m$ are derived for the separation range 200-400\,mas, and $\Delta m>15^m$ for separations >750\,mas using long integrations combined with angular differential imaging and a principle component analysis applied to coronagraphic stellar images.  

\subsubsection{Aperture polarimetry} 

The ZIMPOL polarimetric mode is characterized in detail in \citet{Schmid2018} using zero and high polarization standard stars. The strongest instrumental polarization effect is the residual telescope polarization, which is at the level of $p_{\rm tel}\approx 0.5$\,\% and rotates with the paralactic angle and other, smaller effects are also present. Applying appropriate calibrations provides a polarimetric accuracy of about $\Delta p\approx \pm 0.1$\,\% for low polarization objects and about $\Delta p\approx \pm 0.2$\,\% for bright objects with high polarization $p>1\,\%$. The uncertainty is enhanced for high polarization targets because also polarization cross-talks of the system contribute to the error budget.

\subsubsection{High contrast imaging polarimetry}
\label{sec:zimpol_high_contrast_pol}

ZIMPOL is built for high contrast differential polarimetric imaging and the instrument produced already many science results for polarimetric differential imaging of circumstellar scattered light, mainly dust scattering from circumstellar disks around young stars \citep[e.g., ][]{Garufi16,Stolker2016,Benisty2017,vanBoekel2017}, but also dust scattering in the wind of mass loosing red giants and supergiants \citep{Khouri16,Kervella16,Ohnaka16}.   

\begin{figure}
    \includegraphics[width=0.5\textwidth]{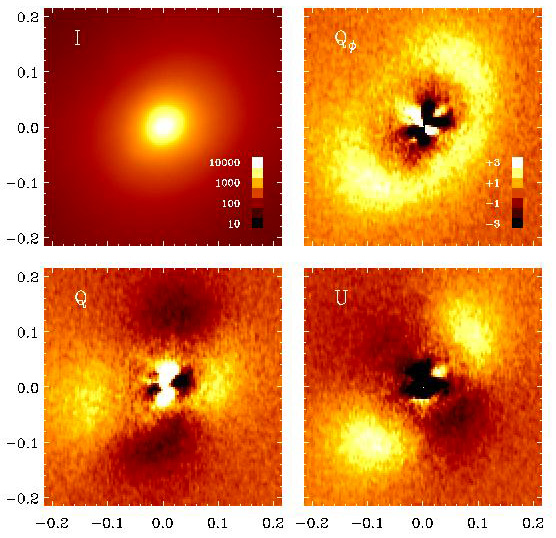}
    \caption{ZIMPOL polarimetry of the circumstellar disk of HD\,100546. The panels show the intensity signal (top left), the Stokes $Q$ and $U$ intensities (lower row) and the polarized flux expressed as $Q_\phi$ in the top right panel. The color scale are given in units of cts per pixel and frame, and all polarimetric images are plotted on the same scale shown in the $Q_\phi$ panel.}
    \label{fig:HD100546}
\end{figure}

As an example, Fig.~\ref{fig:HD100546} shows imaging polarimetry taken in the R\_PRIM-filter ($\lambda_c=626$\,nm) of the circumstellar disk around
HD\,100546 based on a subsample of the data presented in \citet{Garufi16}. The figure shows
only the very central $0.4"\times 0.4"$ region highlighting the high resolution, the
small inner working angle and the fidelity of ZIMPOL differential polarimetric 
measurements which allows to map accurately the polarization signal of the inner disk wall down 
to a separation of $0.05"$. This is possible, because the dynamic range of ZIMPOL polarimetry is large enough to measure faint polarization signals very close to a bright star in non-coronagraphic mode, keeping thanks to the fast modulation the differential residuals of the star well confined inside
a small radius. 

Polarimetric imaging of circumstellar scattered light profits strongly from this high spatial resolution because the measurements of the Stokes intensities $Q=I_{0}-I_{90}$ or $U=I_{45}-I_{135}$ need to separate the positive and negative differential signal regions. If these regions are not well resolved, then significant cancellation between positive and negative signal occurs and the measurable polarization is strongly reduced. 

For extended circumstellar sources the polarimetric performance can be expressed as polarized surface brightness contrast $C_{\rm SBpol}={\rm SB}_{\rm pol}-m_{\rm star}$. Typically, protoplanetary disks have $C_{\rm SBpol}\approx 5\,{\rm mag\,arcsec}^{-2}$. 

Debris disks are fainter and measured surface brightness contrasts for HIP\,79977 are $7.6\,{\rm mag\,arcsec}^{-2}$ at a separation of $0.25''$ or $9.0\,{\rm mag\,arcsec}^{-2}$ at $1''$ \citep{Engler17}. Comparable values are measured for the polarized signal from the dust shell around the symbiotic mira star R Aqr with $5\sigma$ detection limits which are about $3\, {\rm mag\,arcsec}^{-2}$ deeper \citep{Schmid2018}.

A measurement for the polarimetric point source contrast $\Delta m_{\rm pol}=m_{\rm pol}-m_{\rm star}$ has not been reported yet. But deep limits have been determined by the REFPLANET search program of the SPHERE consortium. For example for $\alpha$ Cen A, the achieved $5\sigma$-limit is about $\Delta m_{\rm pol}\approx 17$\,mag for the separation range 0.2\as to 0.5\as and >20\,mag outside 1\as (Hunziker et al. in prep.). The contrast sensitivity in this long observation of 3.3\,h is limited outside of 0.7\as by the photon noise. This means, that the contrast can even be further improved if longer integrations are taken. 

\section{IFS}
\label{sec:ifs}

In this section we describe the SPHERE integral field spectrograph (IFS). Early extensive general descriptions of the SPHERE IFS design can be found in \citet{Claudi2006,Claudi2008,Claudi2010}. The optical design is detailed in \citet{Antichi2008a}, and the mechanical design and control hardware in \citet{DeCaprio2008,DeCaprio2010}. The integral field unit (IFU) principle is discussed in \citet{Antichi2008b,Antichi2009} and details about its construction are in \citet{Giro2008}. The calibration scheme is presented in \citet{Desidera2008}. Integration of the IFS in the laboratory is described in \citet{Claudi2012}. Simulations of IFS results are described in \citet{Mesa2011}, methods for detection and characterization of faint companions with the IFS in \citet{Zurlo2014}, high level laboratory results in \citet{Mesa2015}, early on-sky results in \citet{Claudi2014,Claudi2016}, and the astrometric calibration in \citet{Maire2016}.

\subsection{IFS High-level scientific and technical requirements}

The SPHERE IFS was primarily designed to provide the highest possible contrast for point source detection in the immediate surroundings of the star (0.15\as-0.70\as, with the goal of a contrast of $10^{-6}$\ at 0.5\as) exploiting both ADI and SDI. The possibility to have a low resolution spectrum of every pixel present in the field of view is an obvious advantage for characterization but it was not the primary driver in the instrument design.

IFS was conceived to be used in parallel with IRDIS and to provide the highest contrast even at some sacrifice of the field of view, wavelength coverage, and sensitivity to the faintest targets. Therefore, in IFS, only the detector is at cryogenic temperature, with an upper limit in wavelength of about 1.65\,\mic. While for bright targets ($J < 6$\,mag), the main sources of noise are photon statistics and residuals from speckle subtraction, background noise is the limiting factor for observations of faint targets. We set an upper limit of 20\,e$^-$\ per pixel to background noise, close to the value we expected for the detector read-out noise as originally specified by the constructor (actual readout noise is significantly lower). Since the detector is sensitive up to $\sim$2.5\,\mic, a low enough background noise ($\sim 10$\,e$^-$\ per pixel, depending on ambient temperature) is obtained by a combination of a low pass-filter and of a suitable baffling system. However, background noise is still the limiting factor in high-contrast observations for targets with $J > 9$\,mag.

The 4-d data-cubes ($x$, $y$, time, and wavelength $\lambda$) provided by an IFS enable to combine both ADI and SDI techniques. While ADI can be applied on monochromatic images to remove quasi-static speckles, SDI exploits the smooth variation of speckle properties with wavelength to remove them from the images. To avoid self-cancellation of the signal, SDI works at separations larger than the so-called bifurcation radius \citet{Thatte2007}. To minimize the bifurcation radius and to be compatible with exploitation of the dual band imaging capabilities of IRDIS \citep{Vigan2010}, the SPHERE IFS has two possible configurations with $\lambda_{\rm min}=0.95$\,\mic: Y-J ($\lambda_{\rm max}\sim 1.35$\,\mic) and Y-H ($\lambda_{\rm max}\sim 1.65$\,\mic). The Y-H mode provides a slightly deeper contrast and is effective at shorter separation than the Y-J mode. On the other hand, use of the $H$-band for IFS implies that only the $K$-band is available for IRDIS, which leads to some loss of performances for IRDIS because the K12 filter pair is less efficient than the H23 one to separate small mass companions from background stars, and because the broader wavelength range makes the coronagraph less efficient at short separation. However, this might be paired by the rise of the spectral energy distribution of those sub-stellar objects that have a late L-type spectrum or are very heavily reddened. Having both alternatives available added versatility to SPHERE at low cost.
 
The SPHERE IFS is based on a lenslet integral field unit (IFU). The IFU is located at a focal plane; each lenslet acts as a slit and light from each lenslet is dispersed into a spectrum on the detector. The spatial resolution of the IFS is set by the condition of having a Nyquist sampling of the diffraction peak at 0.95\,\mic that implies that lenslet centers are separated by 0.01225\as. Spectral resolution and field of view (FoV) are the result of a compromise with the detector size (a 2k$\times$2k Hawaii II detector). To minimize the number of detector pixels dedicated to each spaxel, we developed a new optical concept for the IFU \citep[BIGRE][]{Antichi2009}. This left about 35 pixels available for each spectrum along dispersion, setting the two-pixel resolution at $\sim$50 for the Y-J mode and at $\sim$30 for the Y-H mode, dispersion being not exactly constant along the spectrum. A more detailed description of the IFS is given in Appendix \ref{sec:apdx:IFS_description}. 

\begin{figure}
    \centering
    \includegraphics[width=0.4\textwidth]{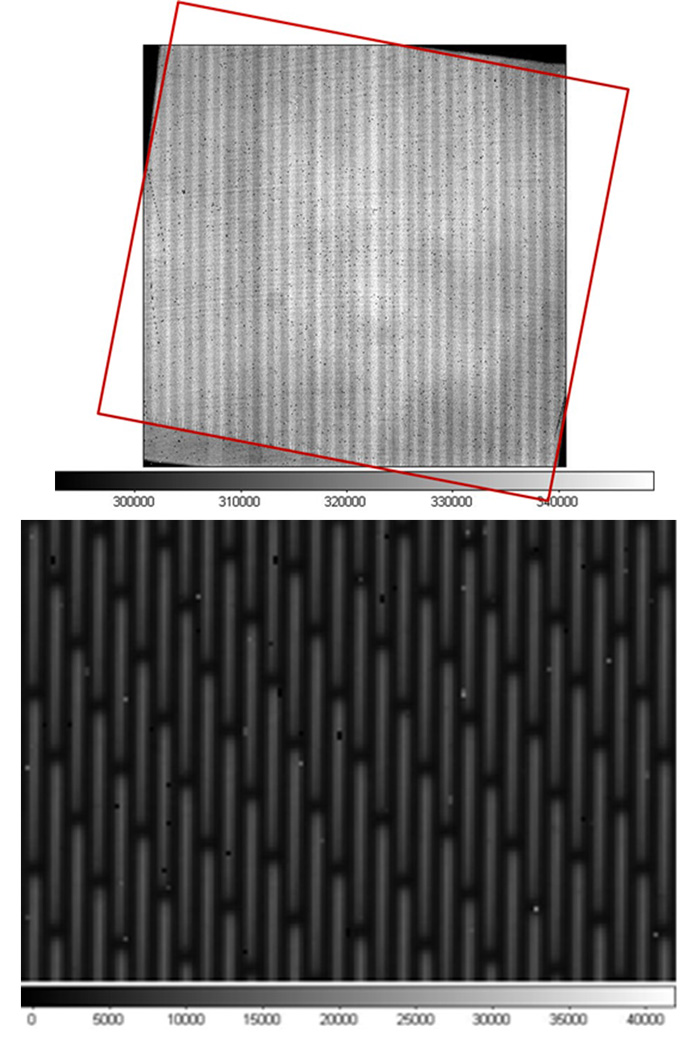}
    \caption{Upper panel: Format of images on the IFS detector. The top image is an overlay of an detector flat field (obtained using the IFS internal calibrating sphere) and of a flat field image obtained through the BIGRE. The whole detector is a square of 2048$\times$2048 pixels, shown with vertical and horizontal sides. The area within the red square represents the projection of the BIGRE on the detector; this is a rectangle, which sides make an angle of 10.5 degrees with the detector; this is the angle between the lenslet array and the spectra, and could not be avoided. Some 9\% of the BIGRE useful area is lost due to vignetting. The black region at the corner of the detector image is the area of the detector for which no internal flat field is possible due to vignetting by the cold filter holder. This is completely outside the area covered by BIGRE, so that there is no field loss due to the cold filter holder.
    Lower panel: portion of an image obtained with a white lamp with IFS in the Y-H mode}
    \label{fig:IFS_detector_image}
\end{figure}

\subsection{IFS Performance and limitations}

IFS was optimized and characterized in laboratory, first at INAF-Osservatorio Astronomico di Padova \citep{Claudi2012} where it was assembled, then at IPAG (Grenoble, France) where it was integrated into SPHERE \citep{Claudi2014,Mesa2015}, and finally on-sky during the commissioning phase at the observatory \citet{Claudi2014,Claudi2016}. In this subsection we briefly discuss some of the main results of these tests. 

An image of a white lamp acquired using the IFS is shown in Figure~\ref{fig:IFS_detector_image}. The lower panel shows a blow-up of a small portion of this image, better showing the individual spectra provided by each lenslet. The FoV is approximately square with a side of $\sim$1.73\as projected on sky and with some vignetting at the edges of the field of view due to the mounting of the cold filter. Since each lenslet samples about 0.01225\as on-sky, there are about 23\,140 spectra on an image. Given the adopted geometry, spectra are aligned along columns on the detector: each spectrum occupies a region of $41\times5.093$ pixels on the detector. 

Each spectrum has a length of 35.4 pixel, and covers the wavelength range 0.96--1.34\,\mic with the Y-J prism, and 0.97-1.66\,\mic with the Y-H prism. The wavelength calibration of the spectra is obtained illuminating the lenslet array with light from four lasers lamps (at 0.9877, 1.1237, 1.3094, and 1.5451\mic) in the calibration unit of SPHERE. An automatic procedure performs the calibration from pixel to wavelength, that is expected to be represented approximately by a cubic relation due to the use of prisms dispersers. We tested the accuracy of this calibration by measuring the wavelength of the laser lines over the extracted spectra, and we found a scatter of $\sim$2\,nm RMS for both the Y-J and Y-H modes, that was the original specification. In the Y-J mode, the median full width at half-maximum (FWHM) of the laser lines are 19.5, 25.3, and 30.4 nm at 0.9877, 1.1237, 1.3094\,\mic. These values correspond to spectral resolution wavelength of 51, 44, and 43 respectively. Given the dispersion at the three wavelengths (9.23, 11.3, and 13.2\,nm/pixels, respectively), the FWHM is between 2.1 and 2.3\,pixels at all these wavelengths, which is the value expected considering the width of the diffraction peak.

\begin{figure}
    \centering
    \includegraphics[width=0.5\textwidth]{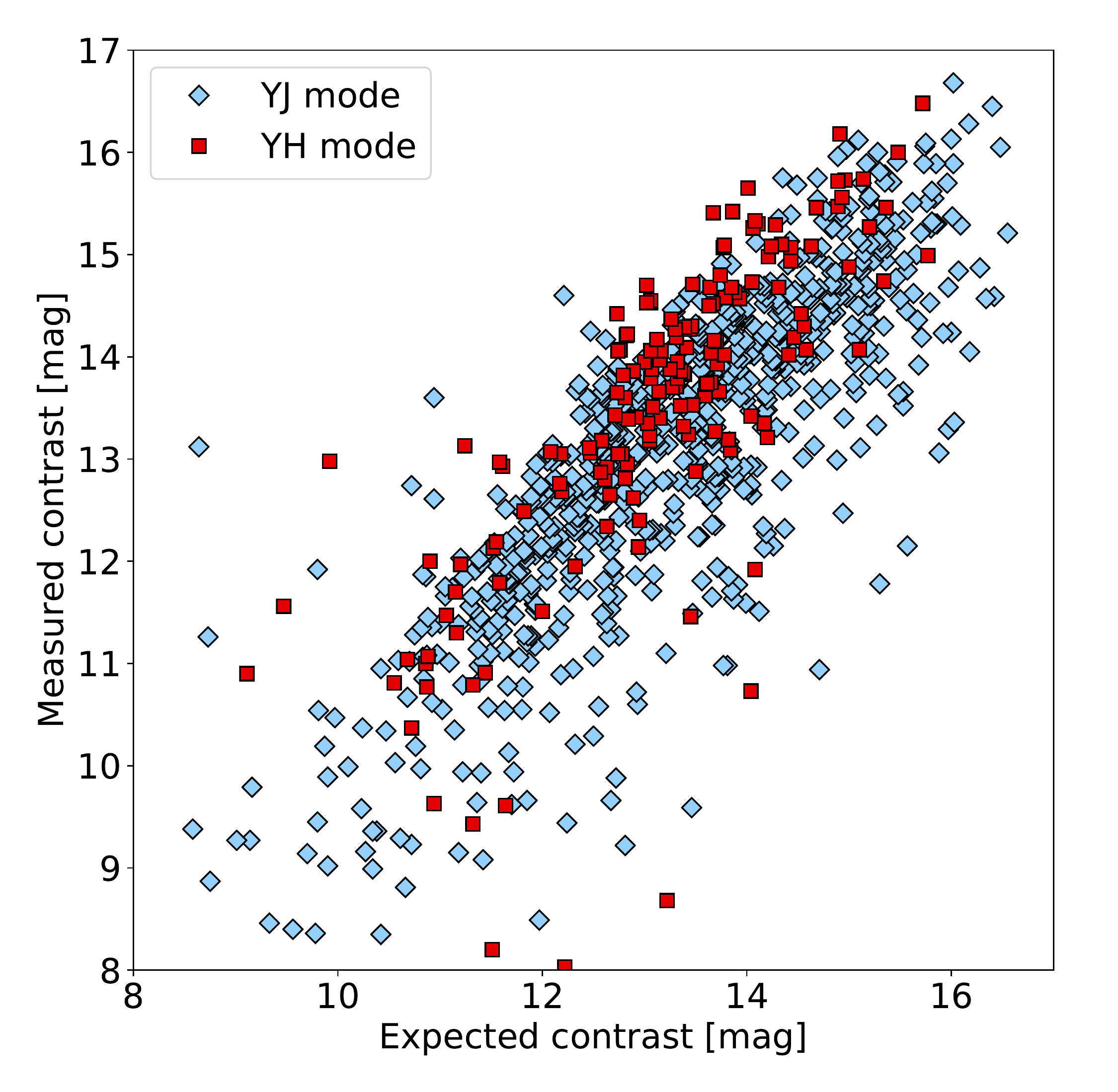}
    \caption{Comparison between measured and expected contrast at 0.5\as from the star in YJ and YH modes. The expected contrast is computed using the noise model for IFS data (Appendix~\ref{sec:apdx:noise_model_ifs}).}
    \label{fig:IFS_contrast}
\end{figure}

The IFS optical transmission was measured by comparing the flux measured in the focal plane, with that measured just in front of the BIGRE unit. The experiment was done using both Y-J and Y-H prisms, including filters and the masks on the intermediate pupil. An appropriate correction factor was included to account for the fraction of light lost when using the sensor because of its limited size. The overall transmission ranges from 53\% to 57\%, with a weak dependence on wavelength. This is in good agreement with the values expected considering losses by the BIGRE design and from the various optical elements.

The relevance of flat field accuracy on the limiting contrast achievable with IFS was carefully examined. We found that the impact of the flat field accuracy is almost negligible even for bright objects if the rms is $10^{-3}$. This confirms that the achieved accuracy obtained without dithering is enough to avoid detector flat fielding limiting detections.

The IFU flat field is a specific calibration used to measure the wavelength-dependent transmission of individual lenslets and the accurate position of the spectra on the detector. IFU flat images are obtained by illuminating the IFS with the SPHERE continuum lamp. Sensitivity of the IFU flat on dithering was obtained by comparing sequences of IFU flats obtained at different dithering positions. Comparison of different IFU flats obtained on different dates shows that without dithering, the RMS accuracy is $2.3\times 10^{-3}$. Accuracy is much poorer when dithering is applied, probably because of the combination of a high sensitivity of the pixel allocation table and imperfect dithering calibration. 

As discussed in \citet{Antichi2009}, a diffraction limited lenslet based IFS should suffer from two types of optical cross-talk: the coherent cross-talk, from interference between monochromatic signals from adjacent lenslets arising already at the spectrograph's entrance slits plane, and the incoherent cross-talk, that is the value of the spectrograph's line spread function (LSF) evaluated at the position of adjacent spectra. The coherent cross-talk should be proportional to the square root of the LSF spatial profile as imaged at the detector plane, while the incoherent cross-talk should be simply proportional to it. Hence, at large separations from a fixed spectrum incoherent cross-talk dominates over the coherent cross-talk, while it is the opposite at short separations. BIGRE was specifically designed to minimize both kinds of cross-talks by carefully apodizing the spectrograph's LSF. The final measured values at the distance to the closest lenslet are $(6.9\pm 0.7)\times 10^{-3}$ and $(3.7\pm 0.7)\times 10^{-3}$ for the coherent and incoherent components, respectively.

\begin{figure}
    \centering
    \includegraphics[width=0.5\textwidth]{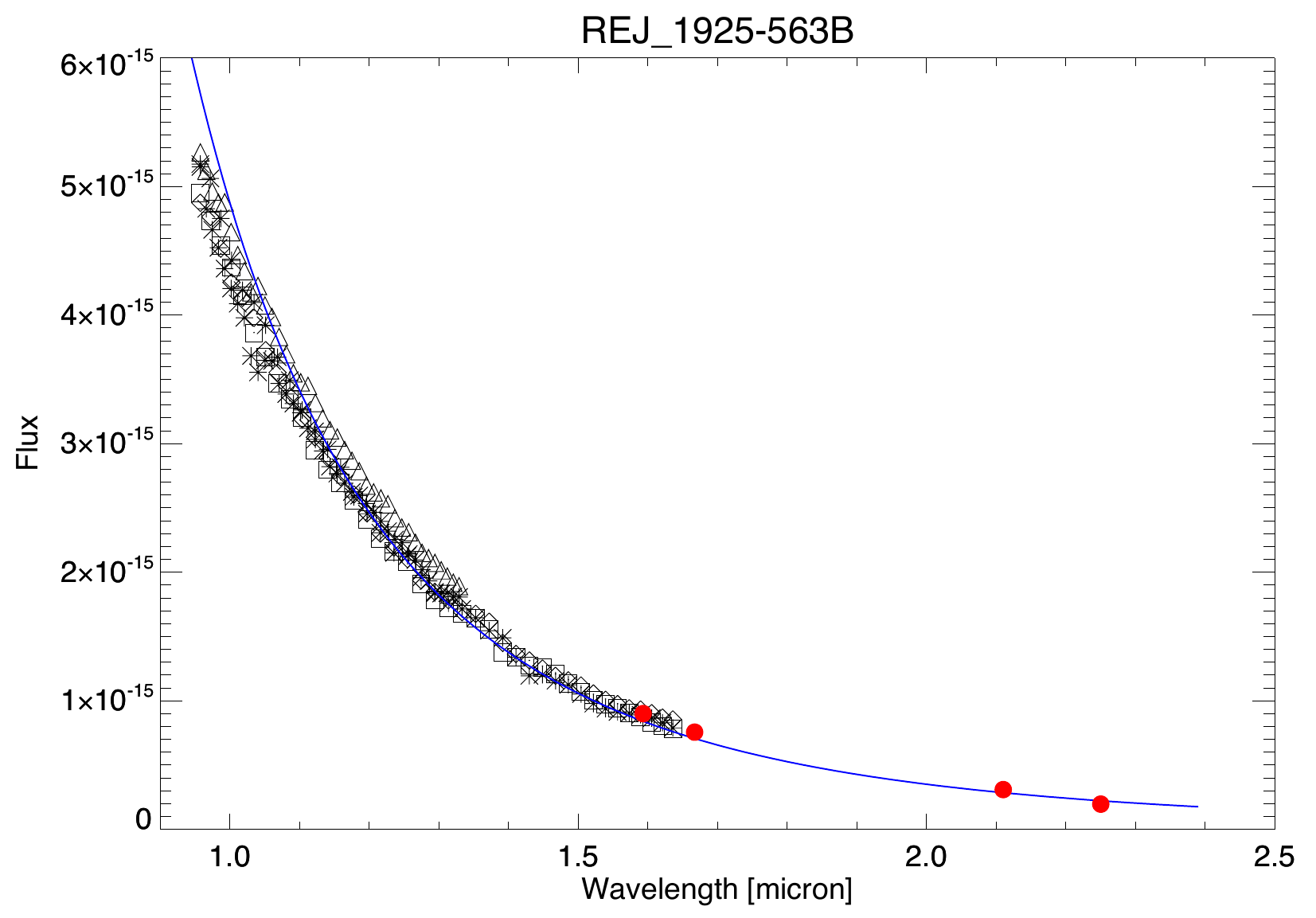}
    \caption{Spectra of the hot white dwarf companion to REJ1925-563 obtained with the SPHERE IFS. Black symbols are data from IFS, red ones from IRDIS (H23 and K12). Different symbols are spectra obtained at different epochs. The solid line is a Rayleigh-Jeans fit to the observed spectra}
    \label{fig:IFS_spectrum}
\end{figure}

\subsection{On-sky performance of IFS}
\label{sec:ifs:on-sky_results}

SPHERE data with IRDIS and IFS are usually normalized using the flux calibration, which is obtained by observing a star offset out of the focal plane coronagraphic mask, in combination with a neutral density filter \citep[e.g., ][]{Vigan2015}. This assumes knowledge of stellar magnitudes in J, H, and Ks (generally from 2MASS), and stability during the observation (both atmospheric extinction and Strehl ratio). 

Better photometric accuracy can be achieved by acquiring science images with satellite spots. These satellite spots (often erroneously called ``waffle spots'') are generated by introducing a 2-d periodic modulation on the HODM, which creates satellite spots that are replicas the stellar PSF, scaled in intensity. The variations in intensity of the satellite spots, either because of Strehl or transmission variations, are expected to be correlated to the intensity variations of the faint companion's PSF. Flux in satellite spots is proportional to the square of the waffle intensity on the HODM and can then be tuned in the observing template to have an intensity comparable to that of the companion.

To show the gain of this procedure we considered a sequence of 16 images of PZ~Tel acquired during COM3. The overall sequence took about 20 minutes, which is much less than the expected rotation periods of both the star and the brown dwarf companion \citep{Neuhauser2003}. The zero point of photometry is here defined by the average of the four satellite spots. The scatter around mean values can be attributed to fluctuations in sky transmission and Strehl ratio (i.e. AO correction). For this particular sequence, such fluctuations are not very large, yielding to a scatter of about 0.025\,mag rms. There is good correlation between the variation of intensities of the satellite spots and companion images. We may then use the average intensity of the four spots to correct the photometry of PZ~Tel~B, which reduces the scatter to below 0.02\,mag. The residual scatter for the photometry of PZ~Tel~B is clearly much higher than the photon noise. It might be attributed to speckle noise at the position of both the companion and of the waffle spots. We conclude that with IFS it is possible to obtain photometric sequences accurate to $\pm$0.02\,mag. A more detailed analysis on the use of satellite spots for photometry monitoring can also be found in \citet{Apai2016} for HR~8799.

The main purpose of SPHERE is to provide high contrast images at very short separations. Although recent work have showed the limitations of this approach \citep{Mawet2014,Jensen-Clem2018}, we quantify this capability using the 5-$\sigma$ contrast level as a function of separation in the present work. We also only present here results homogeneously obtained from pupil-stabilized observations obtained with the APLC coronagraph on a large number of targets, and analyzed with PCA applied on 4-d data-cubes using simultaneously ADI and SDI \citep{Mesa2015}. Cancellation effects are considered using corrections estimated from fake companions injection, which are scaled versions of the stellar PSF images. We considered results obtained using a range of subtracted modes (from 10 up to 150 for field rotation larger than ten degrees, and from one to sixteen modes for smaller field rotations). The area around the star was divided into rings of 0.1\as-width, and for each ring we considered the residual image that yielded the best contrast. Over the years, we accumulated over 800 data sets. To interpret them, we constructed a noise model considering four terms (calibration, photon noise, thermal background, and read out noise). The final expected contrast C (in mag) is obtained by combining the various noise sources. Details about this model are given in Appendix~\ref{sec:apdx:noise_model_ifs}.

Figure~\ref{fig:IFS_contrast} compares the contrast at 0.5\as from the star expected using these equations with actual observations. As it can be seen, in spite of its simplicity, the model captures the main dependencies on the observing conditions and stellar magnitude. At 0.5\as the contrast achievable with SPHERE IFS is limited by photon noise on bright sources and by thermal background for the faintest ones. At short separations ($<0.2\as$), residual uncorrected speckles dominate over other source errors, and the IFS is limited to contrast values of 11 mag at $<0.15\as$. This is due to residual low order aberrations and the low-wind effect \citep[see][]{Milli2018}.

The best $5-\sigma$\ contrast at 0.5\as we were able to obtain with SPHERE IFS is about $2.5\times 10^{-7}$ (16.48 mag) for a J=4.7 mag star observed in good but not exceptional sky conditions, seeing of 0.6\as and coherence time of 6.9\,ms. In roughly 50\% of the cases we obtained a contrast better than $10^{-6}$\ for stars with $J < 6$\,mag observed under median or good conditions. 

IFS can also be used to extract (low resolution) spectra of faint companions. This potentiality is illustrated by Figure~\ref{fig:IFS_spectrum} that shows the spectra of the hot white dwarf companion to the main sequence star REJ1925-563, in addition to the photometry points extracted from IRDIS data. As expected, this spectrum is very well reproduced by a Rayleigh-Jeans curve. As already mentioned, IFS has been designed to be used in combination with IRDIS in the so-called IRDIFS mode. The early scientific results obtained by the combination of the two instruments will be discussed in the forthcoming Sect.~\ref{sec:irdifs_results}.

\section{IRDIS}
\label{sec:irdis}

\subsection{High-level scientific and technical requirements}

The IRDIS differential imaging camera and spectrograph provides imaging, spectroscopy, and polarimetry in two parallel channels, covering a wavelength range from 0.95 to 2.4\,\mic over a wide FoV (11\as$\times$11\as in imaging, 10\as in spectroscopy) with a spatial sampling of 12.25\,mas/pixel (Nyquist-sampled at 0.95\,\mic). This multi-purpose instrument is divided into four observing modes namely dual-band imaging mode \citep[DBI;][]{Vigan2010}, Dual Polarimetric mode \citep[DPI;][]{Langlois2010b}, Long-slit spectroscopy mode \citep[LSS;][]{Vigan2008}, and classical imaging mode (CI). The main science case that drives the IRDIS specifications is the exoplanetary survey as illustrated in Table \ref{tab:IRDISmodes} but complementary specifications have been accommodated to ensure wider scientific returns in particular for circumstellar disk, close stellar environment and planetology. These wide range of scientific results are illustrated in the following sections describing the different modes.

\begin{table*}[t]
    \centering
    \caption{IRDIS observing modes and high-level specifications.}
    \begin{tabular}{cccc} \hline
    Mode & Science case                & Wavelength coverage and resolution & Contrast \\ 
	\hline \hline		
    DBI  & Companions detection        & $Y$, $J$, $H$, $Ks$                & $10^{-4}$ at 0.1\as \\
         &                             & Dual-band (R=30)                   & $10^{-5}$ at 0.5\as \\
    \hline
    DPI  & Planet formation            & $Y$, $J$, $H$, $Ks$                & $10^{-4}$ at 0.1\as \\
         &                             & Broad-band (R=5)                   & $10^{-5}$ at 0.5\as \\ 
         &                             &                                    & for 30\% polarized circumstellar source\\ 
    \hline
    CI   & Multi-purpose               & $Y$, $J$, $H$, $Ks$                & $10^{-4}$ at 0.1\as \\
         &                             & Narrow-band (R=70-80), broad-band (R=5)   & $3\times10^{-4}$ at 0.5\as \\ 
    \hline
    LSS  & Companions characterization & LRS: $YJHKs$ (R=50)                & $10^{-4}$ at 0.3\as \\ 
         &                             & MRS: $YJH$ (R=350)                 & $10^{-5}$ at 0.5\as \\ 
    \hline 
    \end{tabular}
    \label{tab:IRDISmodes}
\end{table*}

\subsection{Sub-system description}
 
\begin{figure}
    \centering
    \includegraphics[width=0.5\textwidth]{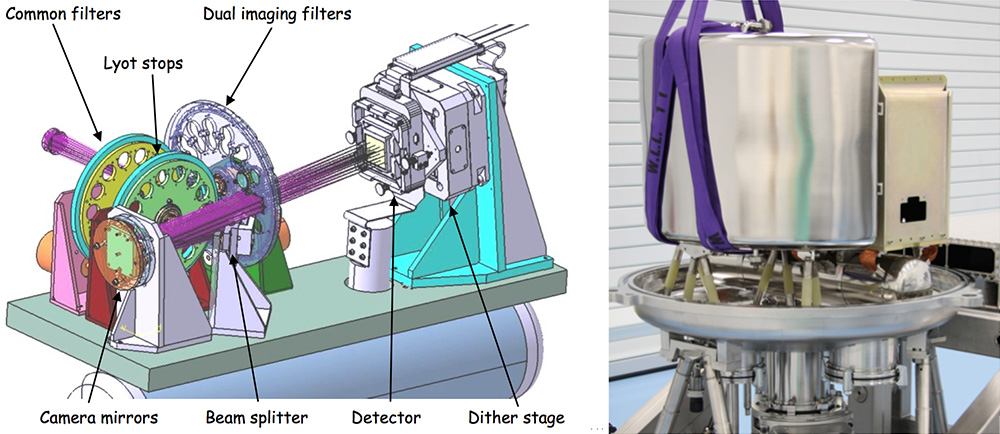}
    \caption{IRDIS opto-mechanical layout (left), and picture of the IRDIS cryostat during integrations (right).}
    \label{fig:IRDIS_layout}
\end{figure}

The opto-mechanical implementation of IRDIS is shown in Fig.~\ref{fig:IRDIS_layout} and described in more details in \citet{Dohlen2008}. A beam splitting plate associated with a mirror separates the main beam in two parallel beams. Three wheels are provided within the cryogenic environment. The first common filter wheel carries wide-band (WB) used in LSS mode, and broad-band (BB) and narrow-band (NB) filters for classical imaging. Then Lyot stop wheel carries all the Lyot stops for coronagraphy, as well as the prism and grism coupled to a slightly undersized circular Lyot stop used in LSS. Finally, a second filter wheel carries the dual-band (DB) imaging filter pairs and polarizers located down-stream of the beam-separation unit. Two parallel beams are projected onto the same 2k$\times$2k Hawaii II-RG (see Sect.~\ref{sec:apdx:detectors}) with 18\,\mic square pixels, of which they occupy half of the available area (2k$\times$1k images are produced). The detector itself is mounted on a two axis piezo motor translation stage to allow dithering for flat-field improvement and to minimize the effect of bad pixels. All of the above opto-mechanical system is contained within a cryostat and maintained at a temperature of 78\,K to limit thermal background.

Because of the high contrast imaging applications, all of the IRDIS filters required very high integrated blocking flux capability. The initial specifications were below 1\% for integrated out-of-band flux vs. integrated in-band flux, and $<10^{-4}$ for the out-of-band transmission for all filters. However, due to technical difficulties, these specifications have been relaxed to 10\% for the NB filters. For the the most demanding combination, which is the DB filters (used in combination with blocking filters), the final specification reaches 0.1\%. 

Detailed descriptions of IRDIS, its observing modes and the different steps of its integration and testing have been presented in numerous works in the past \citep{Dohlen2008,Dohlen2008b,Dohlen2008c,Dohlen2010,Langlois2010a,Langlois2010b,Vigan2012b,Madec2012,Vigan2012b,Langlois2014,Vigan2014}.

\subsection{Dual-band imaging mode}

\subsubsection{Implementation}

The main mode of IRDIS is the dual-band imaging mode \citep[DBI;][]{Vigan2010} designed to detect and characterize planetary companions down to the Jupiter mass around nearby young stars. This mode provides images in two neighboring spectral channels with minimized differential aberrations. Dual imaging separation is done using a beam-splitter combined with a mirror, producing two parallel beams, which are spectrally filtered before reaching the detector using dual-band filters with adjacent bandpasses corresponding to sharp features in the expected planetary spectra. Both center wavelengths and the widths of these filters where optimized using synthetic exoplanetary model spectra. The main filter pair, H23, has been optimized to be centered around the CH$_{4}$ absorption band in the $H$-band that was expected for substellar companions with \Teff<1200\,K at the time of the instrument design.

Differential aberrations between the two beams are critical for achieving high contrast \citep[e.g., ][]{Racine1999,Marois2005}. IRDIS achieves less than 10\,nm differential aberrations between the two channels \citep{Dohlen2008b} and, as a consequence, allows for a high-contrast gain using SDI processing. For the achievement of such high-contrast performances, it is also mandatory to keep errors due to instrumental effects, including common path aberrations, at a very low level (<$\sim$50\,nm) and as stable as possible so that SDI can be combined efficiently with ADI. It is worth mentioning that at short separation, that is within the AO control radius, the ultimate performance is set by the photon noise from the speckle themselves if the speckle suppression from ADI+SDI post processing is perfect, which is not currently the case \citep[e.g., ][]{Galicher2018}.

\begin{figure*}
    \centering
    \includegraphics[width=1\textwidth]{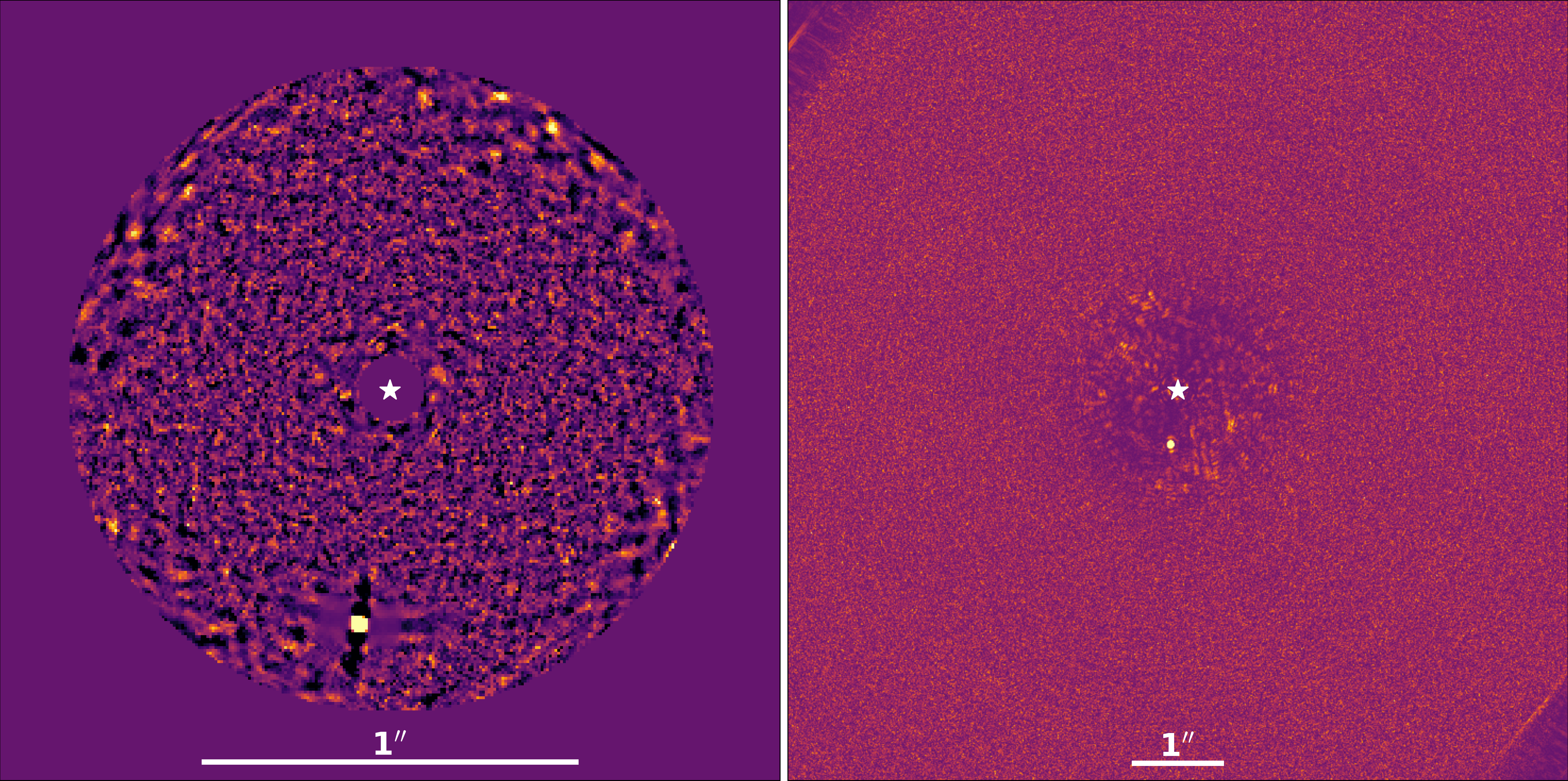}
    \caption{Example of IFS (Y-J, left) and IRDIS (H2 filter, right) images of the spectro-photometric calibrator HD\,114174 obtained by using in ADI for 1\,h observing time. The IRDIS images are processed with cADI, while the IFS ones are processed with ASDI PCA.}
    \label{fig:IRDIS_IFS_Images}
\end{figure*}

\begin{figure}
    \centering
    \includegraphics[width=0.5\textwidth]{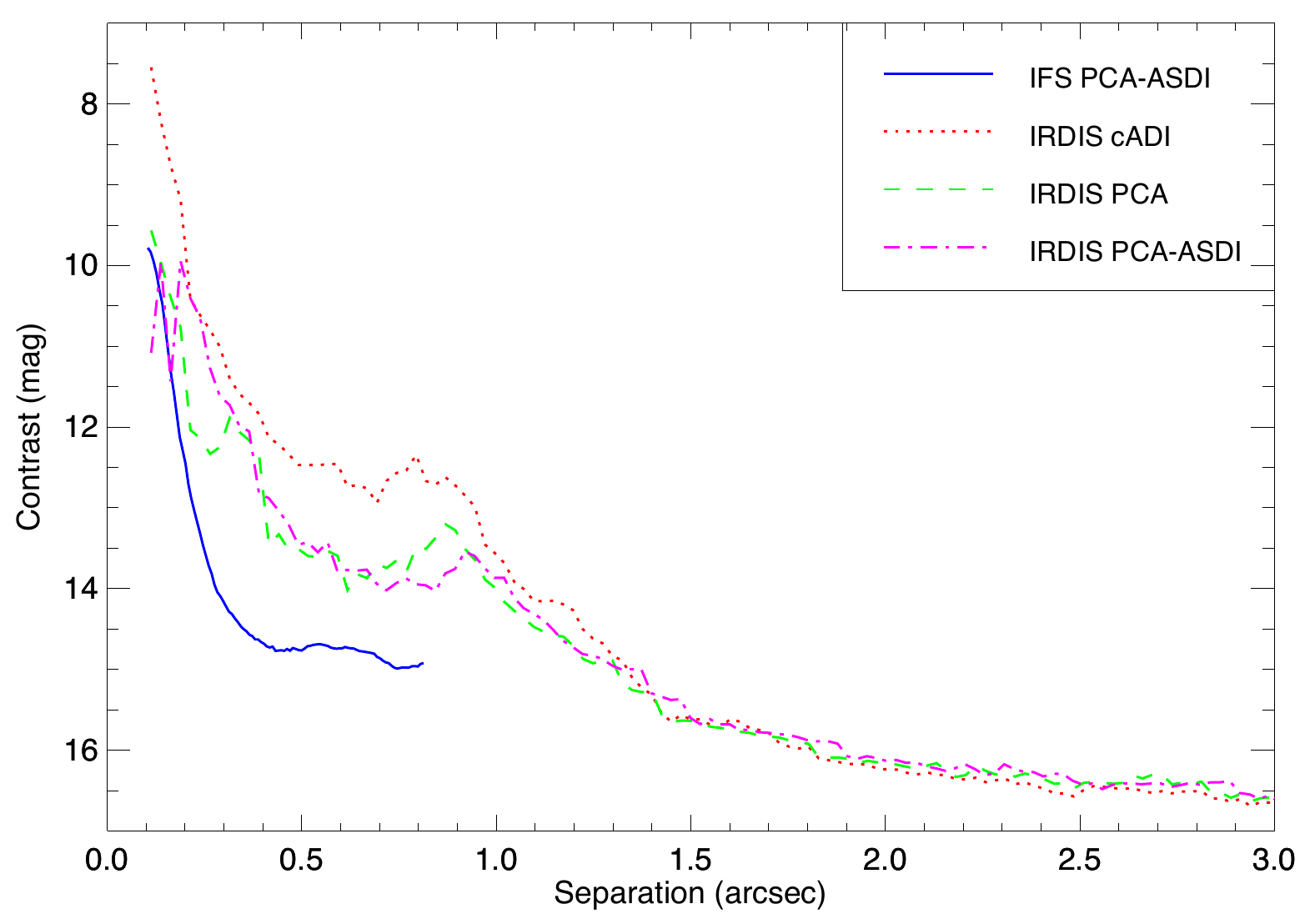}
    \caption{Typical measured 5$\sigma$ contrast achieved by IFS (Y-J) and IRDIS in DBI H23 mode in ADI and ADI+SDI in 1\,h observing time (cADI and PCA post-processing).}
    \label{fig:IRDIS_IFS_contrast}
\end{figure}

\subsubsection{Performance and limitations}

\begin{figure}
    \centering
    \includegraphics[width=0.5\textwidth]{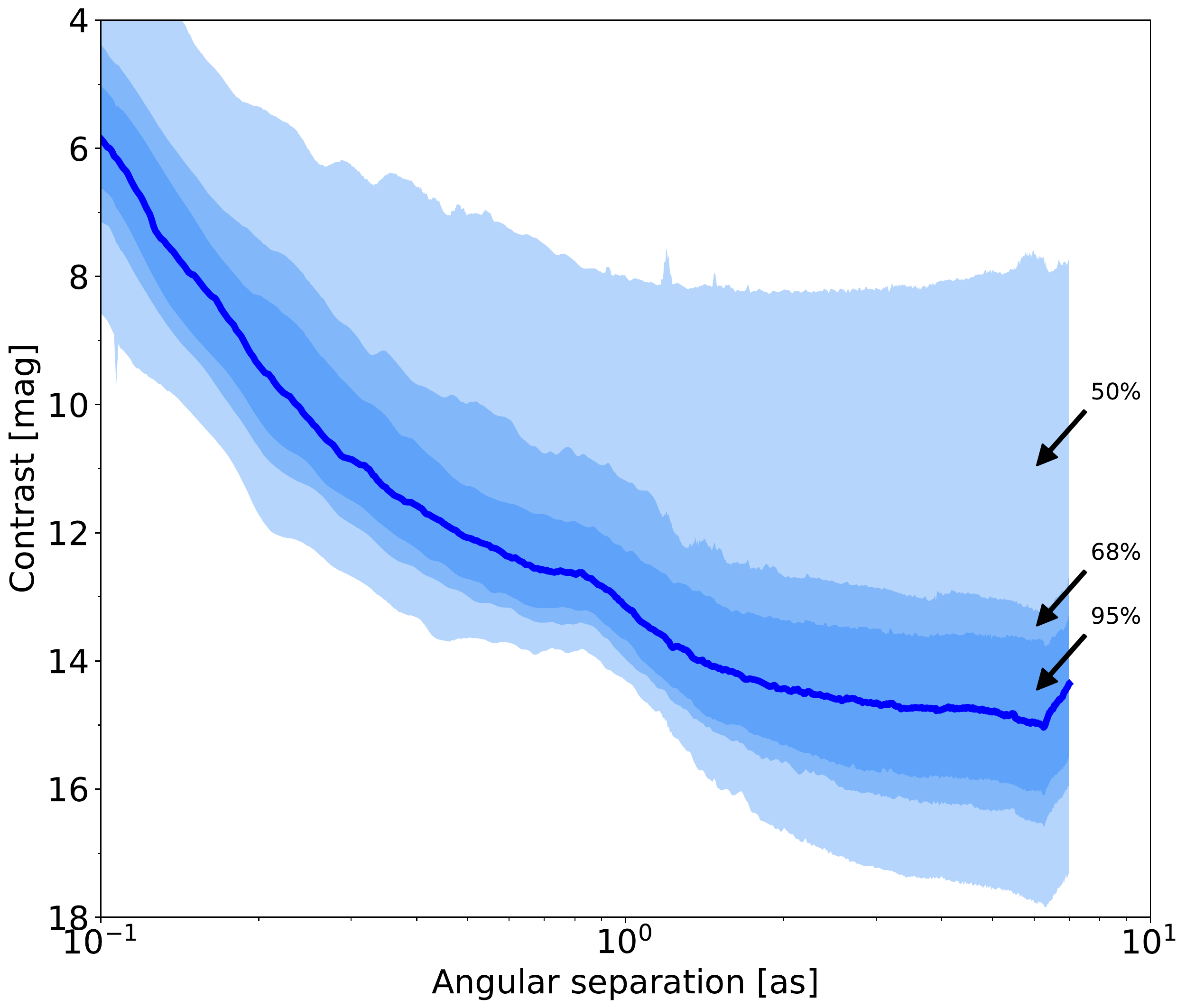}
    \caption{Median 5$\sigma$ contrast achieved by IRDIS in DBI H23 mode (H2 ADI only) in 1h30 observing time on average with T-LOCI post-processing. The plain blue line represents the median, and the different shades of blue background reprensent the 95\%, 68\% and 50\% completeness intervals.}
    \label{fig:IRDIS_contrast}
\end{figure}

The intrinsic optical quality of IRDIS does not appear to be a limitation for the contrast performance on-sky. In particular, the extremely low differential aberrations do not currently limit the performance of SDI, which is in fact limited by the chromaticity of the aberrations in the CPI in the $H$-band. In the $K$-band, the sky and instrument thermal emission difference between the two filters in the K12 pair result in a slightly decreased sensitivity (typically 0.5 to 1 magnitude) in contrast compared to the other filter pairs at separation greater than 1.0\as-1.5\as.

Thanks to its wide FoV, IRDIS is used as the reference for on sky astrometric calibration using multiple stellar systems and stellar clusters \citep{Maire2016}. This calibration  provides measurements of the pixel scale and the position angle with respect to the north for both IRDIS and IFS, as well as the distortion for the IRDIS camera. The IRDIS distortion is shown to be dominated by an anamorphism of 0.60\% between the horizontal and vertical directions of the detector, that is 6 mas at 1\as. The anamorphism is produced by the cylindrical mirrors in the CPI hence is common to all three SPHERE science subsystems (IRDIS, IFS, and ZIMPOL), except for the relative orientation of their field of view. The current estimates of the pixel scale and north angle for IRDIS in H23 coronagraphic images are 12.255\,mas/pixel and -1.75\,deg respectively \citep{Maire2016}.

\subsubsection{Results}
\label{sec:irdifs_results}

The performances of the DBI mode are illustrated on a typical case on Fig.~\ref{fig:IRDIS_IFS_contrast}, which compares the detection limits obtained with different post-processing techniques (ADI, SDI+ADI). Figure ~\ref{fig:IRDIS_contrast} also shows an illustration of the contrast range obtained with IRDIS DBI for a large range of targets and atmospheric conditions achieved by the SPHERE SHINE survey \citep{langlois2018}. Very recently IRDIS DBI has captured an unprecedented series of high contrast images allowing to redetect the exoplanet $\beta$~Pictoris b on the northeast side of the disk at a separation of only 139\,mas from its parent star \citep{lagrange2018}. 

IRDIS and IFS are designed to be used in parallel for the survey mode of SPHERE (see Sect.~\ref{sec:tradeoffs} and \ref{sec:global_system}). The complementarity between the two instruments is illustrated in Figures~\ref{fig:IRDIS_IFS_Images} and \ref{fig:IRDIS_IFS_contrast}, which respectively compare the IRDIS and IFS images and contrast limits obtained on HD\,114174, a star with a white dwarf companion used as a spectro-photometric calibrator in the SPHERE SHINE survey. The two instruments are also highly complementary in terms of spectro-photometric capabilities as illustrated in Fig.~\ref{fig:IFS_spectrum}: in \texttt{IRDIFS-EXT} mode, they enable covering the $Y$-, $J$-, $H$-, and $K$-band in a single observation, providing a high-level of spectral content for subsequent analyses.

The recent new exoplanet detections achieved by SPHERE around HIP\,65426 and PDS\,70 \citep{Chauvin2017,Keppler2018} illustrate the IRDIS DBI capabilities at very high-contrast for the detection of point sources, but IRDIS has also proven very efficient to detect new circumstellar disks \citep{Lagrange2016b,Sissa2018}. In terms of characterization, the capabilities of IRDIS DBI on its own have been demonstrated in \citet{Vigan2016}, but it becomes most efficient when combined with the IFS for companions in the central part of the SPHERE FoV \citep[e.g., ][]{Zurlo2016, Samland2017,Delorme2017,Mesa2018,Cheetham2018}. Examples of what can be obtained at very small separations ($\leq$0.1\as) in IRDIFS mode can be found on the study of HD\,142527\,B \citep{Claudi2019}. The DBI mode is also of particular interest to discriminate exoplanets from background contaminants using the color magnitude diagram placements. For that purpose H23 and J23 filter pairs are the most efficient combinations as illustrated in \citet{bonnefoy2018}. Finally, the remarkable stability and versatility of SPHERE reflects in the possibility to combine multi-epoch and multi-mode observations. An example of this can be found in the study of the environment of HD\,169142 by \citet{Gratton2019}.

\subsection{Long-slit spectroscopy mode}

\subsubsection{Implementation}

\begin{figure}
    \centering
    \includegraphics[width=0.5\textwidth]{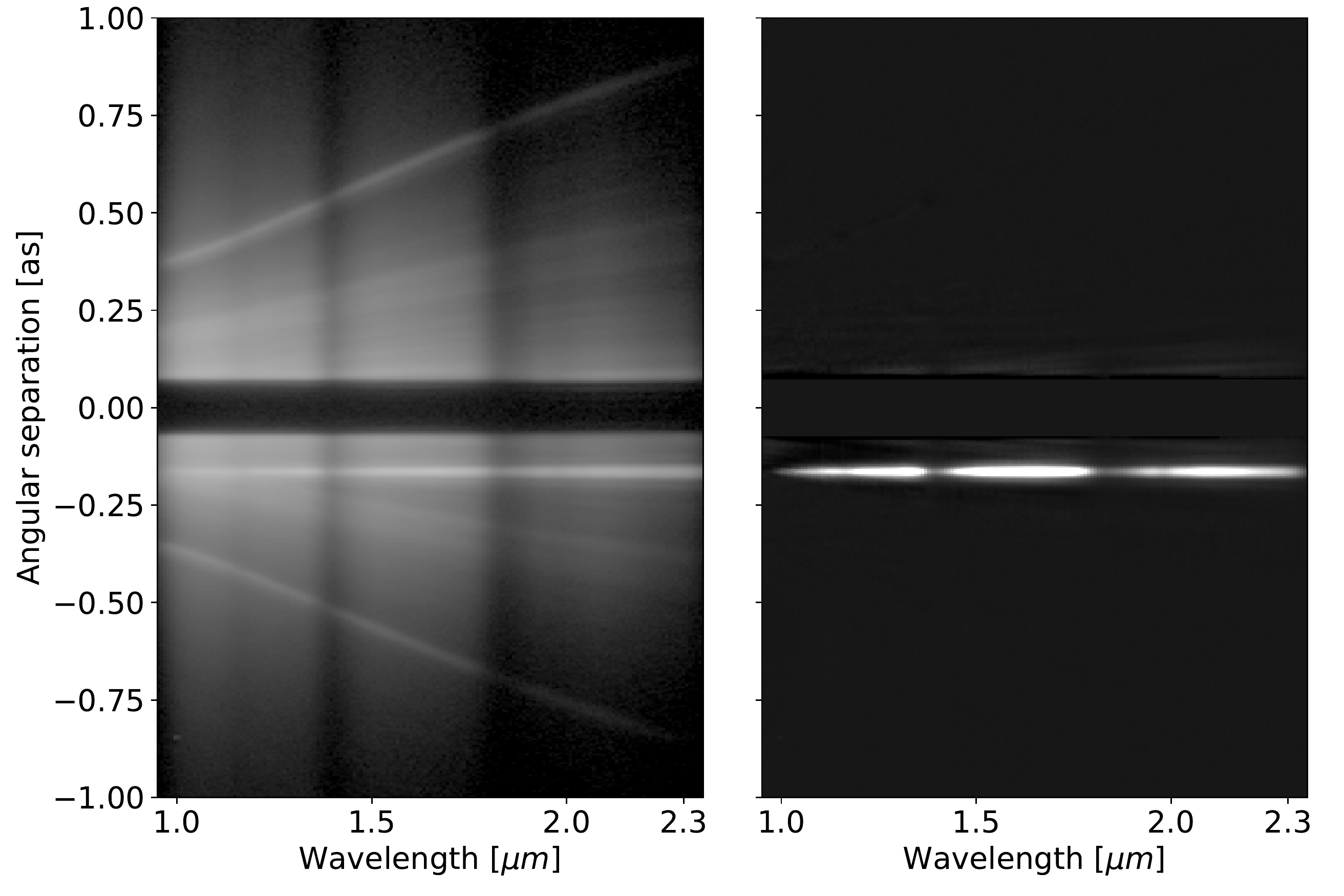}
    \caption{PZ~Tel LRS data before (left) and after (right) speckle subtraction. The left image is the spectrum after pre-processing using the DRH pipeline (see Sect.~\ref{sec:drh}). The spectrum of the companion PZ~Tel~B is visible as a straight line surrounded by speckles at an angular separation of $\sim$0.5\as. The obscured part between $\pm$0.2\as corresponds to the position of the opaque coronagraphic mask. The contrast of the companion is $\sim$5.4 mag in $H$-band \citep{Biller2010}. The right image present the data after stellar halo and speckles subtraction using the SDI approach described in \citep{Vigan2008}.}
    \label{fig:PZtel_LSS}
\end{figure}

The IRDIS long-slit spectroscopy (LSS) mode has originally been designed as a means of performing detailed spectral characterization of companions detected in DBI mode \citep{Vigan2008}. This mode offers an efficient combination of long-slit spectroscopy with coronagraphy. The spectral coverage in LSS is either 0.95--2.32\,\mic (YJHKs) with a resolving power of R$\sim$50 or 0.95--1.65\,\mic (YJH) with R$\sim$350, providing the so-called low-resolution (LRS) and medium-resolution (MRS) spectroscopy modes. A LRS dataset obtained on PZ~Tel~B is presented in Fig.~\ref{fig:PZtel_LSS}. In practice, the slits and opaque coronagraphic masks have been merged into a single device that is placed in the coronagraph wheel of the CPI. Three different combinations of slit widths and coronagraphic mask sizes are provided, but the most widely tested and used combinations are the 0.12\as-wide and 0.09\as-wide slits with a mask of radius 0.2\as. For the spectral dispersion, two dispersive elements are placed in the IRDIS cryostat in the Lyot stop wheel: a prism for the LRS mode and a grism for the MRS mode. The dispersive elements are combined with a circular pupil mask (92\% of the pupil size), which serves as a Lyot stop. The combination of the three slits and the two dispersive elements provides 4 four different configurations. Using the LSS mode is necessarily done in field stabilized mode to maintain the object of interest inside the slit during the observations.

\subsubsection{Performance and limitations}

IRDIS, with its unique LSS mode that can reach resolutions up to 350 in YJH, is a particularly to characterizing directly-imaged giant planets through near infrared spectroscopy but the overall contrast performance is limited at very small angular separation because the coronagraph initially used in this mode is not optimal. In particular, the rudimentary Lyot stop included with the dispersive elements does not provide any optimization regarding the presence of the telescope central obscuration, which results in strong diffraction residuals close to the coronagraphic edge. The new stop-less Lyot coronagraph \citep[SLLC;][]{N'Diaye2007} that has been implemented during the reintegration of SPHERE in Paranal significantly improves the sensibility, by one magnitude in the 0.2\as-0.5\as range, as demonstrated in \citet{Vigan2013,Vigan2016b}.
 
The spectral resolution for point-like sources in LSS is set by the diffraction limit rather than the slit width. The central star halo (the star PSF core is hidden behind the coronagraphic mask) can be considered as an extended object in the sense that the speckle field of the star will fully cover the slit, decreasing the effective resolution of its spectrum. For unresolved point sources, their wavelength calibration will be impacted by the centering accuracy of the object inside of the slit width. Any decentering of the target induce a systematic shift of the wavelength for the object with respect to that of the star located behind the coronagraph and the spectro-photometric calibrator. The centering of the star behind the coronagraph is dealt with during the acquisition sequence, and can be considered accurate to $\sim$20\,mas. Then the centering of a companion inside the slit is directly related to the knowledge of it position angle around its parent star, and which is used to orientate the CPI derotator in the appropriate position. However, the \texttt{POSANG} parameter of the LSS observing template must be corrected by a static value for optimal centering:
\begin{equation}
    \textrm{POSANG}_{template} = \textrm{PA}_{sky} + 1.75\degree
\end{equation}
 
The wavelength calibration of the spectra is obtained by illuminating the slit with light from four lasers lamps (wavelength of 0.9877, 1.1237, 1.3094, and 1.5451\,\mic) located on the common path calibration arm, which typically provides a wavelength calibration accuracy of $\sim$5\,nm in the MRS mode.
 
A dedicated pipeline \citep[SILSS][]{Vigan2016ascl} has been developed specifically to analyze IRDIS LSS data, which combines the standard ESO pipeline with custom IDL routines to process the raw data into a final extracted spectrum for the companion. Ultimately the performance in contrast of the IRDIS LSS mode is set by the speckle subtraction. Because observations are performed in field stabilized mode, the speckle subtraction mostly relies on SDI techniques adapted to LSS data \citep{Vigan2008}. However, these techniques are usually heavily biased spectrally, so that they are rarely used in practice \citep{Vigan2012}. For bright (5-6\,mag contrast) close (<0.8\as) companions, or very distant companions (>2.5\as), the simple subtraction of the symmetric halo profile usually provides an accurate enough speckle subtraction \citep[e.g., ][]{Maire2016,Bonavita2017}. For faint (>6\,mag contrast) close (<0.8\as) companions, a powerful strategy recently used on HIP\,64892~B is the use of a sequence with data obtained both with the companion inside the slit and just outside the slit \citep{Cheetham2018}. The out-of-slit data are used to build a PCA reference library that is used to subtract the speckles in the inside-of-slit data.
 
\subsubsection{Results}

\begin{figure}
    \centering
    \includegraphics[width=0.5\textwidth]{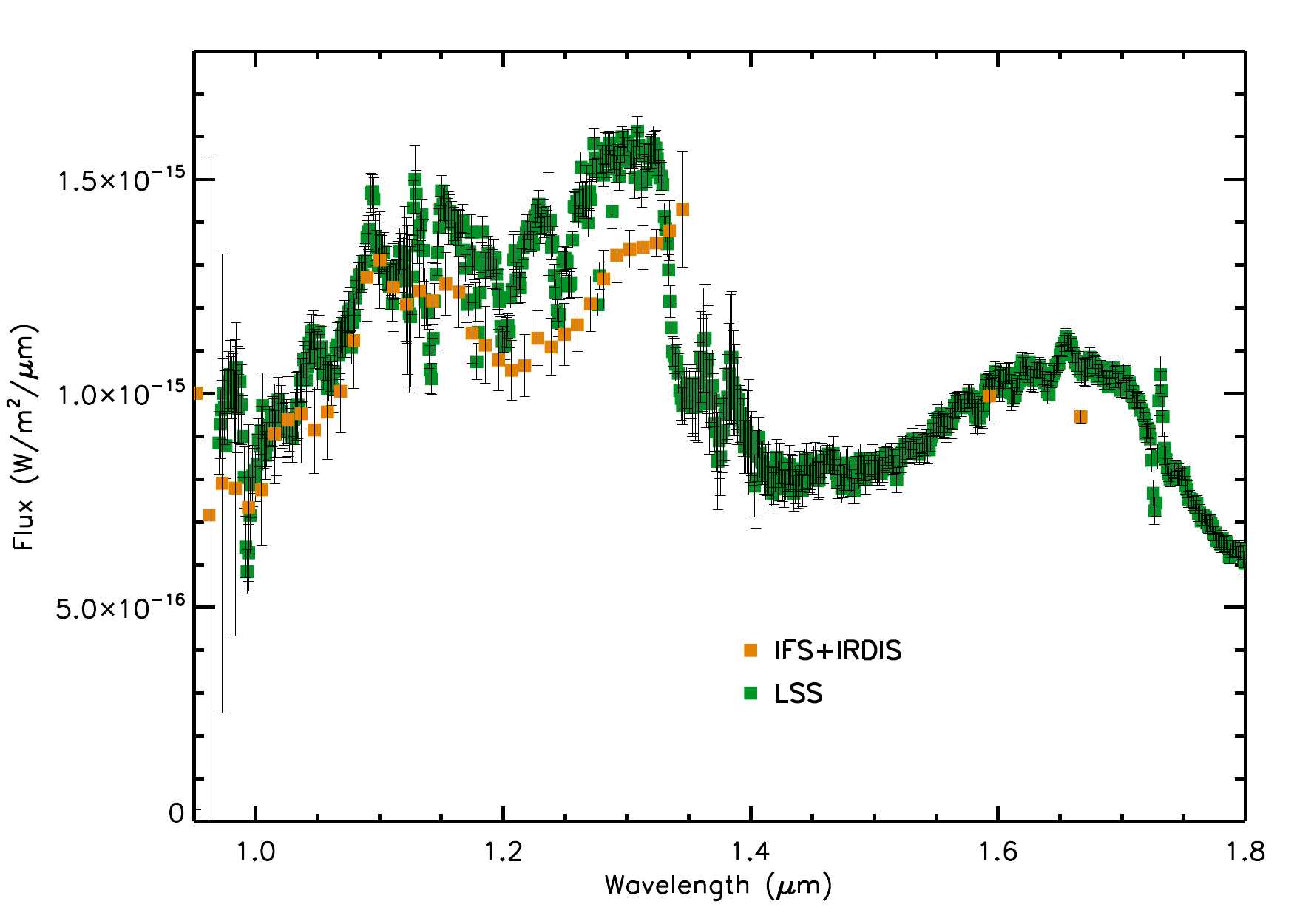}
    \caption{Comparison between the spectrum of HR\,3549~B obtained using IFS (orange squares) and IRDIS MRS mode (green squares).}
    \label{fig:HR3549_LSS}
\end{figure}

The LSS mode is particularly useful for the characterization of moderately faint companions as illustrated by Fig.~\ref{fig:HR3549_LSS}, or to take advantage of IRDIS’s larger FoV. Since the commissioning of SPHERE, the LSS mode has been widely used to perform the characterization of 2MASS 0122-2439~B \citep{Hinkley2015}, PZ~Tel~B \citep{Maire2016}, HR\,3549~B \citep{Mesa2016}, HD\,284149~b \citep{Bonavita2017} or HIP\,64892~B \citep{Cheetham2018}, and several on-going programs are focused on fainter companions closer to their parent star. An example of MRS data obtained on HR\,3549~B is presented in Fig.~\ref{fig:HR3549_LSS}.

\subsection{Dual-polarimetry imaging mode}

\subsubsection{Implementation}

IRDIS DPI has been designed to investigate the reflected light from circumstellar disks. Since circumstellar disks light is partially linearly polarized by the reflection of the star light on its surface, the dual polarimetry imaging mode \citep[DPI;][]{Langlois2010b} allow to recover the intensity and the angle of polarization leading to morphology and dust size studies of these disks. The main goal of IRDIS DPI in this scientific area is to unveil details of the structure of the disks in the inner regions with m$_{disk/arcsec^2}- m_{star} > 6$ at 0.5\as separation. This mode has been used with success to study known disks and to discover new ones \citep[e.g., ][]{Benisty2015,langlois2018b}. In a large number of cases the IRDIS DPI mode has enable access to indirect hints for the possible presence of exoplanets inferred from the presence of gaps, spirals, and shadows \citep{Benisty2017,Maire2017,vanBoekel2017}. In one case a first direct detection of a polarized companion outside of a resolved circumbinary disk around CS\,Cha has been achieved with this mode \citep{ginski2018}. 

The basic principle of high-precision polarization measurements includes polarization modulation using a half-wave plate (HWP) located at the entrance of the SPHERE instrument. This polarization modulator associated to two crossed polarizers located inside the IRDIS cryostat converts the degree-of-polarization signal into a fractional modulation of the intensity signal, which is then measured by a differential intensity measurement between the two temporal measurements. The sum of the two simultaneous IRDIS images is proportional to the intensity while the normalized difference measures the polarization degree of one Stokes component. 

The key advantages of this technique are that images for the two opposite polarization modes are created simultaneously, both images are recorded on different part of the detector, there are only small differential aberrations between the two images corresponding to opposite polarization, and the differential polarization signal is not affected by chromatic effects due to telescope diffraction or speckle chromatism.

The DPI mode provides an efficient means to remove the unpolarized speckles from starlight that are the dominant cause of limiting high contrast sensitivity. At its simplest level, a dual-channel differential imaging polarimeter is a device that splits an image into two orthogonal polarization states. This is achieved in IRDIS DPI through the use of a beam splitter and a set of orthogonal polarizers that separates an incoming beam into two orthogonal polarization states, while introducing an angular deflection between the two beams. A measurement of the Stokes Q parameter is simply obtained through a difference of the left and right channels. Such a subtraction (Difference: +Q), since these ordinary and extraordinary images are taken simultaneously, reliably removes the atmospheric halo and the effects of unpolarized common wavefront aberrations. However, to eliminate the bulk of the remaining aberrations (not common to both channels) that persist in this difference image, the polarization incoming is also modulated in a sequence through the rotation of a HWP. The sequence shall be fast enough compared to the instrumental polarization evolution. Subtracting these in turn gives a -Q image, which is obtained by swapping the positions of the polarization states and subtracting the two channels. After the subtractions, any astrophysical object will now possess negative counts in the image, but those non-common path aberrations will have the same sign and spatial characteristics present in the +Q image. Subtracting the -Q image from the +Q image (ideally) eliminates the non-common path aberrations effects. 

In summary, IRDIS allows obtaining such observations with a relative polarimetric accuracy of $< 10^{-2}$, over the wavelength range from 0.9-2.3\,\mic, with a detector FoV of 11\as$\times$11\as, using broad-band or narrow-band filters, and with both coronagraphic and non-coronagraphic modes.

\subsubsection{Performance and limitations}

IRDIS DPI is currently among the most powerful instrumental modes to perform polarimetric high-contrast imaging. Due to design choices, its performance is strongly dependent on the observation strategy, as will be illustrated in de Boer et al. (in prep) with the observations of TW Hydrae. The polarimetric cross-talk in IRDIS DPI can cause the polarimetry efficiency to drop toward 10\% in $H$- and $Ks$-band, while the efficiency remains above 60\% in $Y$-band and well above 90\% in $J$-band. The measured cross-talk is also responsible for an offset in the measured polarization angle in $H$ and in $Ks$ to a larger extend. Both the effects are fully calibrated and can be corrected for by the use of a Mueller matrix model. This model has already been used successfully in several cases to correct for the variations in efficiency and polarization angle offset due to cross-talk observed in the various datasets \citep{vanHolstein2017,Pohl2017}.

Optimal results can be obtained from IRDIS DPI observations when two important considerations are taken into account: 1) adjusting the observation strategy beforehand as described in de Boer et al. (in prep) to minimize a loss in efficiency; and 2) applying the Mueller matrix model described in van Holstein et al. (in prep.) to correct the data for instrumental and telescope polarization and cross-talk. Such compensation lead to increase polarimetric measurement accuracy, as illustrated \cite{vanHolstein2017} with observations of the HR\,8799 system.

\subsubsection{Results}

The prime objective of the IRDIS DPI mode is the discovery and study the circumstellar environments themselves as well as post signs they can provide on the presence of exoplanets. The challenge consists in the very large contrast of luminosity between the star and the planet, at very small angular separations, typically inside the AO control radius. With such a prime objective, it is obvious that many other research fields will benefit from the large contrast performance and polarimetric capability of the IRDIS DPI mode: proto-planetary disks, brown dwarfs, evolved massive stars, AGN, etc. For instance, young stellar objects retain material from their formation process in the form of remaining parental cloud circumstellar disks, possibly jets and, at a later stage, debris disks. The thermal radiation from accretion and debris disks is easily detected in the mid-IR, but scattered light, in particular close to the star provides many additional constraints on the dust properties and disk structure. While current ground-based observations of proto-planetary disks are very difficult, large progresses have been achieved thanks to the IRDIS DPI capabilities.

The detection of these disks in polarimetry is very precious to set new constraints on numerical modeling. In particular, the increase in sensitivity from IRDIS DPI, complementary to ADI intensity measurements, has led to high-scientific return \citep{Keppler2018,Pohl2017}. The polarimetric study of the transition disk around the young star PDS\,70 \citep{Keppler2018}, which is of particular interest due to its large gap hosting a planet in formation, and led to the new detection of an inner disk not extending farther than 17\,au (0.14\as). Another important result has been achieved on NGC\,1068 using IRDIS DPI data which show strong evidence that there is an extended nuclear torus at the center of NGC\,1068 \citep{gratadour2015}.

\subsection{Classical imaging mode}

\begin{figure}
    \centering
    \includegraphics[width=0.5\textwidth]{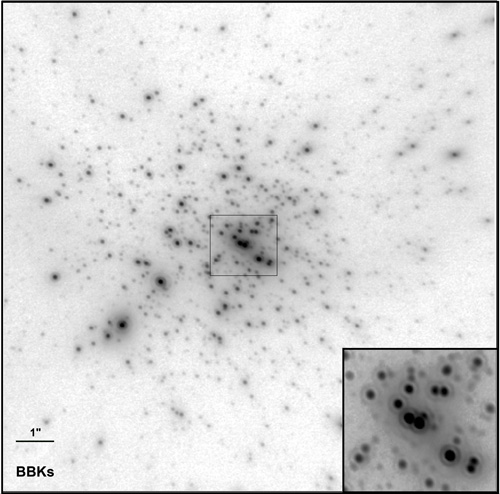}
    \caption{Broadband $Ks$ image demonstrating the PSF quality achieved by SPHERE using IRDIS CI mode without coronagraph in a field-stabilized observation. The bottom right inset displays a zoom image of the square box located at the center of the R136 open cluster \citep{Khorrami2017}.}
    \label{fig:R136}
\end{figure}

The IRDIS Classical imaging modes is a multi-purpose mode that benefits from extreme adaptive optics correction over the 10'' IRDIS FOV as well as the low level of non common path aberrations of SPHERE and IRDIS. For this size of FOV, the effect of anisoplanetism has been measured at shorter wavelengths using observations of the core of the young massive star clusters, but can be most of the time negligible in $K$-band \citep{Khorrami2017} as shown on Fig.~\ref{fig:R136}. This mode that can be used without coronagraphy has produced scientific results in for various science cases (\citealt{Soulain2018,Marsset2017,Sicardy2015}. In particular, IRDIS Classical imaging allowed to detect around the emblematic dusty Wolf-Rayet star WR104 a 2\as circumstellar dust extension including a spiral pattern due to sub-micron grain size due to the rapid growth of the dust nuclei \citep{Soulain2018}. Moreover, the stellar candidate companion previously detected by the HST has been confirmed and characterized by these unique observations. The IRDIS classical imaging mode has also been used for several asteroid studies, leading to shape reconstruction \citep{2015Viikinkoski,Marsset2017,Vernazza18}. These studies largely benefit from the high resolution and high Strehl providing much more detailed images than previous AO corrected images from other instruments.

\section{Instrument control, operations and data reduction}
\label{sec:instrument_control_operations}

\subsection{Instrument control system}
\label{sec:control_software}

The SPHERE instrument is controlled through a dedicated control network (Sect.~\ref{sec:control_netword}) that interconnects the different elements and workstations, and using the instrument software (Sect.~\ref{sec:instrument_software}) that enables to control all the sub-systems in a consistent and reliable way.

\subsubsection{Control network}
\label{sec:control_netword}

\begin{figure*}
	\centering
	\includegraphics[width=0.7\textwidth]{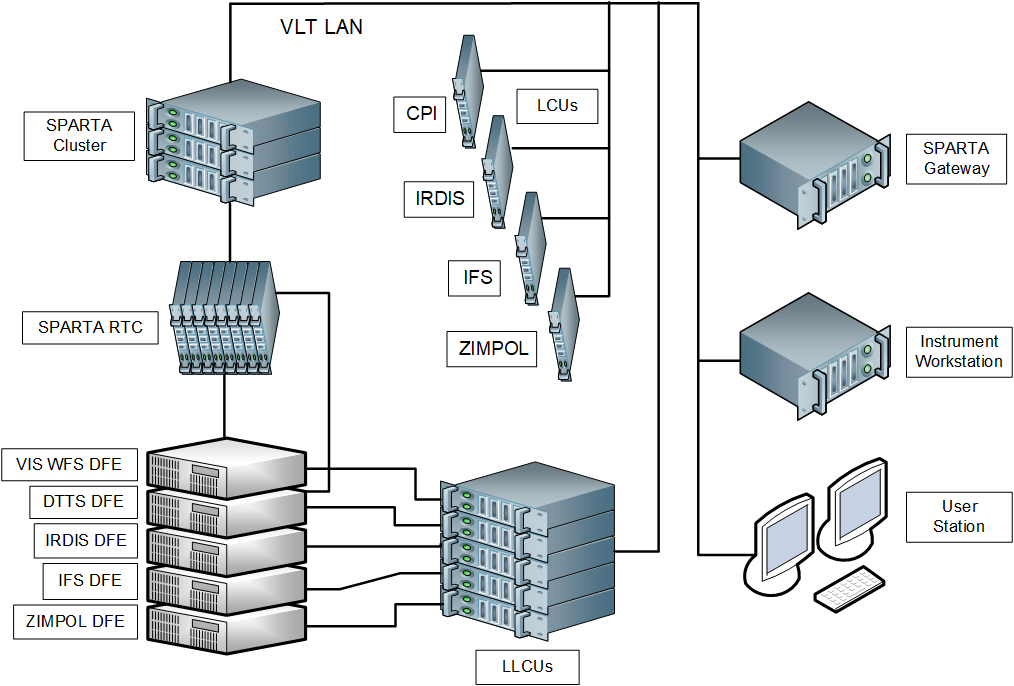}
	\caption{Conceptual view of the SPHERE control network.}
	\label{fig:ctrlnetwork}
\end{figure*}

SPHERE is a complex instrument, comprising two active mirrors, the SPARTA system \citep{Suarez2012}, two wavefront sensors, and three science detectors, all controlled through ESO New General Detector Controller \citep{Baade2009}, more than fifty motorized functions as well as fifteen calibration sources, and a number of sensors and process controllers. This complexity is reflected in the control network architecture (Fig.~\ref{fig:ctrlnetwork}) and is also made apparent by the large number of templates implementing operational procedures.

The instrument control network is based on a distributed system of:

\begin{itemize}
	\item Local Control Units (LCUs), single board computers running the VxWorks operating system and in charge of managing all instrument functions except detectors. In SPHERE there is one LCU per science channel plus one devoted to the control of functions common to the whole instrument;
	\item Linux LCUs (LLCUs), which act as bridges toward the detectors front-end electronics;
	\item the SPARTA system, composed by a real-time computer (RTC) and a cluster, which handle the hard- and soft-real time adaptive optics computations, and the SPARTA workstation, which acts as a gateway toward the rest of the control network;
	\item the instrument workstation (IWS), where the overall coordination software resides and templates are executed.
\end{itemize}

\subsubsection{Instrument software}
\label{sec:instrument_software}

The SPHERE instrument software (INS) is in charge of controlling all instrument functions, coordinate the execution of exposures, and implement all observation, calibration, and maintenance procedures. It includes on-line data reduction processing, necessary during observations and calibrations, as well as quick-look procedures that allow monitoring the status of ongoing observations. INS also manages external interfaces with (i) the VLT telescope control software (TCS), for presetting to target, getting telescope and ambient information, and off-loading of tip-tilt and focus from SAXO, (ii) the high-level observing software (HOS), to retrieve observing blocks to be executed, and (iii) the data handling system (DHS), for the archival of data files. The SPHERE instrument software architecture follows the standard partitioning for VLT control applications \citep{Raffi1997} and has been described in detail in dedicated papers \citep{Baruffolo2008,Baruffolo2012}.

INS fully supports all the observing modes of SPHERE: IRDIFS, in which IRDIS and IFS observe in parallel, IRDIS alone, which includes dual-band, dual-polarimetry imaging, long-slit spectroscopy, and classical imaging sub-modes, and finally ZIMPOL, which includes two polarimetric and one imaging sub-modes. When observing in \texttt{IRDIFS} mode, exposures are performed in parallel in the two science channels and are completely independent.

All SPHERE observation, calibration, and maintenance procedures are implemented in the form of templates, as required for all VLT instruments. Currently, INS includes about 140 templates, which, for the most part, are maintenance and calibration templates. The observer is directly concerned with a small subset: target acquisition, observation, and some calibration templates.

Target acquisition templates are provided for all mode and sub-mode combinations. The acquisition procedure includes: presetting of the telescope to target, acquisition of the telescope guide star, automatic setup of the AO loops and performance check, and starting of the tracking devices (derotator, ADCs, polarization components), if relevant for the observation and according to the tracking law selected by the user, or inherent to the selected observing mode. For focal-plane coronagraphic observations, setting of the focal mask is only performed in the acquisition template, which also takes care of fine-centering and focusing of the target on the coronagraph. The coronagraphic device then remains the same for all observing templates in the same OB, thus preserving centering and focus. In addition, if the same coronagraph is used from one OB to the next, the centering and focus are conserved to enable a significant gain of time in the target acquisition sequence. Since each acquisition template is devoted to a specific instrument mode, in an observing block it must be followed by observing templates for the same mode.

IRDIFS and IRDIS observing templates allow to acquire target images and ancillary data which are useful for proper data reduction. When preparing an OB, the user can specify a list of exposure types to be performed choosing among:

\begin{itemize}
	\item \texttt{OBJECT}: on-axis coronagraphic observations of the target;
	\item \texttt{STAR-CENTER}: this causes the application of a periodic modulation on the SAXO HODM, which results in the creation of four satellite spots, well outside the coronagraphic mask. These spots allow a precise derivation of the target position behind the coronagraph and possible monitoring of the astrometry and relative photometry (see Sect.~\ref{sec:ifs:on-sky_results});
	\item \texttt{FLUX}: this allows to acquire an image of the stellar PSF by moving the target off the coronagraph and inserting a neutral density filter. The measurement is performed without moving the coronagraphic mask and using the SAXO tip-tilt mirror as actuator. In this way, when moving the source back to origin, the centering accuracy is maintained.
	\item \texttt{SKY}: observation of the sky background, performed by offsetting the telescope according to a user-specified pattern. In this sequence all the SAXO control loops are opened.
\end{itemize}

The ZIMPOL polarimetric observing templates differ mainly for the field derotation mode: either no derotation, for higher polarimetric sensitivity and accuracy, or with derotation, to avoid smearing of faint targets that require longer integration times. A dedicated imaging template is also provided, which does not use polarimetric components, thus resulting in higher throughput, and offers the possibility of stabilizing the pupil to enable angular differential imaging. Similar to the templates for IRDIFS and IRDIS, the ZIMPOL observing templates allow the user to specify a list of exposure types to be performed, as described above, with the exception that sky observations are not offered.

During the course of the execution of a science template, SPARTA collects information on the observing conditions that is then stored in a separate FITS file and archived by the instrument software. Such files can be then be retrieved from ESO science archive to obtain the estimated wind and seeing during the observation, as well as images of the stellar PSF in the $H$-band, as recorded by the differential tip-tilt sensor camera, thus not occulted by the coronagraph. These images, in particular, allow to check the quality of the PSF and its temporal variations during an observation.

SPHERE INS contains about seventy calibration templates, a good fraction of which are devoted to the calibration of the adaptive optics module and are not of concern for the observer. Execution of daytime calibration templates, for the production of data required for proper processing of the science frames acquired during an observing night, is performed automatically by INS and is driven by parameters extracted from the science data files themselves. For instance, dark or background frames are acquired using the same integration time as the science frames acquired the night before, flats are taken using the same filters, etc. Templates for nighttime calibrations, besides those performed during the observing templates described above, are also provided. However, most of the required calibrations are performed by the Paranal Observatory so the observer is only concerned with few templates, for calibrations which are optional.

\subsection{Operations \& calibrations}
\label{sec:operations_calibrations}

Regular operations of SPHERE started in ESO period P95 in April 2015, and consists of visitor and queue scheduled service-mode observations like any other VLT instrument.

The Paranal environment does not require strong restrictions on the use of SPHERE. To preserve the HODM, operations are stopped when the relative humidity inside the instrument exceeds 50\%, and this occurs on about 20 days per year, mostly during the altiplanic winter. SPHERE is robust against turbulence conditions, such that it can be operated in more than 90\% of all turbulence conditions over Paranal, corresponding to a seeing better than 1.4\as and a coherence time greater than 1\,ms. This makes visitor-mode observing efficient, and avoids frequent changes from one instrument to another in service mode. 

SPHERE observers will continue to use solely the seeing to constrain the required   atmospheric conditions for their observations in service mode until April 2019. However, many studies have pointed out that the coherence time also strongly influences the quality of the AO correction \citep{Milli2017,Madurowicz2018,Savransky2018,Cantalloube2018}. Therefore starting in April 2019, for service mode, SPHERE observers will define both seeing and coherence time constraints to better ensure that the system reaches the required performance for their science. This new scheme allows a better match between science requirements and instrument performance, especially for faint targets for which the AO loop runs at a lower frequency of 600\,Hz or 300\,Hz instead of the standard 1380\,Hz. The Exposure Time Calculator\footnote{\url{https://www.eso.org/observing/etc/}} is a tool that gives the expected instrument performance in terms of contrast as a function of the atmospheric constraints and the target properties. More details on the atmospheric constraints can be found in the SPHERE User Manual\footnote{\url{https://www.eso.org/sci/facilities/paranal/instruments/sphere/doc.html}}. 

From period P95 to P101 (April 2015 to September 2018), SPHERE observations were equally split between visitor mode and service mode, essentially due to the large number of nights allocated to the SPHERE consortium through Guaranteed Time Observations (GTO), which is almost always carried out in visitor mode. This balance is expected to change in the coming years with a larger fraction of service mode runs. In service mode, two large programs over 100 hours are on-going. The overall science time reaches 86\%, the rest being split between night-time calibrations, technical time or technical down-time. This value is comparable to that for other VLT instruments despite the added complexity of the extreme AO system. 

The calibration plan involves some night-time calibrations limited to spectro-photometric standards and astrometric fields for IRDIS, IRDIFS, and ZIMPOL, as well as unpolarized and highly polarized standards for ZIMPOL, taken on a monthly basis (spectra-photometric standards) or a three-month basis (astrometric and polarimetric standards). The remaining calibrations are obtained during the day with the internal calibration unit. These include flat fields and backgrounds (biases for ZIMPOL) taken on a daily basis, distortion maps for all subsystems, polarimetric flats, and polarimetric modulation efficiency for ZIMPOL taken on a weekly basis. 

In addition to the data required to calibrate the science data, technical calibrations are also obtained after the end of the night with the following objectives: 

\begin{itemize}
    \item perform some sub-system functional checks (e.g., motors, actuator speed, HODM voltage supply)
    \item calibrate the AO system (see also Sect.~\ref{sec:saxo}): offset voltages, interaction matrices, and reference slopes for the various loops. Reference slopes are only re-computed every two weeks. New values are automatically checked against pre-defined thresholds and only saved if they pass this quality control. 
    \item perform end-to-end tests to detect problems before the following night. Artificial turbulence is injected at the level of the HODM and ITTM using offset voltages, and the Strehl is automatically measured and compared to a reference value.  
\end{itemize}

These technical tests take about one hour to complete. Next, the telescope and instrument are handed over to the Maintenance and System Engineering group in charge of regular engineering activities. The day astronomer or operation specialist then check if the data obtained during the night have been calibrated. They make sure calibrations have been taken using an automatic calibration completion tool
\footnote{\url{https://www.eso.org/observing/dfo/quality/SPHERE/reports/CAL/calChecker_SPHERE.html}} and validates their quality using an automatic quality-check tool\footnote{\url{https://www.eso.org/observing/dfo/quality/SPHERE/reports/HEALTH/trend_report_IRDIS_DARK_med_HC.html}}. Both tools are common to all VLT instruments. For the SPHERE instrument startup, an additional small set of technical tests, lasting about 10\,min, is run before the start of the night, to recompute some temperature-dependent calibrations, to adjust for changes during the day, such as the pupil alignment and the shape compensation of the HODM through the adaptive toric mirror.  

\subsection{Data reduction \& handling}
\label{sec:drh}

\begin{table}
    \caption{SPHERE pipeline science recipes}
    \label{tab:DRH:obsmodes}
    \centering  
    \begin{tabular}{rl}
        \hline\hline
        Observing mode         & Recipe                              \\
        \hline
        \multicolumn{2}{c}{IRDIS}                                    \\
        \hline
        Dual-band imaging      & \texttt{sph\_ird\_science\_dbi}     \\
        Classical imaging      & \texttt{sph\_ird\_science\_imaging} \\ 
        Long-slit spectroscopy & \texttt{sph\_ird\_science\_lss}     \\
        Dual-band polarimetry  & \texttt{sph\_ird\_science\_dpi}     \\
        \hline
        \multicolumn{2}{c}{IFS}                                      \\
        \hline
        All modes              & \texttt{sph\_ifs\_science\_dr}      \\
        \hline
        \multicolumn{2}{c}{ZIMPOL}                                   \\
        \hline
        Polarimetry P1         & \texttt{sph\_zpl\_science\_p1}      \\
        Polarimetry P2 \& P3   & \texttt{sph\_zpl\_science\_p23}     \\
        \hline
    \end{tabular}
\end{table}

\subsubsection{The SPHERE pipeline}

The SPHERE pipeline is a subsystem of the VLT data flow system (DFS). It is used in two operational environments: the ESO data flow operations (DFO) and the Paranal science operations (PSO). In these environments, it is used for the quick-look assessment of data, the generation of master calibration data, the reduction of scientific exposures, and the data quality control. Additionally, the SPHERE pipeline recipes are made public to the user community, to allow a more personalized processing of the data from the instrument.

The pipeline is, like all ESO pipelines, organized in plug-ins called {\textit recipes}. A recipe usually comprises one or more reduction steps, operates on a set of input files containing both raw-data and pre-fabricated calibration products, and produces a dedicated output either for scientific use or to serve as input for other recipes. In this way a cascade of recipes is formed for each observing mode. The SPHERE pipeline\footnote{\url{https://www.eso.org/sci/software/pipelines/}} is based on version 6.6 of the ESO's common pipeline library (CPL). Recipes operate as plug-ins to one of the several front-end execution tools provided by ESO, which can be used to launch and execute the various recipes. The traditional EsoRex\footnote{\url{https://www.eso.org/cpl/esorex.html}} \citep{ESO2015} and Gasgano\footnote{\url{https://www.eso.org/gasgano}} \citep{ESO2012} tools are both included in the pipeline distribution or can be downloaded separately from ESO's website. In addition, the more recent Reflex system \citep{Freudling2013} is now available to execute recipes and reflex workflows are now part of the SPHERE pipeline package\footnote{\url{https://www.eso.org/sci/software/esoreflex/}}.

The fundamental organization of the SPHERE pipeline follows the three focal plane instruments and their primary observing modes. Science recipes for the individual observing modes are summarized in Table~\ref{tab:DRH:obsmodes}. Additionally, a few higher level data analysis recipes exist for retrieving signals from data taken in ADI mode with algorithms, such as ANDROMEDA \citep{Cantalloube2015,Mugnier2009}, principal components analysis \citep{Soummer2012,Amara2012}, and others. However, as such methods evolve fast and the pipeline itself is rather static due to the complex environments it is used in, these recipes never proved popular with the user community. Instead, higher-level analysis is regularly being done in a dedicated environment implemented at IPAG, the SPHERE data Center (see below).  

At the time of writing, the SPHERE pipeline publicly offers 41 recipes. It is regularly used to perform basic calibrations such as flat-fielding, bad pixel correction, background subtraction. More specific calibrations like re-centering, combination of interlaced ZIMPOL frames, or extraction of IFS spectra are also part of the pipeline. For subsequent analysis with user provided tools, all science recipes provide re-centered and calibrated duplicates of the raw-data cubes as an additional or optional output. This so-called pre-reduced output is actually the most heavily used feature of the pipeline.

\subsubsection{Further data handling tools}

Handling data of a SPHERE observation comprises much more than the processing of frames coming from the instrument's detectors. Beginning with the preparation of the observations in Phase 2, and ending with the analysis of large quantities of data in the context of the SHINE survey \citep{Chauvin2017}, a number of tools have been developed and deployed within the SPHERE consortium which are partially available to the general user. The two main tools are SPOT and the Data center.

SPOT is a scheduling tool developed to facilitate the preparation of observations for targets contained in a large database as is for example the case for the SHINE survey \citep{Lagrange2016}. SPOT can be fed with scientific priorities of each target plus a set of constraints such as required field rotation, or air mass or date restrictions. It will then produce both long and short term schedules plus final observing blocks ready for processing with ESO's P2 tool.

The SPHERE data center \citep[SDC;][]{Delorme2017} hosted at IPAG in Grenoble is used to run the full reduction and analysis of SPHERE data from both consortium GTO and open time data\footnote{\url{https://sphere.osug.fr/spip.php?rubrique16&lang=en}}. In order to facilitate this task, a full implementation of the pipeline as described above is available at SDC.  In addition, SDC is applying a custom conversion routine to the pre-reduced output to enable feeding the cubes to a full suite of analysis routines, the SpeCal pipeline \citep{Galicher2018}. Additionally, the SDC provides services for instrument monitoring and is able to connect to a number of data bases. SDC services are available to non-consortium P.I.s on demand, and SDC is aiming to provide analysis of all non-proprietary SPHERE data in the future.

\subsubsection{Community developments and future developments}

The field of high-contrast data analysis is progressing fast, and new approaches are regularly being developed in the community. A few examples include: a fast python pipeline for SPHERE/IRDIS data developed by P. Scicluna et al. (priv. communication), an adaptation of the CHARIS pipeline \citep{Brandt2017} to IFS data undertaken by M. Samland (priv. communication) or easy-to-use Python pre-processing environments for IRDIFS data\footnote{\url{https://github.com/avigan/VLTPF}}, a patch covariance method dedicated to ADI post-processing \citep{Flasseur2018}. The team at ETH Zurich has also developed an IDL-based pipeline for ZIMPOL data reduction for imaging and imaging polarimetry. Development that bring significant improvements in the quality of reduced data can be implemented rapidly in the SPHERE-DC. Incorporation into the official ESO pipeline is a bit more complex - new tools are more likely to appear in the context of the Reflex tool as a stand-alone actor than as a full-fledged pipeline plugin.

\section{Conclusions and prospects}
\label{sec:conclusions}

SPHERE is a highly optimized instrument dedicated to --but not limited to-- observing circum-stellar environments to look for, and study, young giant exoplanets and disks. The development of the instrument has faced many technological challenges but overall the performance is well within the original specifications and the number of scientific results based on SPHERE is steadily increasing\footnote{More than 120 papers at the time of writing, see the \href{http://telbib.eso.org/?instrument\%5B\%5D=SPHERE}{ESO Telescope Bibliography database}}. It is particularly important to note that all the proposed observing modes have been used and have produced high-quality results, which, in retrospect, demonstrate that the original design choices were entirely justified.

After four years of operations, the exoplanet search with SPHERE and GPI has yielded three new detections. Although disappointingly low, this number is in agreement with state-of-the-art predictions of population synthesis models \citep[e.g., ][]{Mordasini2017} that predict few giant planets in the 10-100\,au range, while the bulk of (scattered) giant planets would start to dominate in the 1--10\,au range. If it indeed exists, this population of planets could be within reach with SPHERE in the NIR provided a $\times$3-10 gain in contrast in the 50--200\,mas separation range. To reach such a gain, there are three angles of attack: (1) improve correction and control of non-common path aberrations, (2) coronagraphs that provide a better attenuation at smaller inner-working angle, and of course (3) a faster and more sensitive ExAO system.

One of the first items to address is the compensation of NCPA. SPHERE provides extremely low-order (tip and tilt) NCPA correction in parallel of the observations (on-line correction) thanks to the DTTS. The focus optimization is performed at the beginning of the night assuming that it remains sufficiently stable over a few hours, and higher orders are currently not measured or compensated. Gaining an order of magnitude in contrast will require a much finer calibration and possibly on-line compensation of all measurable NCPAs. While significant efforts have already been put into using the ZELDA wavefront sensor for such a monitoring and compensation (\citealt{N'Diaye2013,N'Diaye2016,Vigan2018b}; Vigan et al. in prep.), there are many other alternatives to perform wavefront sensing and stabilization, either using dedicated physical devices \citep{Por2016,Singh2017,Wilby2017}, specific algorithms \citep{Paul2013,Huby2015,Herscovici-Schiller2018}, active manipulation of the speckles \citep{Martinache2014,Bottom2016,Delorme2016} or a mix of all these. 

Once the NCPAs are under control, it is possible to imagine moving toward coronagraphs optimized for small IWA. Here again there are many different possibilities that all present their respective advantages and drawbacks in terms of IWA, band-pass, manufacturing, polarimetric requirements, etc \citep[e.g., ][]{Guyon2003,Mawet2005,Mawet2009,Kenworthy2010,Snik2012,Carlotti2013,Mawet2013,N'Diaye2016b,Otten2017,N'Diaye2018}. A careful trade-off study will be required to ensure an important gain in contrast while maintaining some of the unique features of SPHERE. In particular most coronagraphs can be optimized for a spectral band-pass up to $\sim$20\%, that is more or less a full NIR band, but very few concepts can enable working over several spectral bands. This means that the unique \texttt{IRDIFS-EXT} mode, which covers simultaneously from 0.95 up to 2.3\,\mic, may not be maintained or offered with all coronagraphic setups in a future upgraded system.

The third pillar to increase the contrast at very small separation is an upgrade of SAXO, which will enable the full potential of the NCPA and coronagraph upgrades. Improving the contrast performance close to the star means reducing the wavefront residuals close to the optical axis. This area is mainly driven by two contributors in the AO error budget: temporal error and noise error \citep[e.g., ][]{Fusco2006}. The goal of an ExAO upgrade consists therefore in decreasing both of these terms. The first idea points toward a faster AO loop, while keeping a $\sim$2 frames delay system. The residual being inversely proportional to the square of the loop bandwidth, increasing the speed of the ExAO by a factor two ($\sim$3\,kHz) will reduce the residuals by a factor 4, hence increasing the contrast by the same amount. Of course this improvement will happen at most during episodes of small coherence time. The second contributor being the noise, a more sensitive WFS would improve the performance close the star. Switching to a pyramid WFS \citep{Ragazzoni1996} therefore seems to be an interesting track to follow because of the increased sensitivity of this type of sensor \citep{Verinaud2004}. With these technical considerations in mind, there are two possible upgrade paths for SAXO. In the first one, the existing WFS and RTC are replaced with an upgraded, faster versions. This has the main advantage of benefiting the complete SPHERE system, both in VIS and NIR, but it implies major modifications of the RTC software and possibly of some hardware (HODM, ITTM, CPI optics), resulting in a possibly long down time for SPHERE. The second upgrade path, more focused on the exoplanet search in NIR, would be the addition of a second stage AO system in the NIR path. It would be composed of an IR pyramid WFS in the $J$-band coupled to a 20$\times$20 or 30$\times$30 HODM and run by a dedicated RTC at high frame rate (3\,kHz or more). This option would provide SPHERE with a factor four gain in contrast performance close to the axis, but would only concern the NIR part of the instrument. The final choice will directly depend on the main science drivers of the upgrade.

In parallel or on top of these necessary upgrades, other important developments are being considered for a SPHERE upgrade. One of them is the possibility to provide access to much higher spectral resolution than currently available. High-spectral resolution techniques have long been thought as a means of boosting the sensitivity in direct imaging \citep{Sparks2002,Riaud2007} with the capacity of disentangling the stellar and planetary signals thanks to resolved spectral lines. These techniques were beautifully demonstrated first on transiting hot Jupiters \citep{Snellen2010}, then on non-transiting planets \citep{Brogi2012}, and finally on directly imaged exoplanets \citep{Snellen2014}. These positive observational results have spurred several new works suggesting that high-contrast imaging coupled to high-resolution spectroscopy could be a key to detecting --and characterizing!-- Earth-like planets \citep{Snellen2015,Wang2017}. Dedicated coronagraph designs specifically optimized to be coupled with fiber-fed high-resolution spectrographs have also recently been proposed \citep{Por2018,Ruane2018}.

Coupling SPHERE with existing high-resolution spectrographs at the VLT has recently been proposed either in the visible with ESPRESSO \citep{Lovis2017} or in the near-infrared with CRIRES+ \citep{Vigan2018a}. The former proposition is focused on the detection of the light reflected from planets orbiting around extremely nearby stars like Proxima Cen~b \citep{Anglada-Escude2016}, but it would require a complete overhaul of SAXO and of coronagraphs in the visible arm. The latter proposition is focused on the detailed characterization of all the known directly imaged exoplanets and does not require any upgrade of SPHERE to be implemented. In both propositions, the global idea is to sample the focal plane coronagraphic image with several single-mode fibers that are used to transmit the light to the spectrographs. Standalone spectrographs \citep[e.g., ][]{Bourdarot2018} optimized from the start for diffraction limited beams in place of ESPRESSO or CRIRES+ are also an alternative, and mini-IFU systems based on fiber bundles have also recently seen important developments that make them attractive possibilities to benefit from both increased spectral resolution and spatial resolution \citep{Por2018,Haffert2018a,Haffert2018b}.

Finally, another idea under consideration is a very fast visible imager that would enable lucky-imaging in ExAO data and possibly provide a significant gain in sensitivity at small separations \citep{LiCausi2017}. An important science niche of such a system would be the detection of accreting young objects that present strong H$\alpha$ emission, for instance LkCa~15 \citep{Sallum2015} or PDS\,70 \citep{Wagner2018}.

In a longer perspective, it is clear that the science driver for exoplanet studies very strongly motivates much further developments. One major goal is a better characterization (high spectral resolution, high signal-to-noise monitoring of orbits and photometry, polarimetry). A second one is the detection capability at closer separation (<100 mas) and better (typically a factor 10 to 100) contrast to reach the planets in reflected light in the habitable zone around nearby M stars. This will most likely be addressed on upcoming extremely large telescopes (ELT). New challenges will include the segmented unfriendly pupils, the atmospheric conditions and the opto-mechanical limited stability of these huge structures. A third major goal will be the detection of such planets around solar-type stars, which means a larger separation for the habitable zone, but also a much higher contrast (>$10^9-10^{10}$). This should certainly involve large space-based telescopes   \citep{HABEX2018,LUVOIR2018} with another set of completely new issues including amplitude and phase error control, chromaticity control at an unprecedented level, coupled with dedicated coronagraphic devices \citep{Ruane2018} and signal processing. The way toward such ambitious goals is definitely defined on the experience gained on current instruments, and intermediate steps to implement, with increasing maturity, new system solutions and technological devices.

\begin{acknowledgements}
SPHERE is an instrument designed and built by a consortium consisting of IPAG (Grenoble, France), MPIA (Heidelberg, Germany), LAM (Marseille, France), LESIA (Paris, France), Laboratoire Lagrange (Nice, France), INAF - Osservatorio di Padova (Italy), Observatoire de Gen\`eve (Switzerland), ETH Z\"urich (Switzerland), NOVA (Netherlands), ONERA (France) and ASTRON (Netherlands) in collaboration with ESO. SPHERE was funded by ESO, with additional contributions from CNRS (France), MPIA (Germany), INAF (Italy), FINES (Switzerland) and NOVA (Netherlands). SPHERE also received funding from the European Commission Sixth and Seventh Framework Programmes as part of the Optical Infrared Coordination Network for Astronomy (OPTICON) under grant number RII3-Ct-2004-001566 for FP6 (2004-2008), grant number 226604 for FP7 (2009-2012) and grant number 312430 for FP7 (2013-2016). This work has made use of the SPHERE Data Center, jointly operated by OSUG/IPAG (Grenoble), PYTHEAS/LAM/CeSAM (Marseille), OCA/Lagrange (Nice), Observatoire de Paris/LESIA (Paris), and Observatoire de Lyon (OSUL/CRAL). AV acknowledges funding from the European Research Council (ERC) under the European Union's Horizon 2020 research and innovation programme (grant agreement No. 757561).
\end{acknowledgements}

\bibliographystyle{aa}
\bibliography{paper}

\appendix

\section{IFS and IRDIS detectors}
\label{sec:apdx:detectors}

Detector optimization included the choice of the read out mode for the Hawaii II-RG detector and minimization of a number of issues related to the detectors. As for IRDIS, the finally adopted read out modes included non destructive read-out for detector integration times (DITs) longer than 3\,s and double correlate read out for shorter DITs. A ramp effect was originally visible in low exposed images. It is likely due to transient capacity effects and it is sensitive to the exposure level causing lack of linearity. It was cured by setting the DIT delay at 0.2 sec. This setting causes a slight loss of efficiency with short DITs. A further annoying effect of Hawaii II-RG detectors is the electronic cross-talk between pixels read simultaneously; this effect is due to the multiplexer, and if not properly taken into account causes electronic ghosts with $\sim 0.5$\% intensities. It was reduced to much lower values by modifying the detector set-up (reducing the bias set up and making the read out slower; this changes the minimum DIT from 0.825 to 1.65\,s). A software correction was also introduced to finally reduce to negligible levels its impact; this correction is included in the standard DRH recipes. With the adopted configuration, the read out noise (RON) depends on the read mode and DIT time: it is 8.1\,e- for the shortest DIT (1.66\,s), and reduces at about 3.2 e- for 64 sec DITs. The gain is 1.87\,e-/ADU. Dark current is very low (0.003\,e-/s/pixel). The average ratio between the effective exposure time and the total time required for running the observation range from 0.66 to 0.84, depending on the length of DITs. Finally, Hawaii II-RG detectors suffer of a quite strong persistence effect from saturated images. Persistence scales down with the inverse of time. Given this functional form, there is no characteristic time and the level of persistence is not negligible even after a quite long time. Saturation of the detector should then be avoided as much as possible. A shutter is mounted at the entrance of the IFS to prevent excessive light to fall on the detector during the calibration procedures. Anyhow, careful examination of the images is required to avoid misinterpretation of signals. Detectors are linear up to about 35\,000 ADU. 

\begin{figure}
    \centering
    \includegraphics[width=0.5\textwidth]{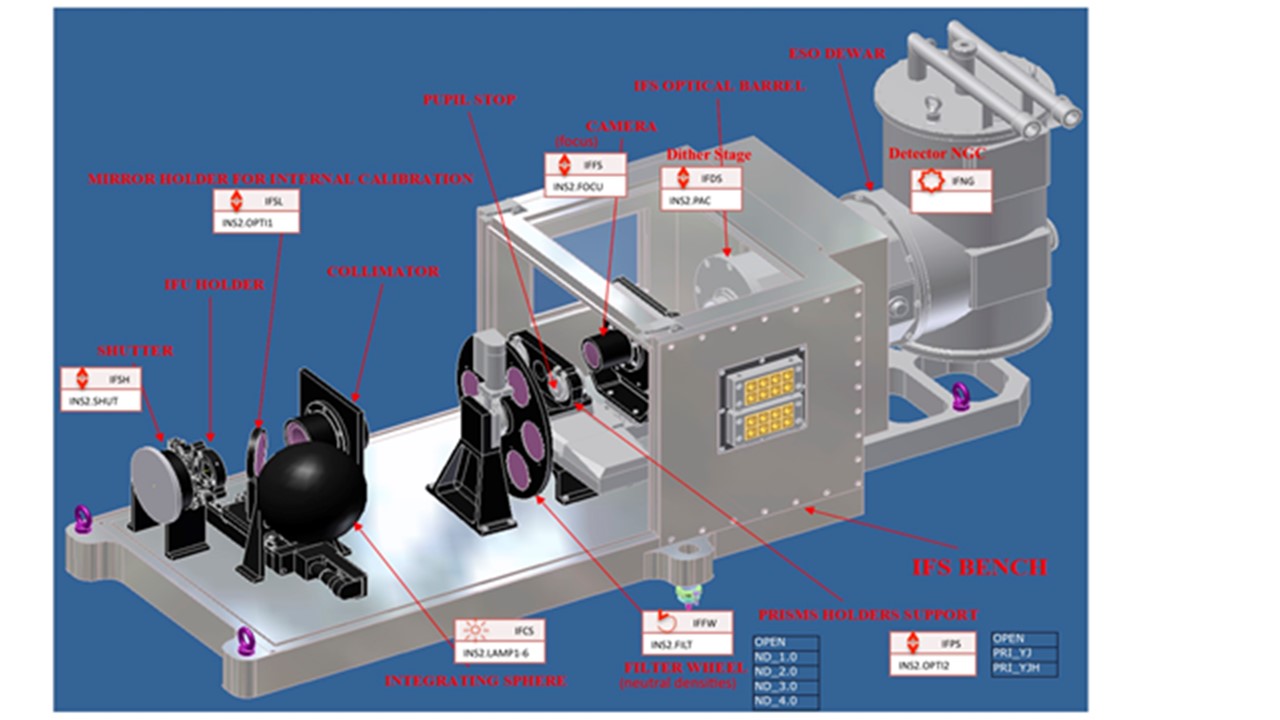}
    \caption{IFS layout with main components labeled}
    \label{fig:IFS_layout}
\end{figure}

\section{Detailed IFS description}
\label{sec:apdx:IFS_description}

An image of the SPHERE IFS is shown in Figure~\ref{fig:IFS_layout}, with the main components labeled. The design shares several of the basic principles of other IFS based on lenslet arrays (e.g., TIGER at the Canadian-French Hawaii Telescope, \citealt{Bacon1995}). The lenslet array is located on a telescope focal plane and each lenslet acts as a slit feeding an afocal system: light is collected by a collimator and is dispersed by a suitable device located in the intermediate pupil. A camera then focus light on the detector; the system magnification is set by the condition of Nyquist sampling the entrance slit on the detector. In most lenslet based IFU, the spectrograph's LSF is a microscopic image of the telescope pupil created by each lenslet. However, in diffraction-limited conditions each of these pupil images is actually an Airy disk determined by the finite size of the single lenslet; in this frame coherent and incoherent cross-talks scale down slowly with distance between lenslets. To reduce both, we conceived a new optical scheme for our IFU, the BIGRE \citep[see][]{Antichi2009}. In the BIGRE, each lenslet is actually itself an afocal system with two active surfaces (practically, two lenslet arrays with lenslets having the same size but different radius of curvature). The system creates an image of the first surface (that is, a focal plane), that is the effective entrance slit of the IFS; the de-magnification factor $K$\ of the lenslet afocal system allows to create room enough for each spectrum on the detector. To further reduce cross-talk, a mask located in the intermediate pupil cuts all high order diffraction rings of the spectrograph's LSF; the best size of the mask corresponds to the first minimum of the Airy disk at this position (at an intermediate wavelength). The final BIGRE LSF profile results to be correctly apodized and achromatic. Since the collimator is designed to be telecentric, the macroscopic intermediate pupil of the IFS is optically conjugated with the intermediate pupil of each microscopic IFU lenslet array; placing a mask on this location - just in front of the disperser - is then more practical. To further reduce cross-talk between lenslets, each one of them is masked to a circular aperture (avoiding the strong diffraction due to corners) by depositing a reflective layer at the edges of the lenslets. To improve efficiency, an hexagonal design was adopted, reducing the amount of masking required. Finally, the array is rotated with respect to the dispersion direction to allow longer spectra to be projected on the detector. This scheme is presented in detail in \citet{Antichi2008b,Antichi2009}, while fabrication details are given in \citet{Giro2008}. In these papers it was shown that this design allows about an order of magnitude lower cross-talk level than in traditional diffraction limited lenslet IFS.

The BIGRE lenslet array was constructed by Advanced MicroOptics Systems, who also took care of the alignment of the two arrays with respect each other to within 1.6\,\mic. It consists of an array of 145$\times$145 hexagonal lenslets (slightly oversized with respect to the detector area) on an INFRASIL substrate, with a pitch (=distance from centers of adjacent lenslets) of 161.5\,\mic providing the required sampling of 0.01225\as on the F/316 beam created by the IFS arm of the common path. The lenslets are masked to circular apertures of 158\,\mic. The lenslets of the first array have focal lengths of 4.58 mm, the second ones of 1.12 mm, providing a magnification factor of $K=4.1$. The BIGRE is rotated by $\sim$10.5\,degree with respect to the rest of the optical elements in order to provide room enough for the spectra on the detector (the spectra are actually accurately aligned along detector columns). The IFS camera and collimator are custom designed S-TiH11 and BaF2 dioptric systems manufactured by SILO, with effective focal lengths of 250 and 422.5 mm, respectively. The magnification is then 1.69 ensuring that the slit is projected onto two detector pixels. The overall optical quality is very high. The cross dispersers are two Amici direct vision prisms (Y-J and Y-H modes) that allows low and quite uniform dispersion \citep{Oliva2003}, as desired for the SPHERE IFS. They were also fabricated by SILO. A macroscopic mask with a diameter of 20.65 mm is located on the intermediate pupil position, to suppress high order diffraction by the microlenses. Filters constructed by JDSU defining the accepted wavelength range for each spectroscopic mode are also located close to this position. All optics are anti-reflection coated. The camera is mounted on a 20 mm slide in order to allow fine focusing on the detector, and on a two-dimension piezo stage by 200$\times$200\,\mic for dithering of the images on the detector. This function, implemented in order to provide best flat fielding accuracy, is actually very rarely used because experience shows that it does not improve significantly results.

The detector is a 2k$\times$2k Hawaii II-RG detector, with square pixels with a side of 18\,\mic. The detector has a very high sensitivity ($> 90$\,\%) over the wavelength range from $\sim$0.6\,\mic to $\sim$2.5\,\mic. It is housed in a custom-made dewar designed by ESO and cooled by a continuous flux liquid nitrogen system. The detector temperature is fixed at 80\,K by an active thermal control system. While the dark current is very low (0.003\,e$^-$/s), the sensitivity of the detector to thermal radiation in the environment needed proper consideration. A cold low-pass filter fabricated by JDSU, with a transmission lower than $10^{-4}$\ for wavelengths longer than 1.65\,\mic and larger than 90\% for shorter wavelengths, is located just 40\,mm in front of the detector; this filter was designed to work on the F/23.7 converging beam provided by the (warm) camera. To reduce the warm solid angle seen by this filter, the dewar window was located far ($\sim 110$\,mm) from the filter itself. Finally, two baffling systems were implemented: a very low-reflectivity cold baffle within the dewar with a special geometrical design minimizing surfaces that may allow light reflected only once or twice to reach the cold filter; and a spherical Narcissus reflecting surface centered on the cold filter and located outside the dewar. This mirror minimizes thermal emission from outside the dewar possibly reaching the cold filter passing through the dewar window. A thermal design of this whole cold filter and baffling system showed that it provides a thermal background of about 11.7\,photons/s/pixel for an ambient temperature of 12$^{\mathrm{o}}$C, that is typical of Paranal. This value was confirmed by measurements. Thermal background of IFS has no significant impact on its performances for sources with $J<8$, causing a loss smaller than 0.1\,mag in its limiting contrast; however, the impact is significant for very faint sources.

Light from a 6-inch Zenith gold inner coated integrating sphere may be inserted in the optical path by means of a 45 degree mirror mounted on a 70 mm slide after the microlens array and before the collimator; this arm is used to obtain full flat field of the IFS detector with various flat field lamps (either colored or white). The exit of this lamp is optically conjugated with the virtual slit plane provided by the IFU. Neutral density filters, fabricated by SILO, mounted on a 5-fold OWIS filter wheel located in the collimated portion of the beam allows tuning the exposure levels of these internal flat fields; these filters are properly inclined with respect to the optical axis to avoid ghosts. Additional calibrations (focus, lenslet flat field, spectrum positions, and wavelength calibration), requiring light passing through the lenslet array, are obtained using the facilities provided by the Common Path calibration arm. These are basic steps for extracting data-cubes from the raw images. Finally, additional on-sky calibrations are needed for precise astrometric and photometric calibrations.

The control electronics design was realized considering the standard specifications of ESO. All moving parts are remotely controlled and a modular approach allows easy maintenance and reliability of the instrument. For details, see \citet{DeCaprio2012}.

The DRH software (see Sect.~\ref{sec:drh}) takes care of the most relevant steps in data reduction: handling of bad pixels, background subtraction, detector flat field, identification of spectra and their extraction, wavelength calibration, and extractions of data cubes in x, y, $\lambda$. This last step is done using first an interpolation from pixels to constant wavelength steps, and then a bi-dimensional interpolation to pass from the hexagonal grid to a Cartesian one. The final product is a matrix of $290\times 290\times 39$\,pixels for both Y-J and Y-H modes, with a spatial scale of $\sim 7.46$\,mas/pixel. Software for fine astrometry (centering, correction for anamorphism, scale and true north) of the final data cubes runs at the Grenoble SPHERE Data Center \citep{Delorme2017}.

\section{IFS noise model}
\label{sec:apdx:noise_model_ifs}

To interpret results about contrast, we constructed a noise model considering four terms (calibration, photon noise, thermal background and read out noise).

The calibration error (we simplify here the complex dependence on angle discussed in the previous sub-section adopting a single power law dependence):
\begin{equation}
    p = 18.31 - \frac{0.62}{s} + 3.0 \log{\frac{a}{60}}.
\end{equation}
    
\noindent The photon noise error:
\begin{equation}
    q = -0.5 J + 18.31 - \frac{0.13}{s} - 1.25 \log{\frac{t_{exp}}{3600}}.
\end{equation}
    
\noindent The thermal background error:
\begin{equation}
    r = 20.7 - J - 1.25 \log{\frac{t_{exp}}{3600}}.
\end{equation}
        
\noindent And the read out noise error:
\begin{equation}
    u = 20.7 + 1.25 \frac{\log{D}}{4.6} - J - 1.25 \log{\frac{t_{exp}}{3600}}.
\end{equation}

In these equations, $s$\ is the separation (in arcsec), $a$\ is the field rotation angle (in degree),
$J$\  is the J-magnitude of the star, $t_{exp}$ is the total exposure time (in s) and $D$ is the
DIT length (in s). 

The final expected contrast C (in mag) is obtained by combining the various noise sources:

\begin{equation}
    C = -2.5 \log\left(\sqrt{10^{-0.8 p} + 10^{-0.8 q} + 10^{-0.8 r} + 10^{-0.8 u}} \, \right) \, - 0.55 \, (\sigma -1) ,
\end{equation}

\noindent where $\sigma$\ is the ESO-DIMM seeing FWHM.

\end{document}